\documentclass[10pt]{article}
\usepackage{multicol}
\usepackage{simpleConference}
\usepackage{arxiv}
\usepackage[backend=bibtex, style=chem-rsc, autocite=superscript]{biblatex}
\let\tempautocite\autocite
\renewcommand{\autocite}[1]{\tempautocite[][]{#1}{}}
\usepackage[version=3]{mhchem}
\usepackage{bm} 
\usepackage{float}
\usepackage{fancyhdr} 
\usepackage{comment}

\usepackage[utf8]{inputenc} 
\usepackage[T1]{fontenc}    
\usepackage{hyperref}       
\usepackage{url}            
\usepackage{booktabs}       
\usepackage{amsfonts}       
\usepackage{nicefrac}       
\usepackage{microtype}      
\usepackage{lipsum}		
\usepackage{graphicx}
\usepackage{float}
\usepackage{subfig}
\usepackage{caption}
\usepackage{xcolor}
\usepackage{longtable}
\newcommand\blfootnote[1]{%
  \begingroup
  \renewcommand\thefootnote{}\footnotetext{#1}%
  \addtocounter{footnote}{-1}%
  \endgroup
}
\usepackage{fontawesome5}
\usepackage{ulem}
\def\dout{\bgroup
 \markoverwith{\lower-0.4ex\hbox
 {\kern-.03em\vbox{\hrule width.2em\kern0.25ex\hrule}\kern-.03em}}%
 \ULon}
\MakeRobust\dout
\usepackage[perpage,hang,bottom]{footmisc}
\makeatletter
\newcommand*{\shifttext}[2]{%
  \settowidth{\@tempdima}{#2}%
  \makebox[\@tempdima]{\hspace*{#1}#2}%
}
\makeatother
\newcommand{\darr}{\shifttext{0.1ex}{$\downarrow$}}
\renewcommand{\thefootnote}{\ifcase\value{footnote}\or \darr  \or \sout{\darr} \or \dout{\darr} \fi}
\usepackage{perpage} 
\MakePerPage{footnote} 

\date{January 5, 2023}
\begin{document}

\title{Optimised Morse transform of a Gaussian process feature space}

\author{Fabio E. A.~Albertani\textsuperscript{$\ast \dagger$}\ , \ Alex J. W.~Thom\textsuperscript{$\ast$}}
\thispagestyle{empty}

  	\maketitle
	\begin{abstract}
Morse projections are well-known in chemistry and allow one,  within a Morse potential approximation,  to redefine the potential in a simple quadratic form.  The latter,  being a non-linear transform,  is also very helpful for machine learning methods as they improve the performance of models by projecting the feature space onto more well-suited coordinates.  Usually,  the Morse projection parameters are taken from numerical benchmarks.  We investigate the effect of changing these parameters latter on the model learning, as well as using the machine learning method itself to make the parameters decision. We find that learning is not necessarily improved by the latter and that general Morse projections are extremely susceptible to changes in the training data.
	\end{abstract}
	\blfootnote{\textsuperscript{$\ast$}\ Yusuf Hamied Department of Chemistry,  University of Cambridge,  Cambridge,  Lensfield Road,  CB2 1EW}
	\blfootnote{\textsuperscript{$\dagger$}\ fa381@cam.ac.uk}
	\setcounter{footnote}{0}
\begin{multicols}{2}

\section{Introduction}
Machine learning,  as in many fields of science,  has revolutionised the way theoretical chemists approach the interpolation of molecular properties.  The many methods encompassed by the machine learning framework provide tools to construct models of the former with great accuracy\autocite{Behler2016,Noe2020,Deringer2021}.  A particular method that has seen success is the Gaussian process (GP) framework which has seen extensive publications in machine learning potential energy surface applications\autocite{Bartok2015,Das2015,Cresswell2016,Cui2016,Uteva2017,Kolb2017,Uteva2018,Dragoni2018, Dai2020}. 
\par
The representation of the molecular geometry is an essential part of the ML building process and has seen many ``solutions'' spring up through the years\autocite{Langer2003}.  When using global or local descriptors of the atomic configuration of the system to build a ``feature space'',  one often uses the internuclear distances as an underlying coordinates.  The latter are often transformed to improve the accuracy of models since ML models do not perform equally when the training data is projected onto different feature spaces.  One known projection of the feature space is the Morse transform of the internuclear distances which often improves one's ability to learn the surface\autocite{Qu2018}.
\par
Given the ability of a GPs to learn the underlying pattern of the target function\autocite{Sacks1989, Book:Rasmussen2005},  it is interesting to consider a GP which can change the underlying function in its optimisation.  This is done by making the distance fed to the kernel (see next section) transform with GP hyperparameters within the kernel itself.
\par
Many more feature space transformations could be considered (these are also not restricted to transformations based on internuclear distances) but we will here discuss the effect of the added ``transformation hyperparameters'' on the GP optimisation process.

\section{Gaussian Processes\label{sec:GAP}}
A Gaussian process is a machine learning regression method and is defined as \textit{a collection of random variables, any finite number of which have a joint Gaussian distribution}\autocite{Book:Rasmussen2005}.  An essential part of a GP model is its kernel function which defines,  over a feature space (the input space of the GP),  a measure of similarity.
\par
There are many possible kernel functions one can defined,  as they only need to adhere to a few simple rules\autocite{Book:Rasmussen2005}.  We use here the Mat\'{e}rn class kernel multiplied by a constant kernel (CK) and summed with a White Kernel (WK) to model noise.  The covariance between two vectors over the feature space,  $\mathbf{X}$ and $\mathbf{X}'$ here,  is given by
\begin{equation}
\mathrm{K}(\mathbf{X}, \mathbf{X}') = \sigma^2\ \frac{2^{1-\nu}}{\Gamma(\nu)}\Bigg(\sqrt{2\nu}\frac{d}{\rho}\Bigg)^\nu K_\nu\Bigg(\sqrt{2\nu}\frac{d}{\rho}\Bigg) + \lambda^2
\label{eq:covariance-Mat\'{e}rn}
\end{equation}
where $\Gamma$ is the gamma function, $K_\nu$ is the modified Bessel function of the second kind of degree $\nu$, $\rho$ are length scales and $d$ is the Euclidean distance in feature space $|\mathbf{X}-\mathbf{X}'|$.  The $\nu$ parameter is not optimised and defines the smoothness of the kernel: a GP with a Mat\'{e}rn kernel of parameter $\nu = n + 0.5$ is $n$-times differentiable\footnote{For example,  with $\nu=2.5$,  one ensures that the GP latent function is physical since both the atomic forces (first derivative) and atomic Hessians (second derivative) are smooth w.r.t.  geometrical changes}.  We also explore an infinitely smooth version of the Mat\'{e}rn kernel with $\nu \to \infty$,  commonly known as the radial basis function (RBF) kernel.
\par
At a set of query points,  forming a matrix $\mathbf{X}_p$ of size $N_p \times N_{\mathrm{features}}$,  a GP model predicts a Gaussian distribution with a mean (sometimes called the latent function),  here denoted $y(\mathbf{X}_p)$,  and a variance,  here denoted $\Delta(\mathbf{X}_p)$,  which is associated to the model confidence.  For a set of prediction points,  $\mathbf{X}_p$,  the predicted distribution are given by\autocite{Book:Rasmussen2005}:
\begin{equation}
\begin{gathered}
y (\mathbf{X}_p) = \mathbf{K}_{pt} \ \mathbf{K}^{-1}_{tt} \ \mathbf{y} \\
\Delta (\mathbf{X}_p) = \mathbf{K}_{pp} \ - \mathbf{K}_{pt} \ \mathbf{K}_{tt}^{-1} \ \mathbf{K}_{tp}
\label{eq:gp-maineq}
\end{gathered}
\end{equation}
where the kernel matrices are subscripted with the matrices they evaluate ($p$ for query points and $t$ for training) and the $ij$\textsuperscript{th} element of the matrix $\mathbf{K}_{nm}$ is given by $\mathrm{K}(\mathbf{X}_{n,i}, \mathbf{X}_{m,j})$.  A common metric,  used by the ML community,  to define the confidence in predictions is the $\Delta_{95\%}$ confidence interval which is given as $y \pm 2 \Delta$ for GPs.
\par
GPs are optimised by finding the most suited hyperparameters for its kernel.  Using a Bayesian approach,  one finds the latter by maximising the log-marginal likelihood (LML) defined as\autocite{Book:Rasmussen2005}
\begin{equation}
\mathrm{LML} = -\frac{1}{2} \mathbf{y}^{\mathrm{T}} \mathbf{K}_{tt}^{-1} \mathbf{y} - \frac{1}{2} \mathrm{log} |\mathbf{K}_{tt}| - \frac{n}{2} \mathrm{log} (2\pi)
\label{eq:lml}
\end{equation}
where $\mathbf{K}_{tt}$,  as before,  is the covariance matrix of the training set to itself.  The terms on the LHS of equation \ref{eq:lml} can be understood as a fit,  a regularisation and a normalisation term respectively.
\par
Practically,  the maximisation is done by minimising $-\mathrm{LML}$ but we will use the term LML as the surface we minimise and the term ``minimum'' as a set of hyperparameters corresponding to a model selected by the GP. 
\par
The LML exploration is done with the GMIN suite\autocite{GMIN,OPTIM,PATHSAMPLE} which allows to give a full description of the minima,  both global and local,  of the surface as well as their connectivity.  In order to visualise the surfaces,  we use disconnectivity graphs\autocite{Becker1997,Wales1998,disconnectionDPS} which represent,  on a $-$LML vertical scale,  minima as vertical lines connected by transition states shown by connecting those lines.
  
\section{Methodology}
If one takes the coordinates to be specified as a vector,  $\mathbf{X}$,  of $N(N-1)/2$ internuclear distances,  then the Morse transformed coordinates form a vector defined as
\vspace{0.2cm}
\begin{equation}
\begin{aligned}
&\mathcal{T}(\mathbf{X}; M=\{ \alpha,\mathrm{X}_0\} ): \\
&\mathbb{R}^{N(N-1)/2} &&\to \: \mathbb{R}^{N(N-1)/2}\\
& \mathrm{X}_{i} &&\mapsto \: \mathrm{exp}\big( -(\mathrm{X}_{i} - \mathrm{X}_0) / \alpha \big)
\end{aligned}
\label{eq:morse-transform}
\end{equation}
\vspace{0.2cm}

where $\alpha$ is the Morse parameter and $\mathrm{X}_0$ is the Morse shift parameter.  In order to simplify the notation,  we will write the Morse transformed vector $\mathbf{X}_J$ as $^\mathcal{T}\mathbf{X}_J$.
\par
The reasoning behind this transform is that an analytical Morse potential becomes quadratic when projected onto the coordinates $^\mathcal{T}\mathbf{X}$.  If one considers non-analytical potentials,  one expects the potentials to closer to quadratic in $^\mathcal{T}\mathbf{X}$ than $\mathbf{X}$.  The simpler PES is better described by a GP since the length scale of the problem becomes more ``unique''.  Despite some specific bonds having chemically derived optimal Morse parameters,  there is not always a straight-forward way to select those parameters.  There are two ways of optimising those parameters: a numerical optimisation to reduce the error on a testing set (the traditional ``best-fit'' approach) and a Bayesian approach with a Morse hyperparameter.  As one does not want the number of hyperparameters to be too large and optimise in a very large space,  we will set Morse hyparparameters to be equal for all feature dimensions.
\par
Taking a basic RBF kernel\autocite{Book:Rasmussen2005},  on a Morse transformed feature space,  the kernel is evaluated as
\vspace{0.2cm}
\begin{equation}
\begin{aligned}
\mathrm{K}(&\tilde{\mathbf{X}}_A=  {}^\mathcal{T}\mathbf{X}_A,  \tilde{\mathbf{X}}_B=  {}^\mathcal{T}\mathbf{X}_B; \boldsymbol{\rho}) = \\
&\mathrm{exp}\Big( -\frac{1}{2} (\tilde{\mathbf{X}}_A - \tilde{\mathbf{X}}_B)^{\mathrm{T}}
\mathbf{P}
(\tilde{\mathbf{X}}_A - \tilde{\mathbf{X}}_B)\Big)\\
&\mathrm{where} \ \mathbf{P} = \mathbf{I}_n 
\begin{bmatrix}
\rho_1^{-2}\\[0.1cm]
\rho_2^{-2}\\[0.1cm]
\vdots\\[0.1cm]
\rho_n^{-2}\\
\end{bmatrix}
\end{aligned}
\label{eq:morserbf-kernel-not}
\end{equation}
\vspace{0.2cm}

where we now used the matrix notation of the RBF kernel and where $\rho_i$ are the length scales along each feature dimension.  We use here the $\tilde{\mathbf{X}}_J$ notation to differentiate the fixed Morse projection of the kernel input (the Morse parameters do not appear in the evaluation of the kernel in the LHS of equation \ref{eq:morserbf-kernel-not}) from the projection taken within the kernel itself,  like in equation \ref{eq:cov-morse-rbf}.
\par
Instead of equation \ref{eq:morserbf-kernel-not},  one can use the internuclear distances,  $\mathbf{X}$ as an input and optimise the Morse parameters inside a ``MorseRBF'' kernel which is evaluated as
\begin{equation}
\begin{aligned}
&\mathrm{K}(\mathbf{X}_A,  \mathbf{X}_B; M,  \boldsymbol{\rho}) =\\
&\mathrm{exp}\Big( -\frac{1}{2} (  {}^\mathcal{T}\mathbf{X}_A -  {}^\mathcal{T}\mathbf{X}_B)^{\mathrm{T}} 
\ \mathbf{P}\ ( {}^\mathcal{T}\mathbf{X}_A - {}^\mathcal{T}\mathbf{X}_B)\Big)\\
&\not \equiv \mathrm{exp}\Big( -\frac{1}{2} (\tilde{\mathbf{X}}_A - \tilde{\mathbf{X}}_B)^{\mathrm{T}}
\ \mathbf{P}\ (\tilde{\mathbf{X}}_A - \tilde{\mathbf{X}}_B)\Big)
\end{aligned}
\label{eq:cov-morse-rbf}
\end{equation}
where $\mathbf{P}$ is the same matrix as the one in equation \ref{eq:morserbf-kernel-not}.  In the MorseRBF approach,  the Morse parameters are hyperparameters of the kernel alongside the length scales.  The interesting approach of the kernel is that it does not,  as is common practice,  optimise to the Morse parameters that minimises the error on the testing set,  the ``best-fit'' approach,  but instead uses a Bayesian approach and a ``statistically relevant'' set of Morse parameters. 
\par
\begin{figure}[H]
	\centering
	\captionsetup[subfigure]{labelformat=empty}
	\subfloat[$\mathrm{X}_0= 0$]{\includegraphics[width=0.17\textwidth]{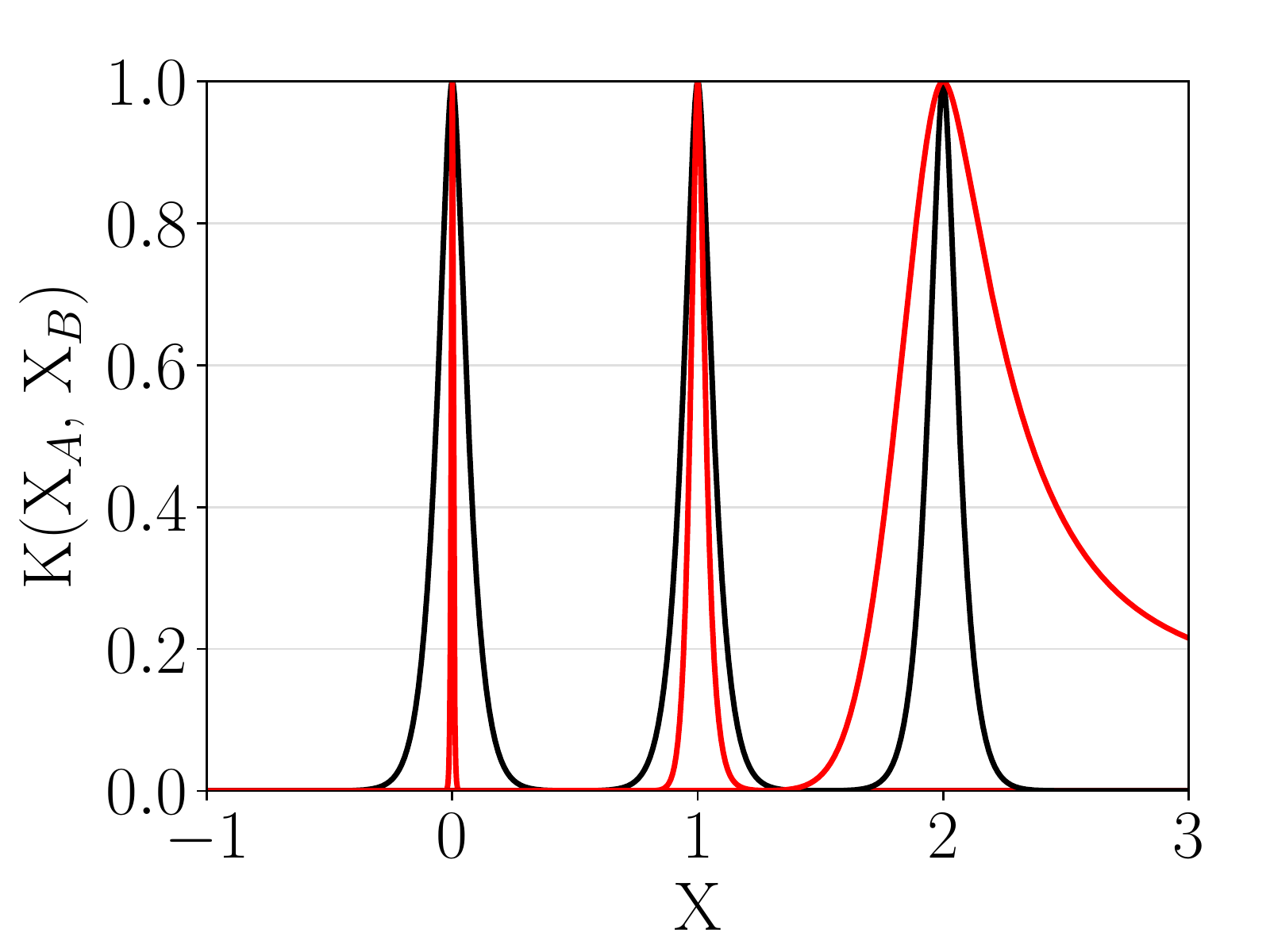}}
	\subfloat[$\mathrm{X}_0=1$]{\includegraphics[width=0.17\textwidth]{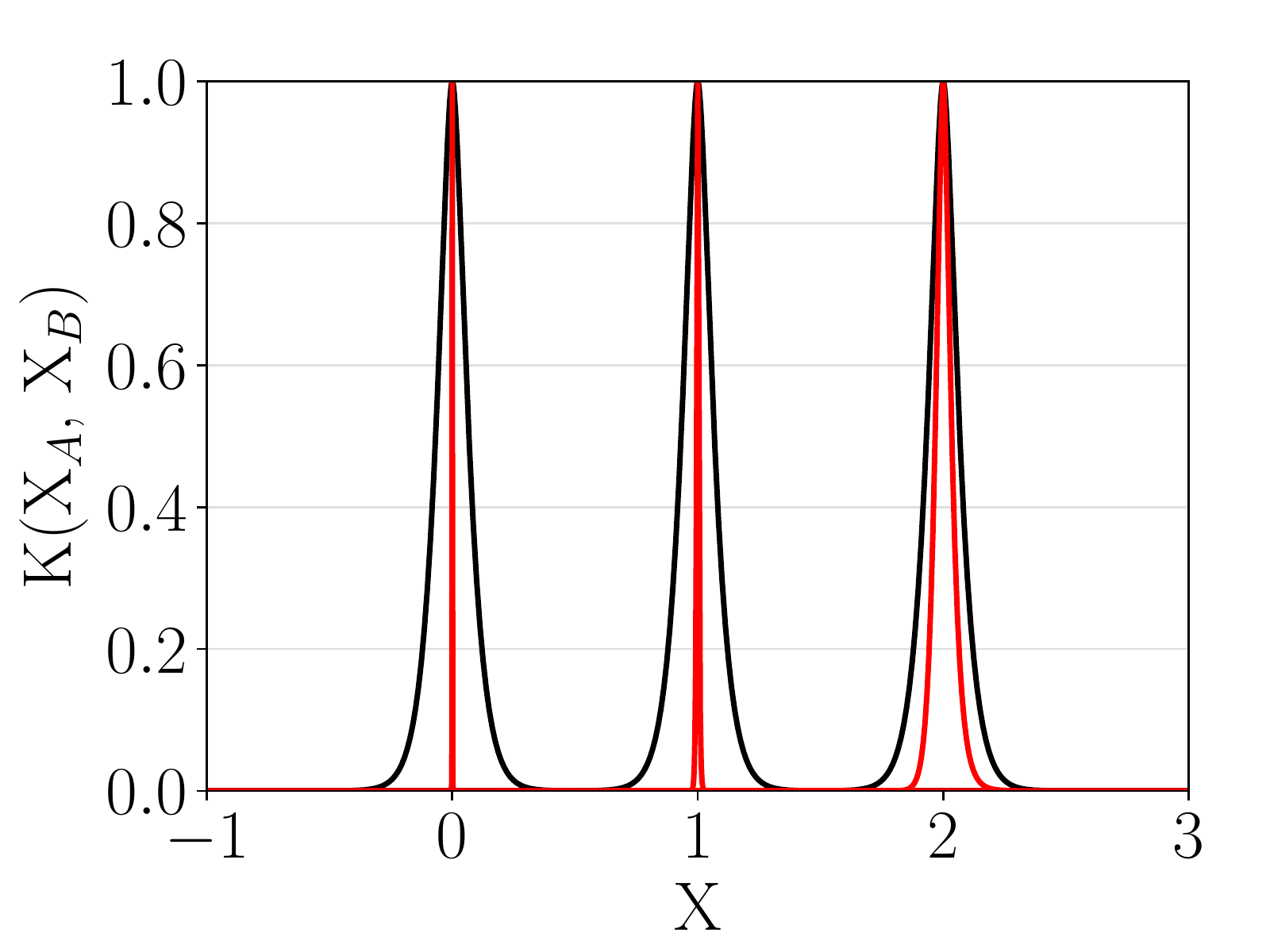}}
	\subfloat[$\mathrm{X}_0=2$]{\includegraphics[width=0.17\textwidth]{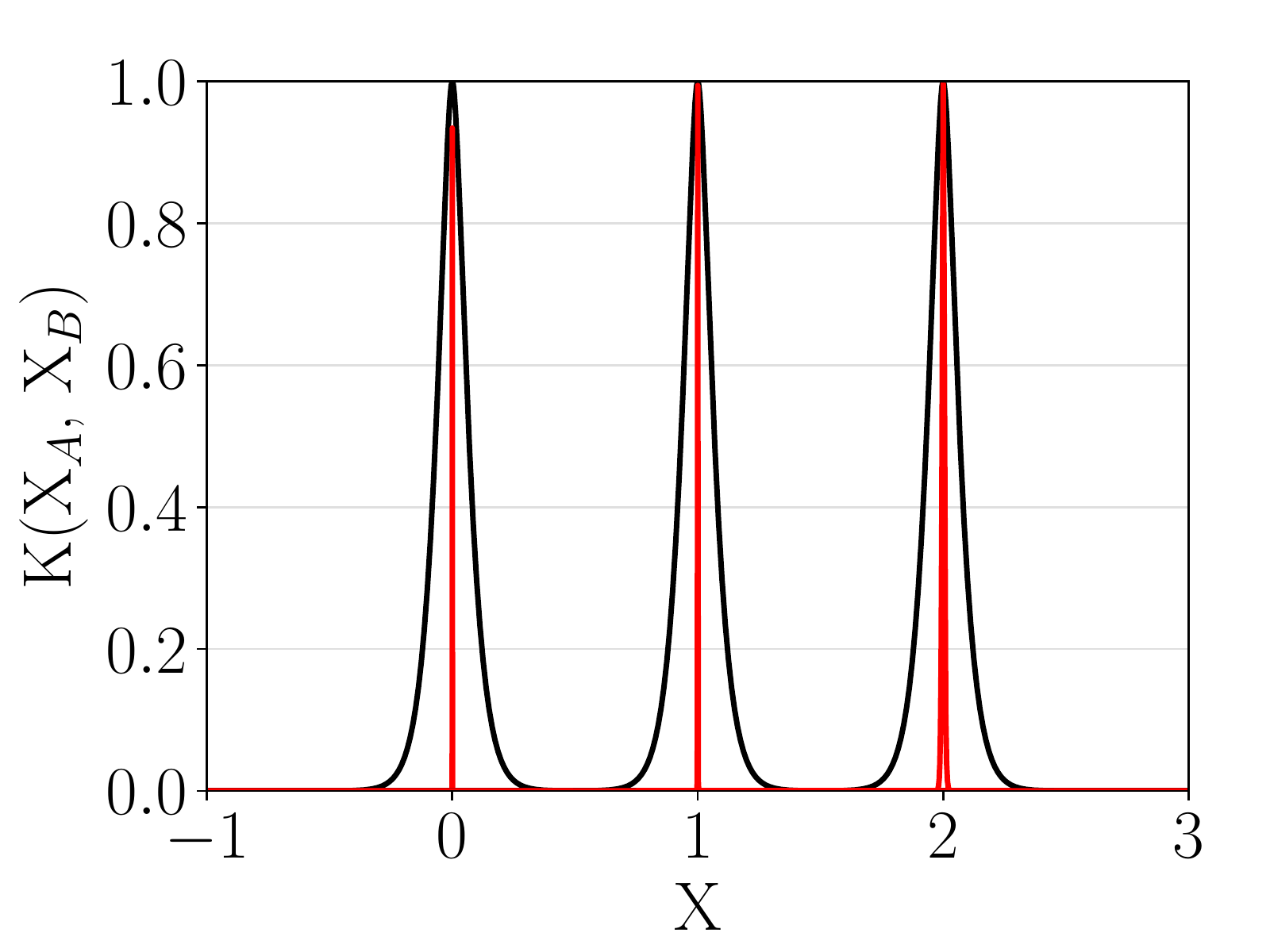}}
	\caption[Morse transformed covariance functions for different Morse parameters.]{Covariance of the Mat\'{e}rn ($\nu=2.5$) kernel (black lines) compared to the MorseMat\'{e}rn kernel (red lines), projected back onto X,  for different $\mathrm{X}_0$ and with $\alpha=2.0$.  The covariance is quite unsymmetrical an the forward influence is greater than the backward influence since the transform expands the dataset at large X values.  The $\mathrm{X}_0$ parameter dampens the strong ``elongation'' of the covariance at small X values and also strongly contracts the covariance extent at X$<\mathrm{X}_0$ where the exponent in equation \ref{eq:morse-transform} becomes positive. }
	\label{fig:cov-alpha}
\end{figure}
\begin{figure}[H]
	\centering
	\captionsetup[subfigure]{labelformat=empty}
	\subfloat[$ \mathrm{X}_0=0$]{\includegraphics[width=0.17\textwidth]{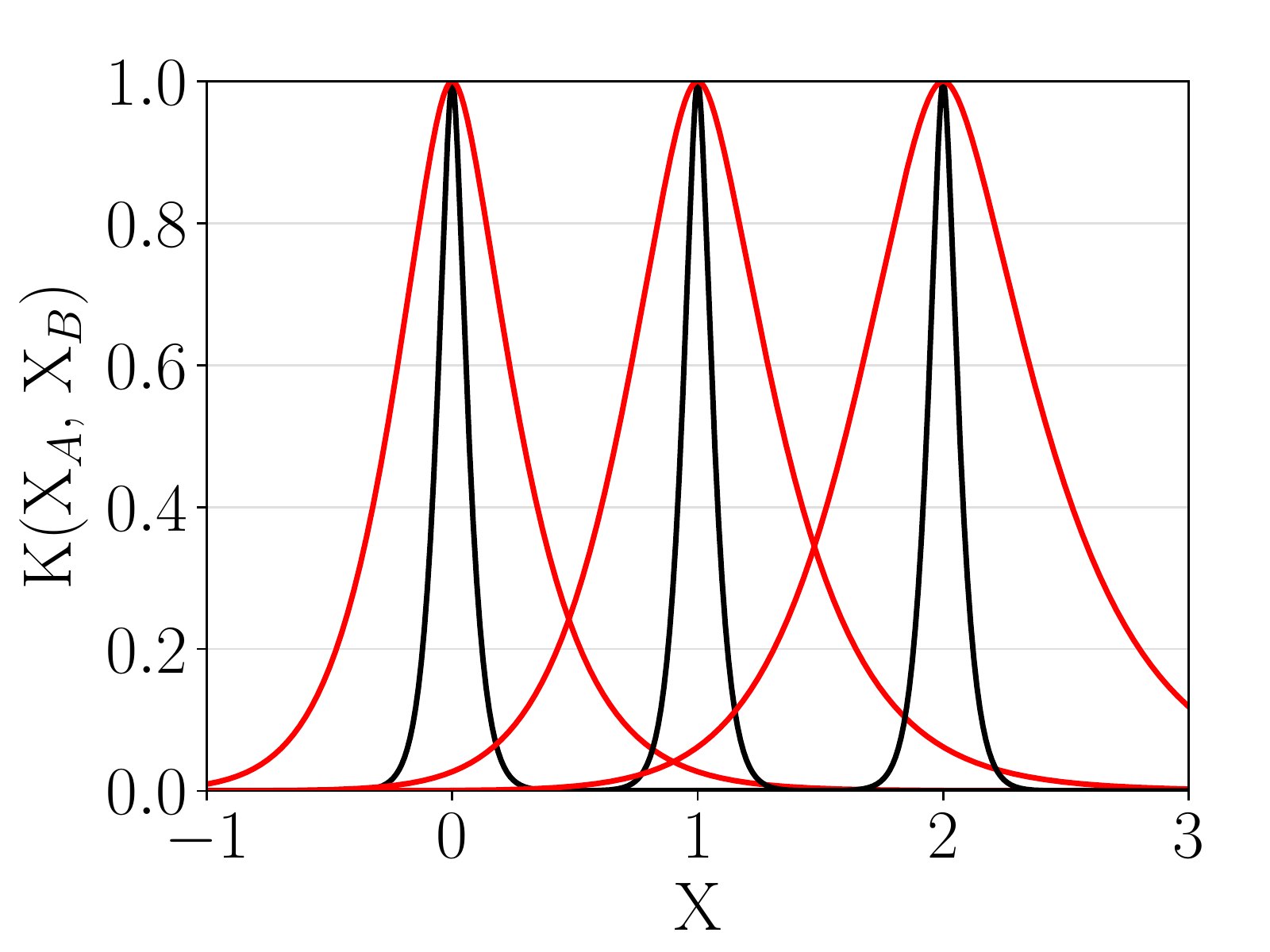}}
	\subfloat[$\mathrm{X}_0=1$]{\includegraphics[width=0.17\textwidth]{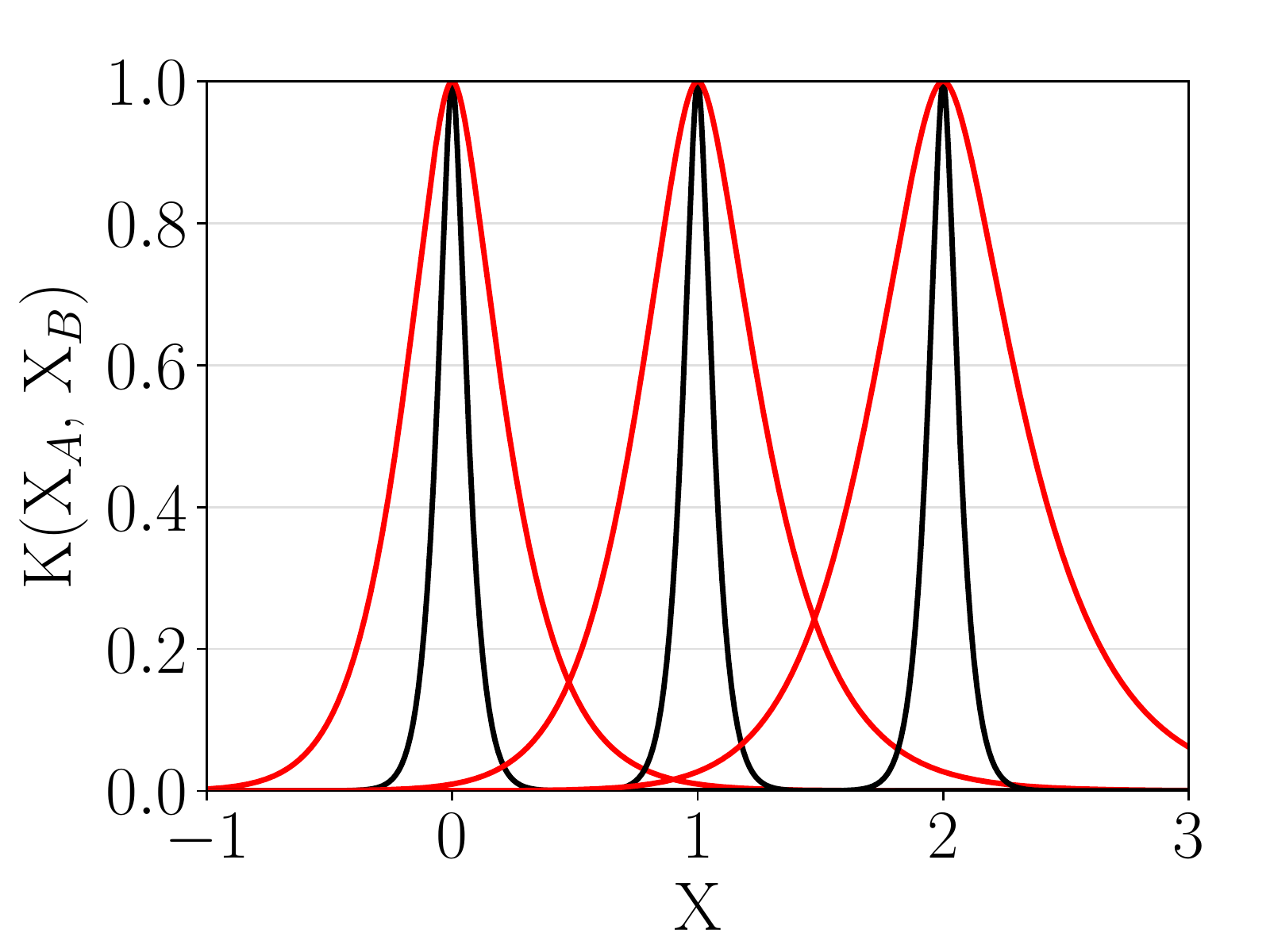}}
	\subfloat[$\mathrm{X}_0=2$]{\includegraphics[width=0.17\textwidth]{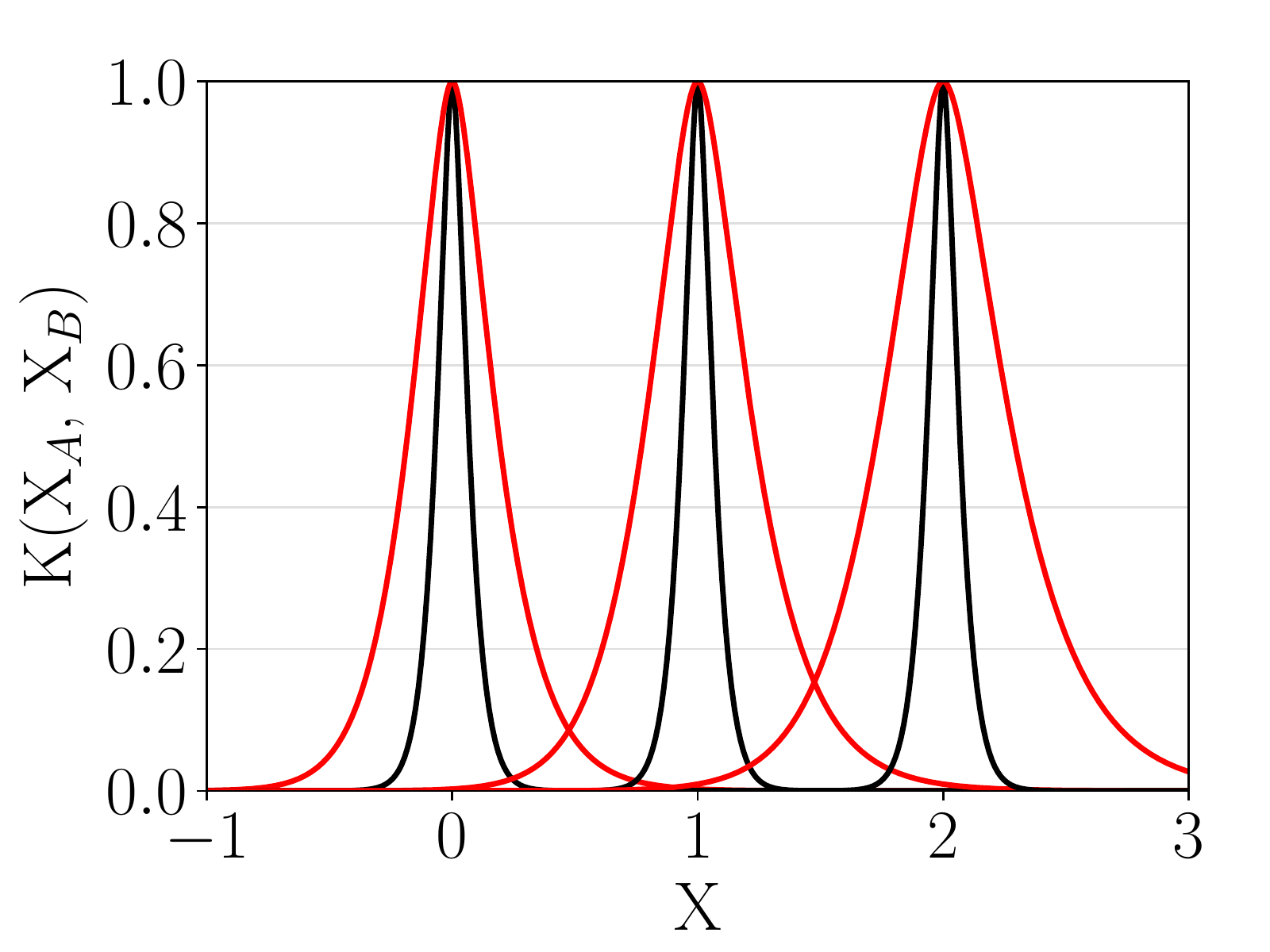}}
	\caption[Morse transformed covariance functions for different Morse parameters.]{Covariance of the Mat\'{e}rn ($\nu=2.5$) kernel (black lines) compared to the MorseMat\'{e}rn kernel (red lines), projected back onto X,  for a larger $\alpha=5.0$.  As opposed to figure \ref{fig:cov-alpha},  the effect of the $\mathrm{X}_0$ does not affect the covariance as much.  The widening seen from the previous figure is just a consequence of the length scale hyperparameter being equal since one cannot associate the Morse transformed length scale to the linear space one. }
	\label{fig:cov-alpha2}
\end{figure}

Regarding the $\mathrm{X}_0$ parameter,  one can see in figure \ref{fig:cov-alpha}-\ref{fig:cov-alpha2} that the covariance function drops very quickly for data points in the region $X<\mathrm{X}_0$ which could potentially lead to a loss of information.  With very compressed kernel length scales around $X<\mathrm{X}_0$,  data will not affect the latent function.  Learning was also performed for those surfaces but were not considered further in this study.
\par
The derivatives of the kernel with respect to each hyperparameter can be obtained analytically.  The derivatives with respect to the length scales are not affected by the Morse transform and are equivalent to simply changing the feature space in the derivatives of the standard RBF kernel.  The derivative with respect to the Morse parameter can also be analytical obtained and is given by
\vspace{0.2cm}
\begin{equation}
\begin{aligned}
&\partial_{\alpha} \mathrm{K}(\mathbf{X}_A,  \mathbf{X}_B) =\\
&\Big( -\frac{1}{2} \partial_{\alpha}\Big[ \big( {}^\mathcal{T}\mathbf{X}_A -  {}^\mathcal{T}\mathbf{X}_B \big)^{\mathrm{T}}\Big]
\ \mathbf{P}\\
&\qquad \big( {}^\mathcal{T}\mathbf{X}_A -  {}^\mathcal{T}\mathbf{X}_B\big)\Big)  \mathrm{K}(\mathbf{X}_A,  \mathbf{X}_B) + \\[0.4cm]
&\Big( -\frac{1}{2} \big( {}^\mathcal{T}\mathbf{X}_A - {}^\mathcal{T}\mathbf{X}_B\big)^{\mathrm{T}}
\ \mathbf{P}\\
&\qquad \partial_{\alpha}\Big[ \big( ^\mathcal{T}\mathbf{X}_A -  {}^\mathcal{T}\mathbf{X}_B\big)\Big] \Big)  \mathrm{K}(\mathbf{X}_A,  \mathbf{X}_B) \\[0.4cm]
= &\Big( \big( \frac{\mathbf{X}_A}{2\alpha^2}  {}^\mathcal{T}\mathbf{X}_A - \frac{\mathbf{X}_B}{2\alpha^2}  {}^\mathcal{T}\mathbf{X}_B \big)^{\mathrm{T}}
\ \mathbf{P}\\
&\qquad \big( {}^\mathcal{T}\mathbf{X}_A -  {}^\mathcal{T}\mathbf{X}_B \big) \Big) \mathrm{K}(\mathbf{X}_A,  \mathbf{X}_B) + \\[0.4cm]
&\Big( \big( {}^\mathcal{T}\mathbf{X}_A -  {}^\mathcal{T}\mathbf{X}_B \big)^{\mathrm{T}}
\ \mathbf{P}\\
&\qquad \big( \frac{\mathbf{X}_A}{2\alpha^2}  {}^\mathcal{T}\mathbf{X}_A - \frac{\mathbf{X}_B}{2\alpha^2}  {}^\mathcal{T}\mathbf{X}_B \big) \Big)  \mathrm{K}(\mathbf{X}_A,  \mathbf{X}_B)
\end{aligned}
\label{eq:morserbf-der2}
\end{equation}
\vspace{0.2cm}

where ${}^\mathcal{T}\mathbf{X}_I$ are Morse transformed vectors of the original $\mathbf{X}_I$ data points (to simplify the notation we did not write the dependency on $\alpha$ and $\boldsymbol{\rho}$).
\par
In a very similar manner to the MorseRBF kernel,  one can define a MorseMat\'{e}rn kernel starting from equation \ref{eq:covariance-Mat\'{e}rn} and, despite analytical definitions of the gradient of the kernel with respect to the Morse hyperparameter being quite complicated,  one can use numerical gradients and optimise the Morse transformed kernel.
\par
To understand the effect of the Morse kernels,  we compare the shape of the kernel functions in the non-transformed space.  Figure \ref{fig:cov-alpha}-\ref{fig:cov-alpha2} shows a Mat\'{e}rn and a MorseMat\'{e}rn,  projected back to the non-transformed dimension,  to give a better insight.


Morse kernels allow the covariance to be unsymmetrical in the internuclear space,  which affects the correlation of data in a very particular way over the feature space.  As shown on figure \ref{fig:cov-alpha},  the forward and backward correlation differ and the extent of that effect depends greatly on the $\alpha$ parameter.  This increased flexibility of the kernel in the model optimisation,  allows to control the long range effect of the training data for PES modelling.  
\par

\section{Results\label{sec:morse-results}}

We use a training set of 48 water geometries calculated UHF/aug-cc-pVDZ energies,  sampled from a Boltzmann distribution using the Metropolis--Hastings algorithm with data up to 0.3 Ha above the equilibrium energy,  using the Q-Chem software\autocite{QChem}.  Firstly,  the training data is projected on the 3 internuclear distances (ID) and Morse transformed according to equation \ref{eq:morse-transform} to create the feature space of the GPs.  Secondly,  the optimisable transformation using GPs with both the MorseRBF kernel and the MorseMat\'{e}rn class of kernel,  we use the twice differentiable kernel with $\nu=2.5$.  Finally,  in order to assess the performance of each latent function,  we define the MAE of predictions on a testing set also sampled from a Boltzmann distribution with data up to 0.2 Ha above the equilibrium energy.
\par

The Bayesian approach optimises the LML($\boldsymbol{\theta}$) which only includes the training data.  This is very different from optimising the MAE($\boldsymbol{\theta}$) which only includes the testing data (we do not explore this surface here).  Since we use GMIN and explore the whole LML landscape,  one can combine those approaches and rank local minima of the LML surface with their respective MAEs.   One then selects the minimum which has the lowest error.  This gives an hybrid approach which optimises the MAE($\boldsymbol{\theta}$ | $\partial$ LML($\boldsymbol{\theta}$) = 0). 
\par
\begin{figure}[H]
	\centering
	\includegraphics[width=0.4\textwidth]{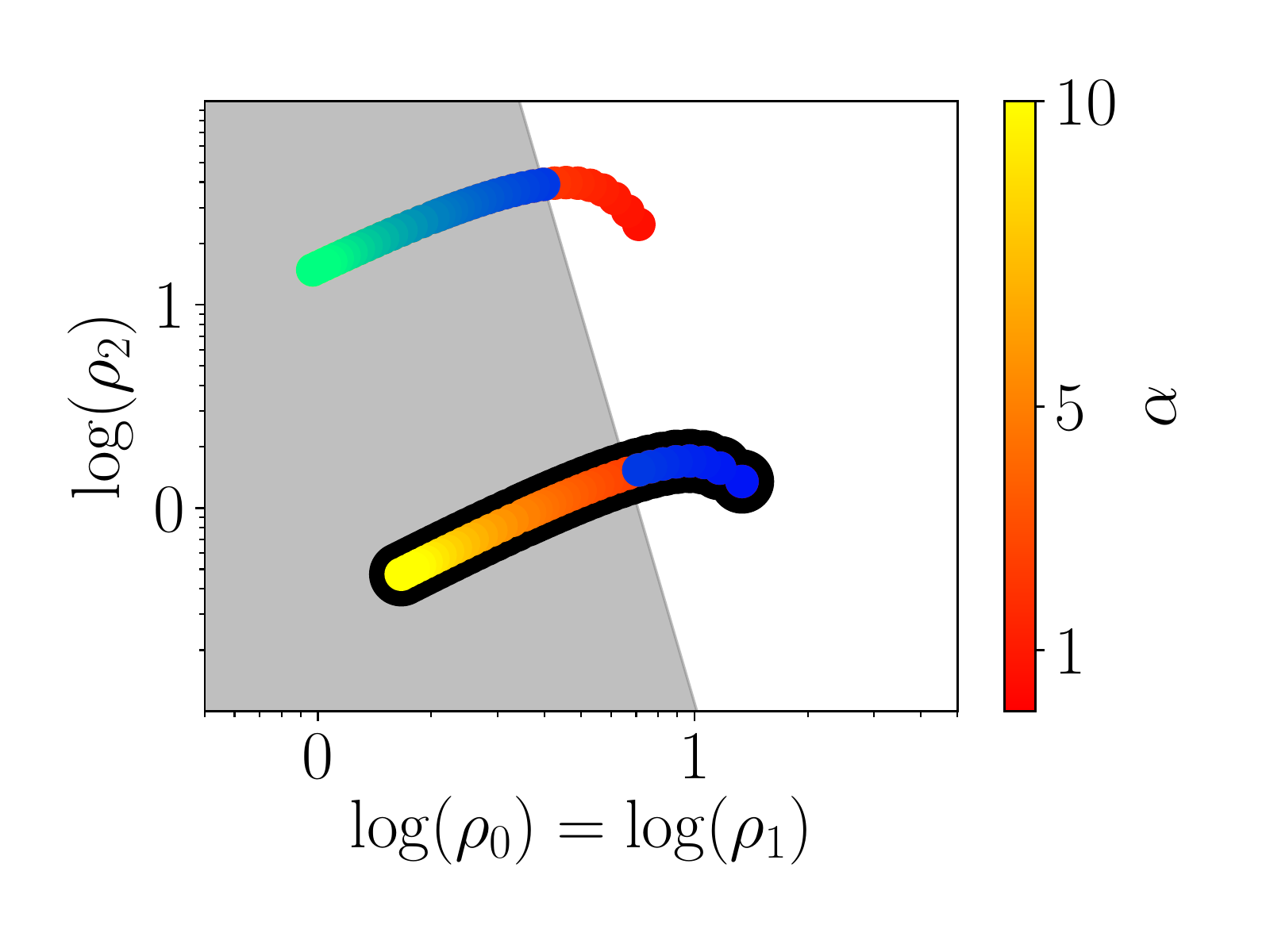}
	\hfill
	\caption[Evolution of hyperparameters for training data projected with different Morse parameters]{Optimised hyperparameters of the Mat\'{e}rn ($\nu=2.5$) kernel along the lengths scales representing the Morse transformed O-H distances ($\rho_0=\rho_1$) and the Morse transformed H-H distance ($\rho_2$) for different minima as well as the MAE (of each respective minima) on a test set.  The blue-green dots represent the lower of the minima on the LML while the red-yellow dots are the second lowest minimum.  The trajectory highlighted in black represents the models with the lowest MAE in both panels.  Around $\alpha=2.0$,  the two modes of selection yield different models (the grey area is plotted to aid clarity of the switch between the two regimes).}
	\label{fig:hyp-MAE-morse}
\end{figure}

The ``best-fit'' approach,  given we use a single Morse parameter,  is a 1D minimisation of the MAE($\alpha$).  One also selects,  for each GP trained with a different Morse parameter,  the LML minimum with the lowest MAE.  We first look at the results of the Mat\'{e}rn ($\nu=2.5$) kernel.  As mentioned before,  the better performing minimum of the LML is not always the global minimum.   For the Mat\'{e}rn ($\nu=2.5$) kernel,  this is seen in figure \ref{fig:hyp-MAE-morse}: from $\alpha=2.2$ it is a worse performing model that is lower on the LML surface while the best performing minimum can be followed.  Even though there is no guarantee that following the better performing minimum on the LML as $\alpha$ changes ensures selection of the best model,  it also seem unlikely that an eventually better performing minimum at a different and larger $\alpha$ could not be followed back to smaller Morse parameters where it disappears (with the exception of $\alpha \to 0$).

\begin{figure}[H]
	\centering
	\includegraphics[width=0.4\textwidth]{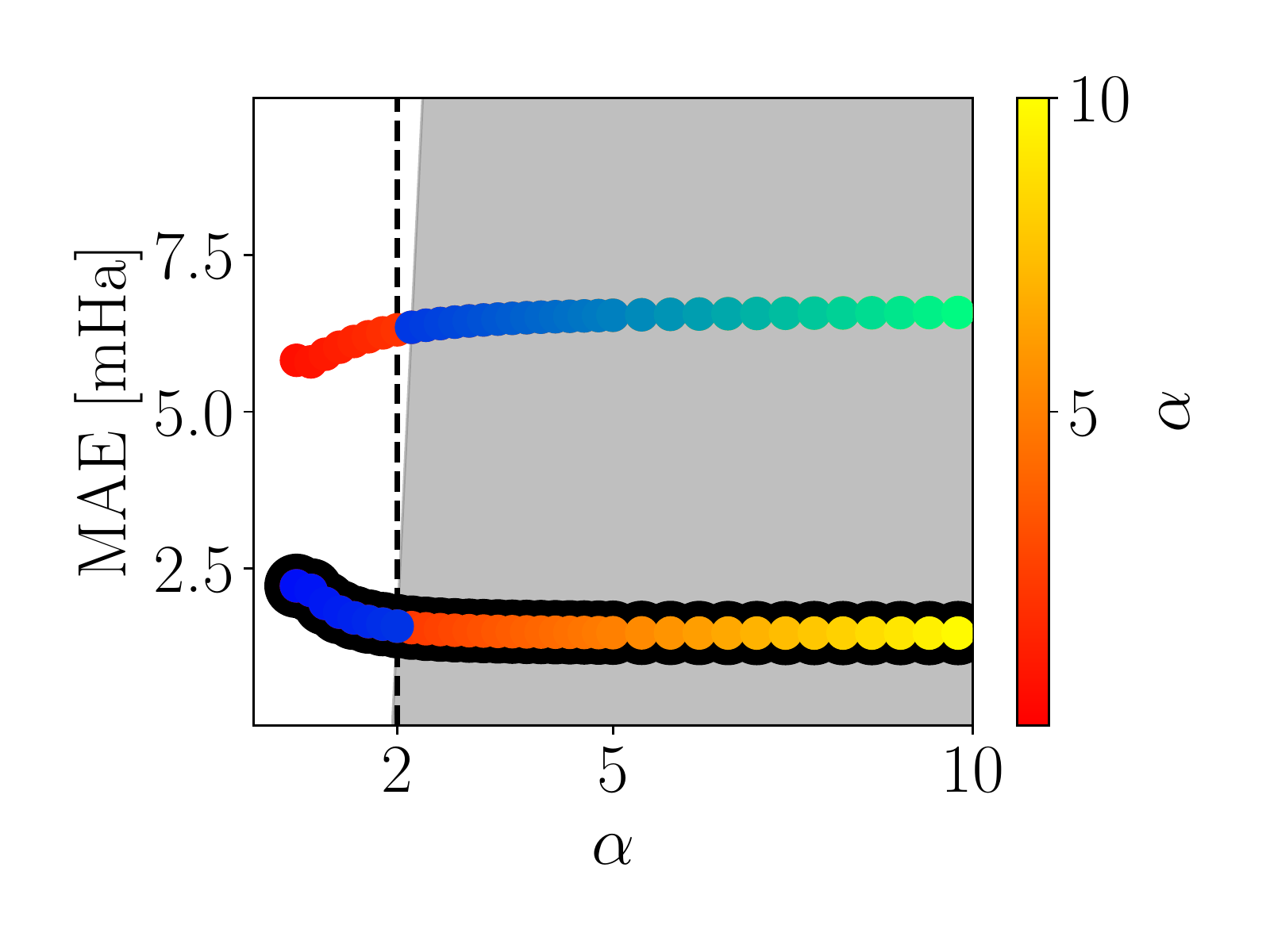}
\end{figure}

The overall behaviour of the trajectories in hyperparameters space of both LML minima represented in panel (b) of figure \ref{fig:hyp-MAE-morse} is expected.   As $\alpha$ increases,  the length scales shorten.  A larger Morse parameter compresses the training data,  shortening the distance between data.  As a consequence,  a constant length scale would flatten the GP latent function.  This is prevented by the data term of the LML,  which causes the minima to move towards shorter length scales.  Moreover,  the minima trajectory can be observed to be almost linear towards larger Morse parameters as the transform of equation \ref{eq:morse-transform} becomes itself more linear since,  in the limit of infinite $\alpha$,  the transform is linear:
\vspace{0.2cm}
\begin{equation}
\begin{aligned}
\mathbf{X}_i \mapsto &\lim_{\alpha \to \infty} \mathrm{exp}(-\mathbf{X}_i/\alpha) =\\
&\lim_{\alpha \to \infty} \Big[ 1 - \frac{\mathbf{X}_i}{\alpha} + \frac{\mathbf{X}_i^2}{2\alpha^2} + \dots \Big] \simeq 1 - \frac{\mathbf{X}_i}{\alpha}
\end{aligned}
\label{eq:morse-trans-limit}
\end{equation}
\vspace{0.2cm}

This opens the question of redefining length scales to reflect the change in $\alpha$ (for example as $\rho_i \to \rho_i/\alpha$) hyperparameter. This does not seem to affect the optimisation process and will not be considered further.
\par
Despite figure \ref{fig:hyp-MAE-morse} showing only two minima,  the GP has multiple minima on the LML surface.  However,  only two minima provided PES models with low MAEs.  Figure \ref{fig:Mat\'{e}rn-disco} shows the disconnectivity graphs of the LML to show the complexity of the surface for the Mat\'{e}rn ($\nu=2.5$) kernel.  It is surprising that the variations are so large despite the training data being unchanged.
\par
Disconnectivity graphs show some surprising variations that are sometimes a simple consequence of TSs being very flat and hard to capture.  This leads to the latter disappearing after small changes to the LML space which has strong consequences on the network of minima that can be shown.  This is the case of the graphs for $\alpha=0.2$ and $\alpha=0.4$ in figure \ref{fig:Mat\'{e}rn-disco} where the latter finds TSs between LML minima more easily. 
\begin{figure}[H]
	\centering
	\captionsetup[subfigure]{labelformat=empty}
	\subfloat[$0.2$]{\includegraphics[width=0.05\textwidth, trim=3.7cm 0 3.7cm 0, clip]{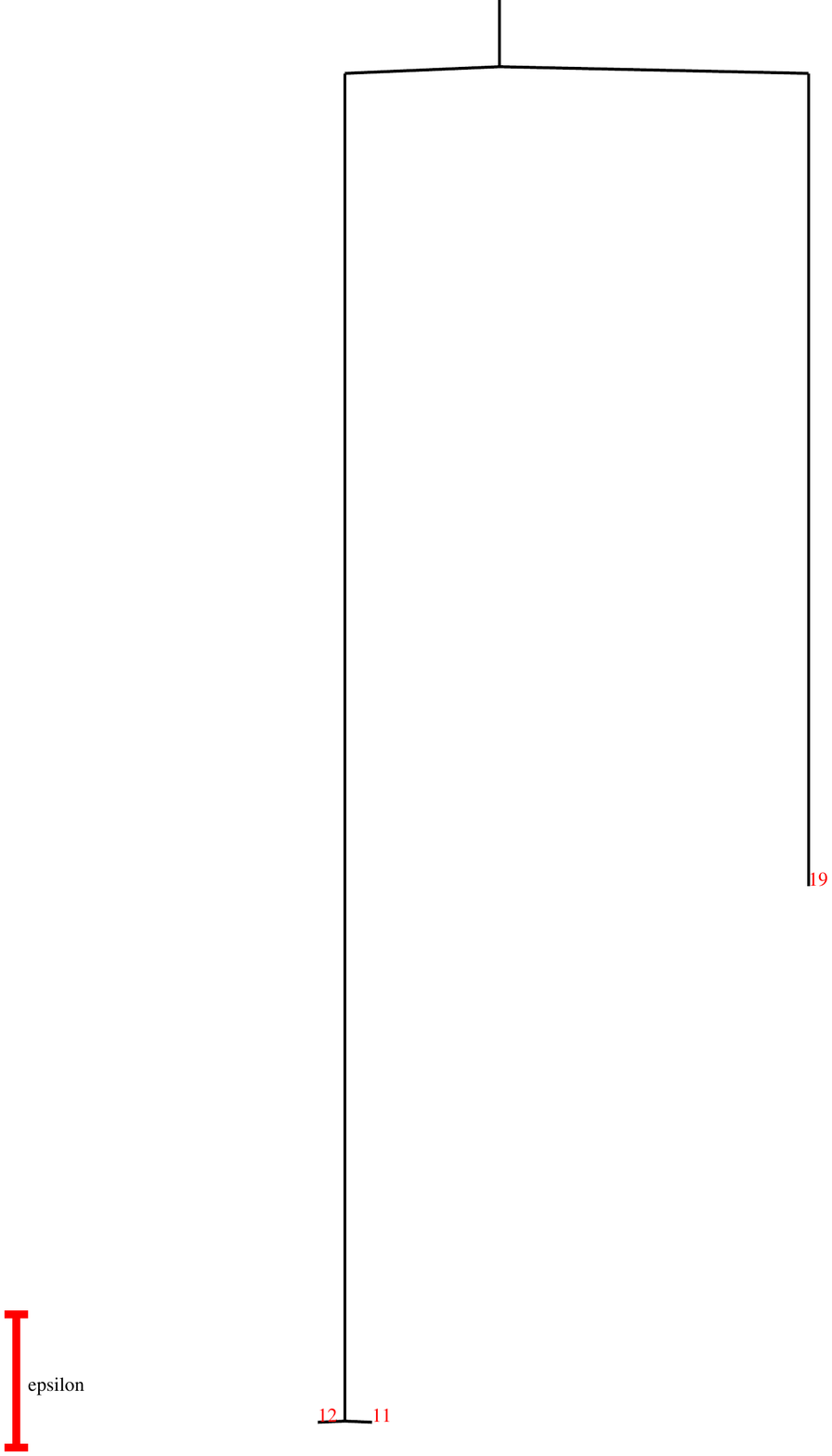}}
	\hspace{0.05\textwidth}
	\subfloat[$0.4$]{\includegraphics[width=0.05\textwidth, trim=3.7cm 0 3.7cm 0, clip]{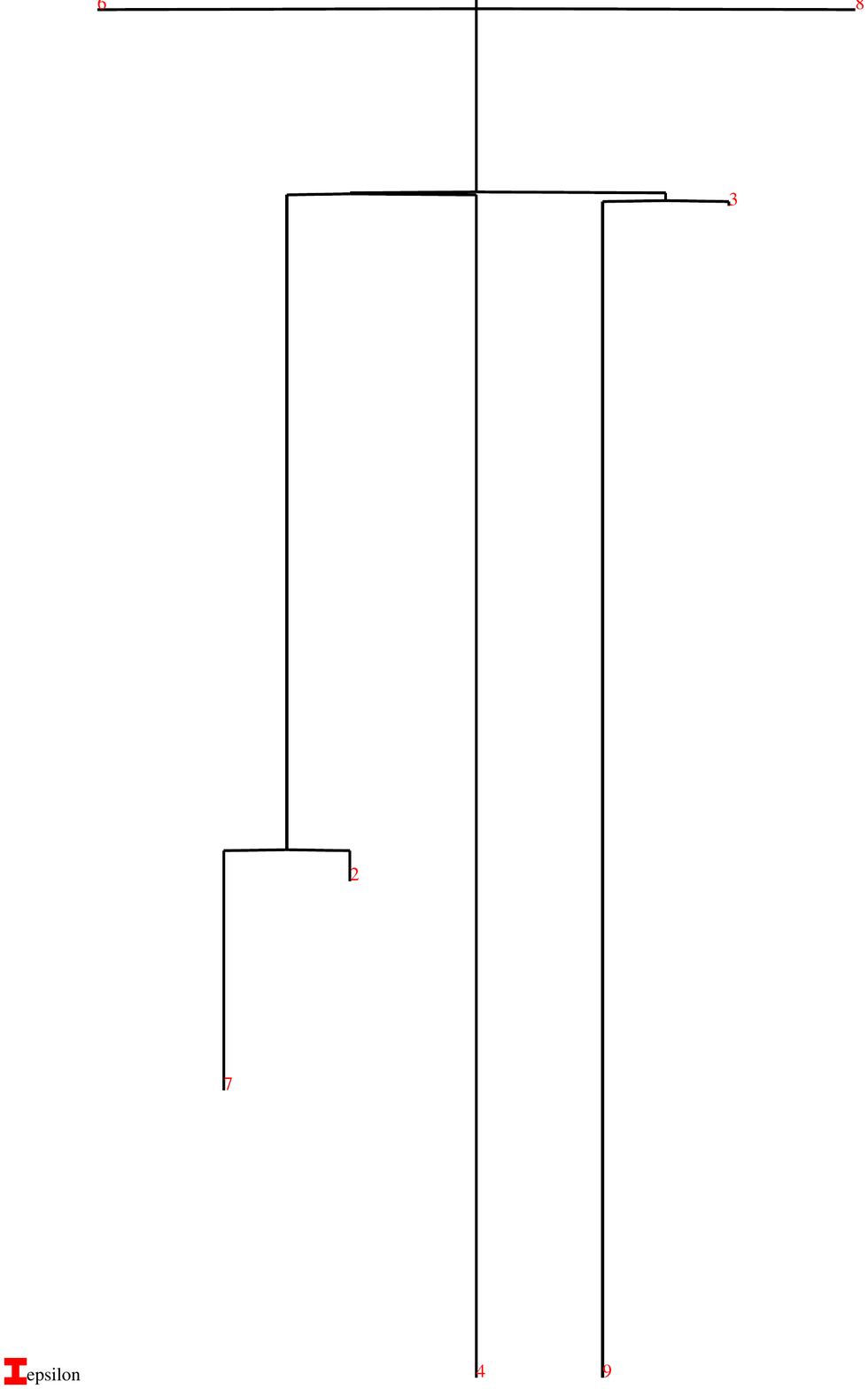}}
	\hspace{0.05\textwidth}
	\subfloat[$1.2$]{\includegraphics[width=0.05\textwidth, trim=3.7cm 0 3.7cm 0, clip]{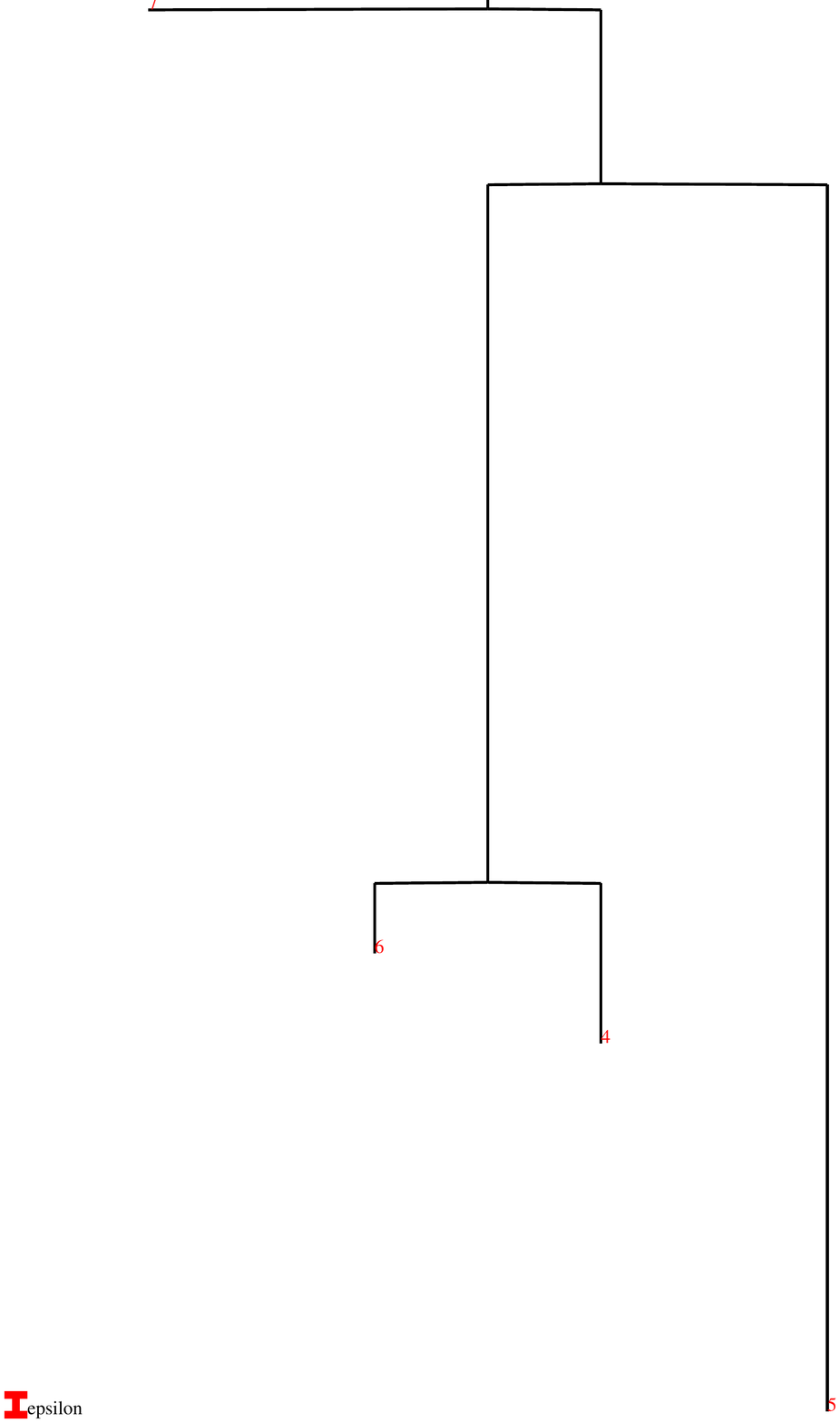}}
	\hspace{0.05\textwidth}
	\subfloat[$1.4$]{\includegraphics[width=0.05\textwidth, trim=3.7cm 0 3.7cm 0, clip]{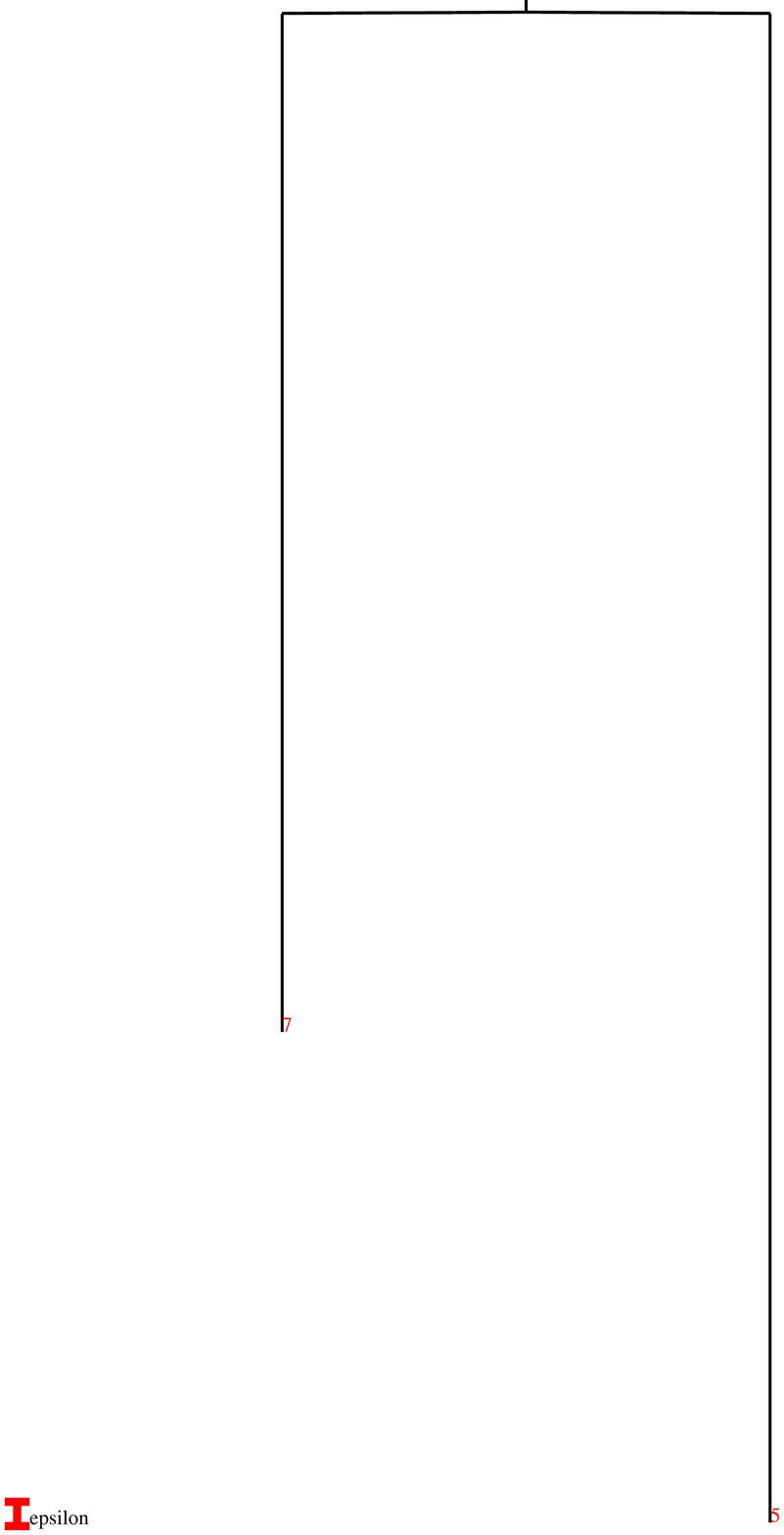}}
	\hspace{0.05\textwidth}
	\subfloat[$2.2$]{\includegraphics[width=0.05\textwidth, trim=3.7cm 0 3.7cm 0, clip]{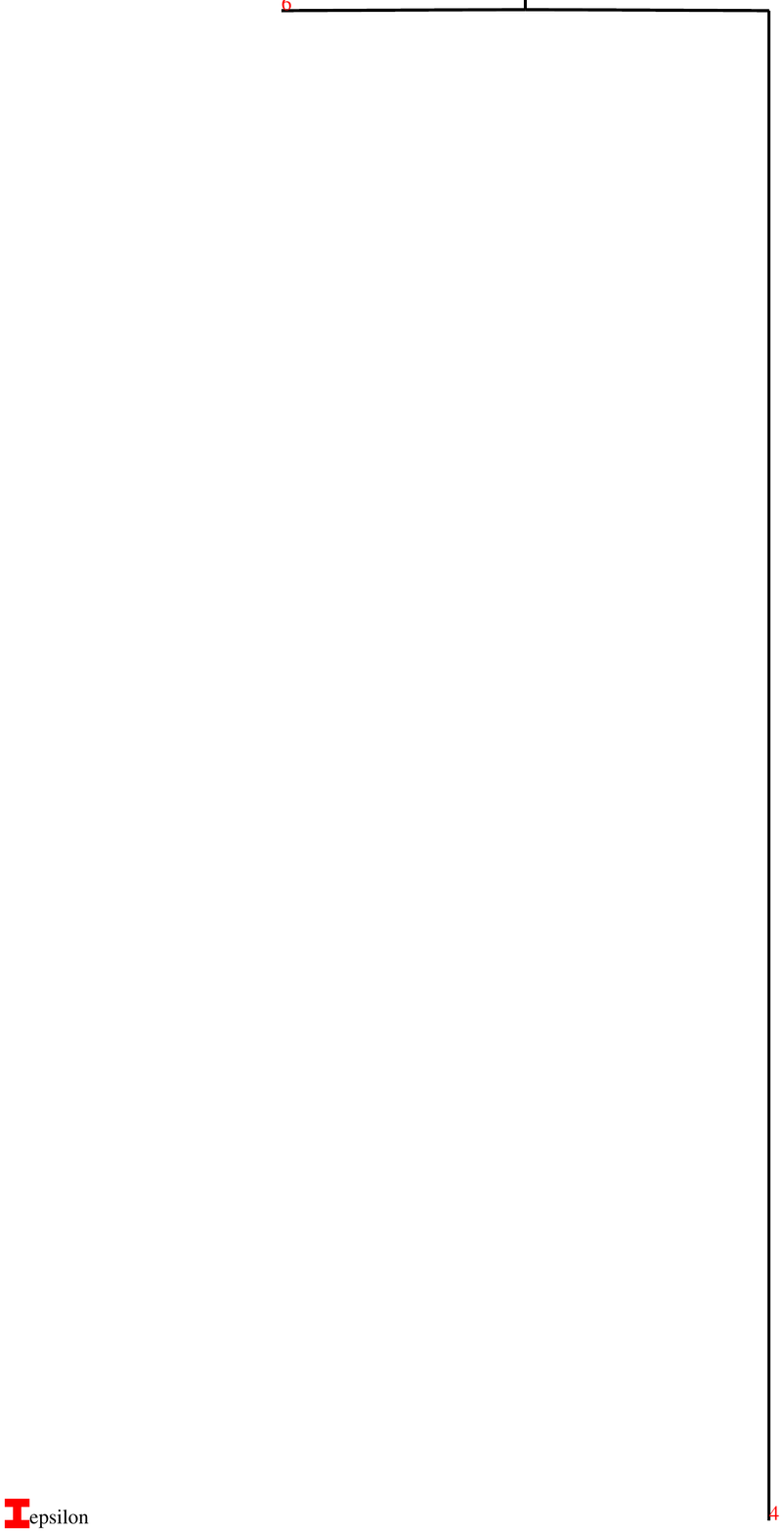}}
	\hspace{0.05\textwidth}
	\subfloat[$2.6$]{\includegraphics[width=0.05\textwidth, trim=3.7cm 0 3.7cm 0, clip]{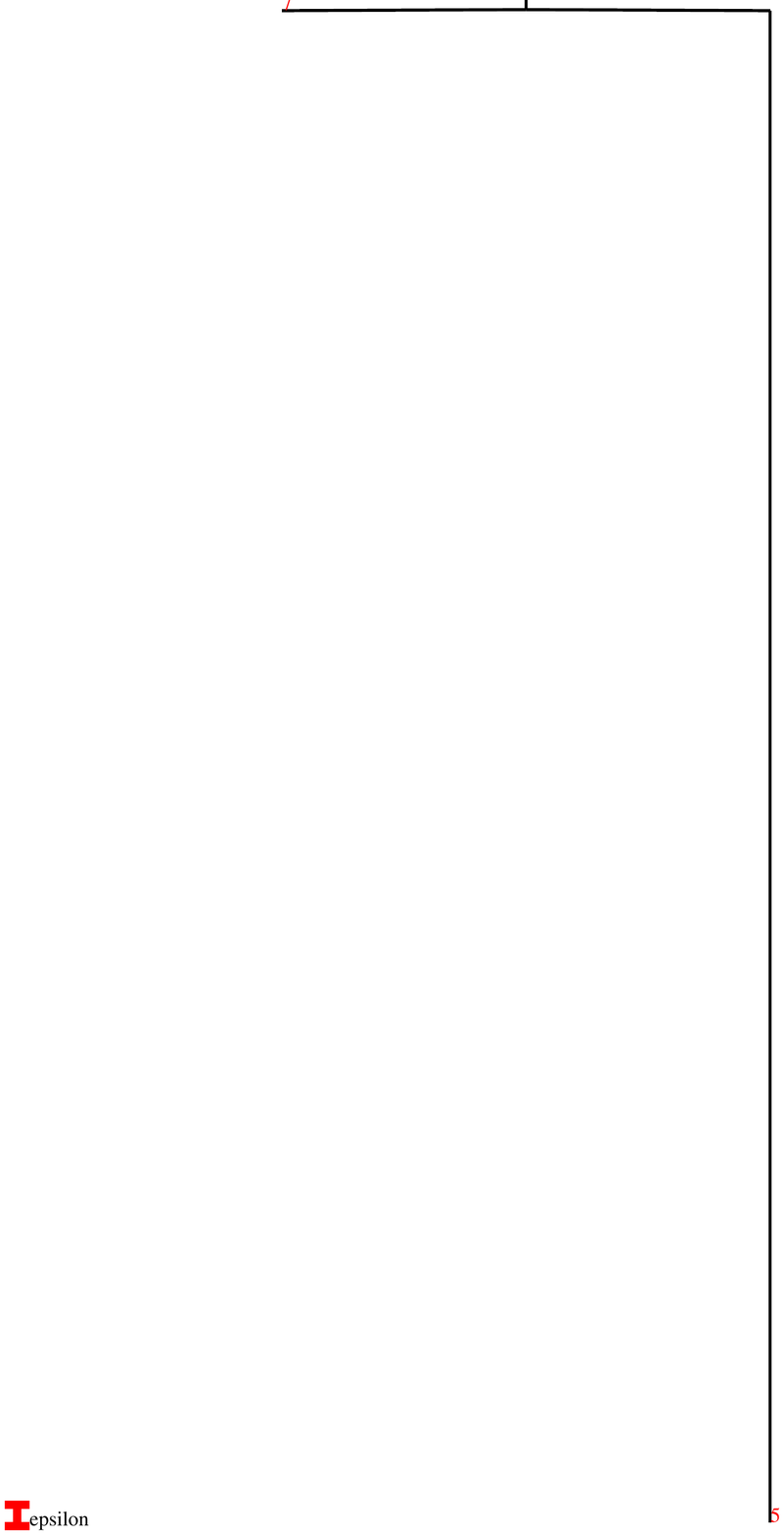}}
	\hspace{0.05\textwidth}
	\subfloat[$2.8$]{\includegraphics[width=0.05\textwidth, trim=3.7cm 0 3.7cm 0, clip]{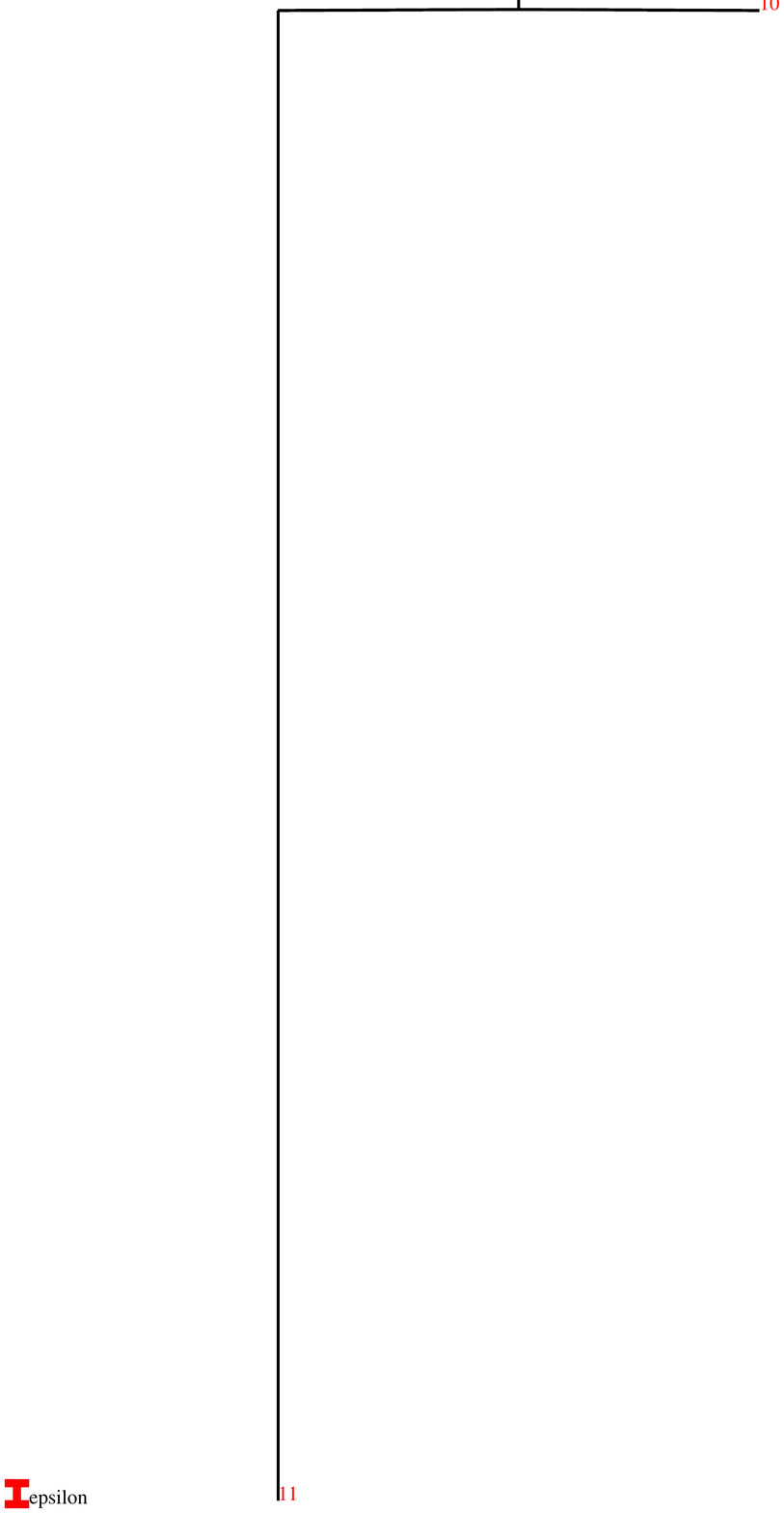}}
	\hspace{0.05\textwidth}
	\subfloat[$3.0$]{\includegraphics[width=0.05\textwidth, trim=3.7cm 0 3.7cm 0, clip]{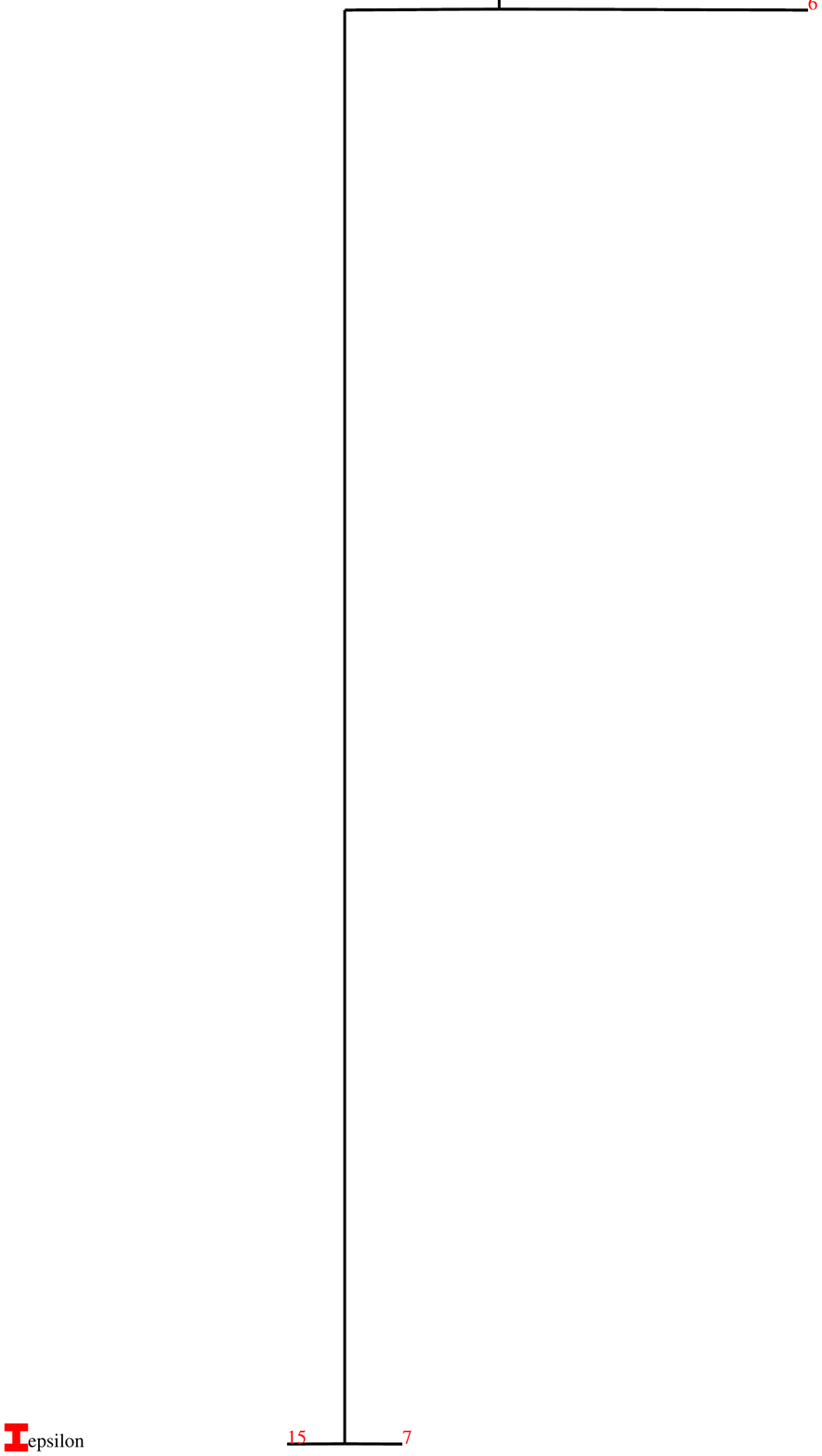}}
	\hspace{0.05\textwidth}
	\subfloat[$3.2$]{\includegraphics[width=0.05\textwidth, trim=3.7cm 0 3.7cm 0, clip]{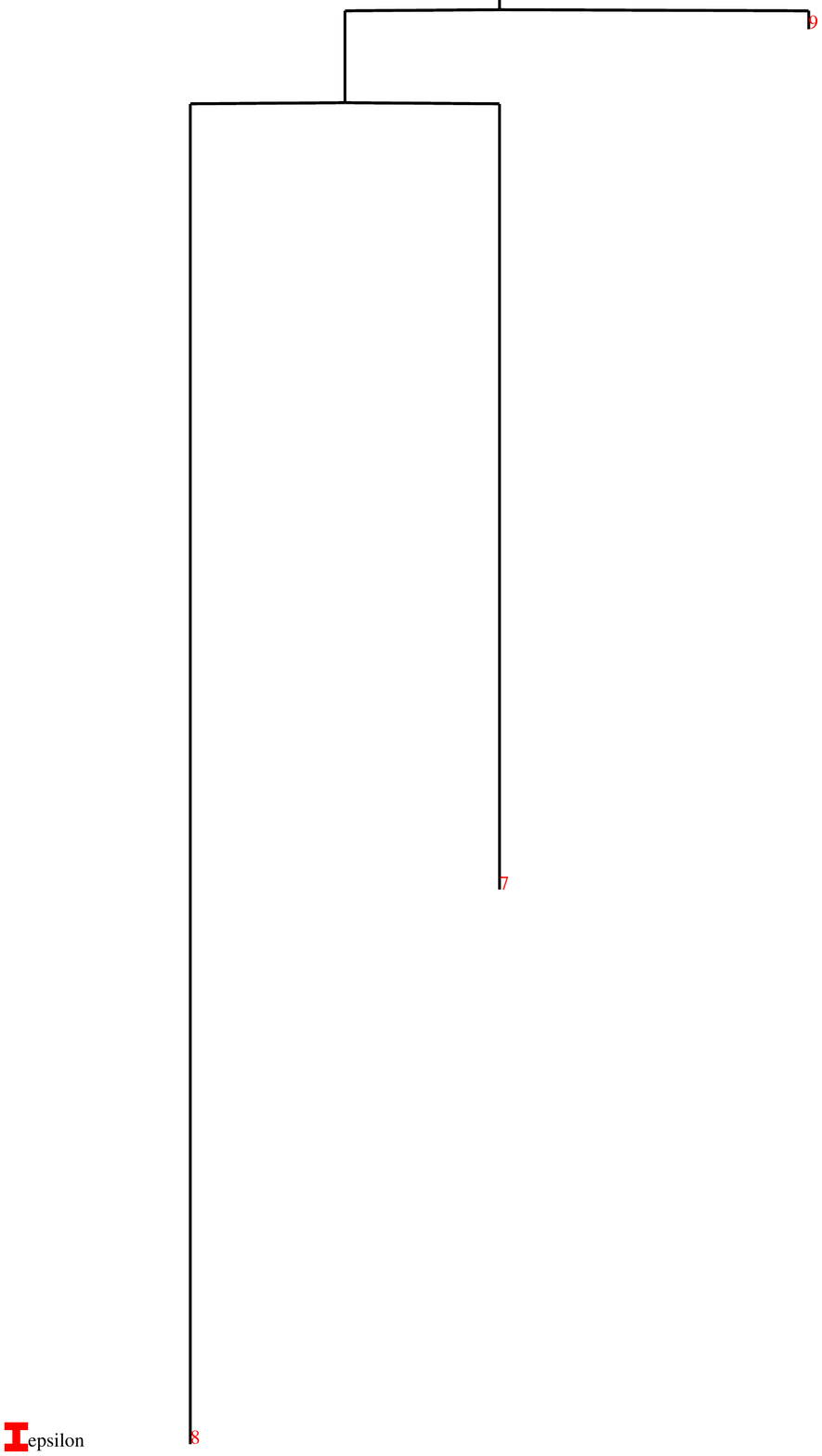}}
	\hspace{0.05\textwidth}
	\subfloat[$3.6$]{\includegraphics[width=0.05\textwidth, trim=3.7cm 0 3.7cm 0, clip]{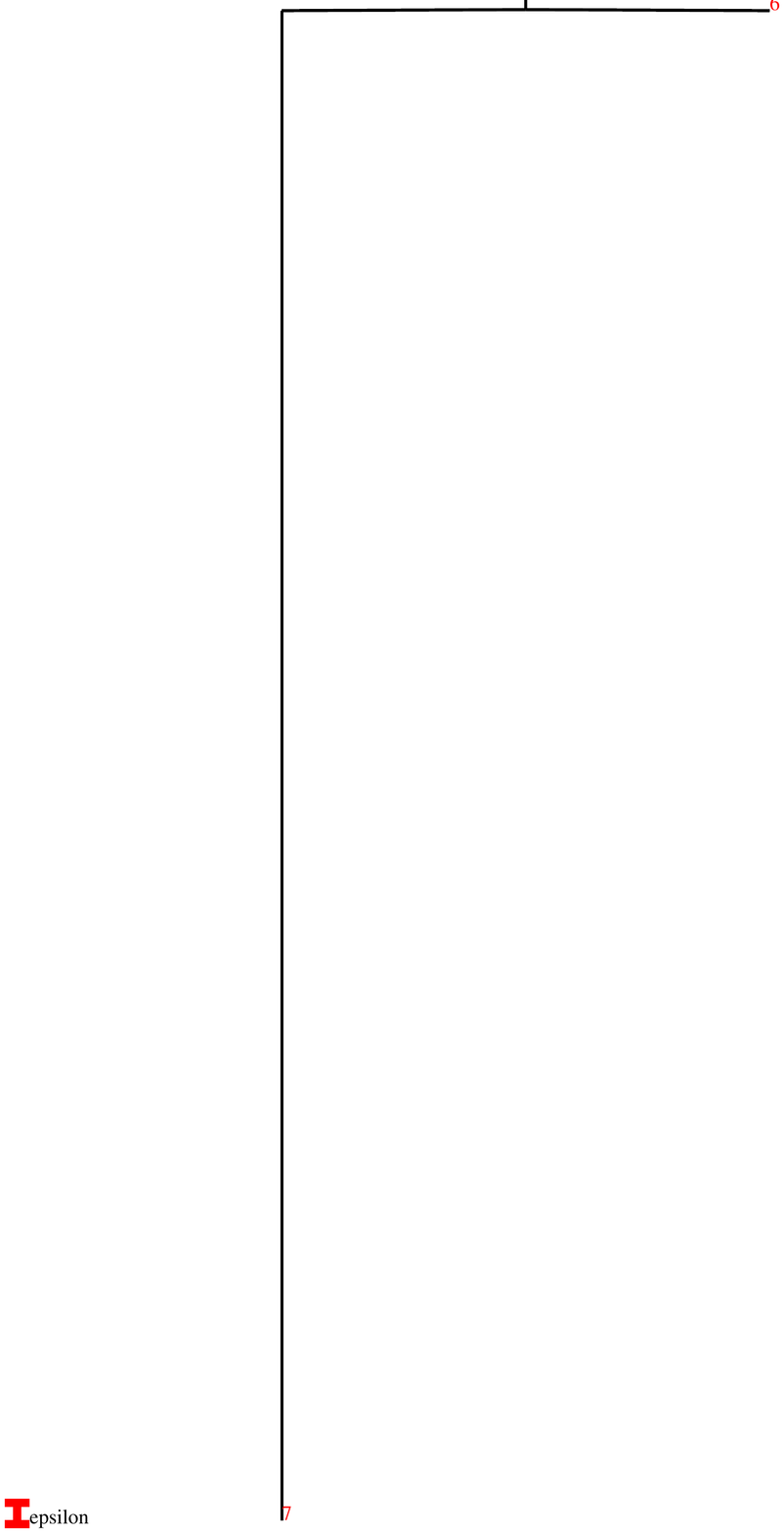}}
	\hspace{0.05\textwidth}
	\subfloat[$4.0$]{\includegraphics[width=0.05\textwidth, trim=3.7cm 0 3.7cm 0, clip]{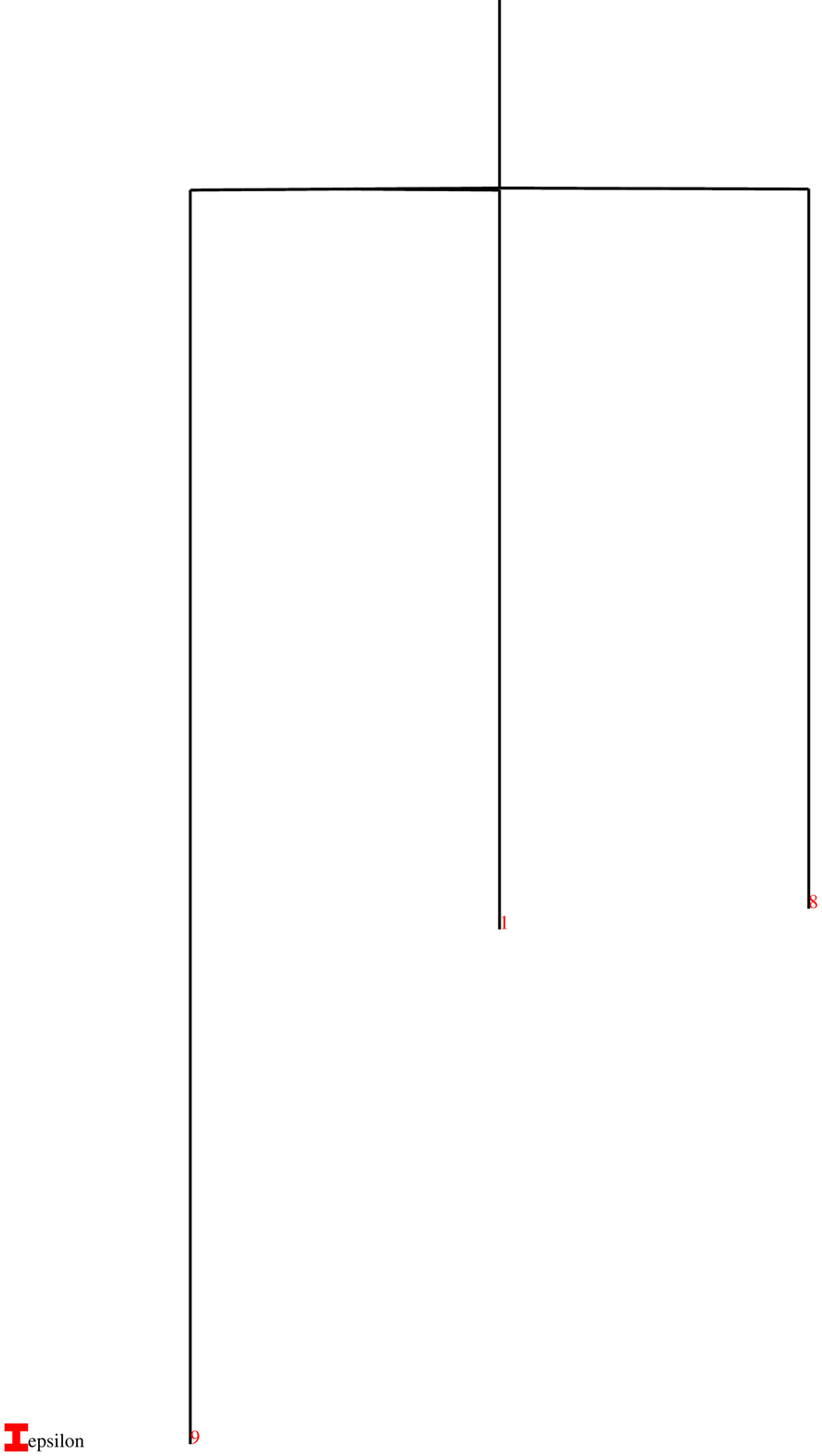}}
	\hspace{0.05\textwidth}
	\subfloat[$4.4$]{\includegraphics[width=0.05\textwidth, trim=3.7cm 0 3.7cm 0, clip]{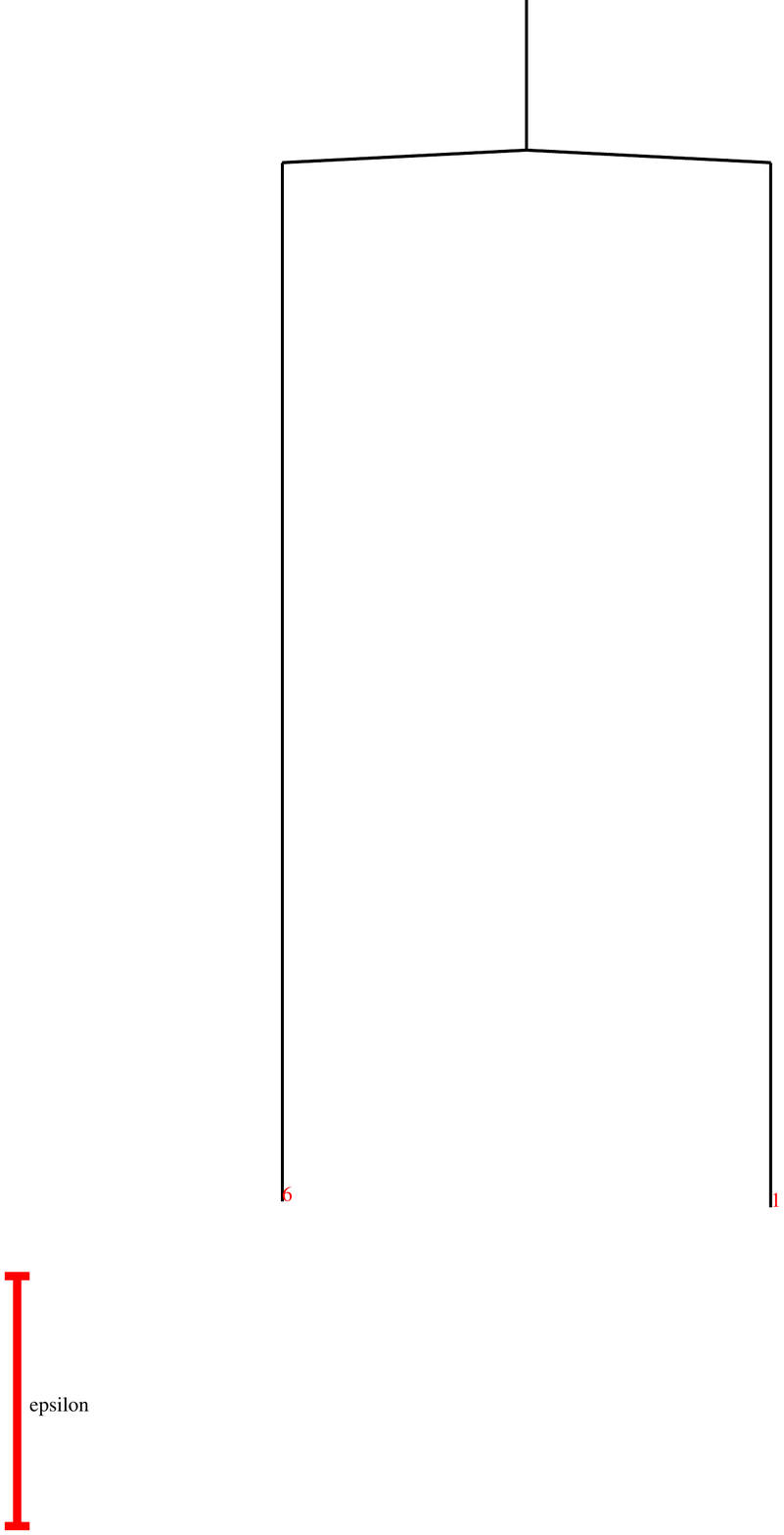}}
	\hspace{0.05\textwidth}
	\subfloat[$4.6$]{\includegraphics[width=0.05\textwidth, trim=3.7cm 0 3.7cm 0, clip]{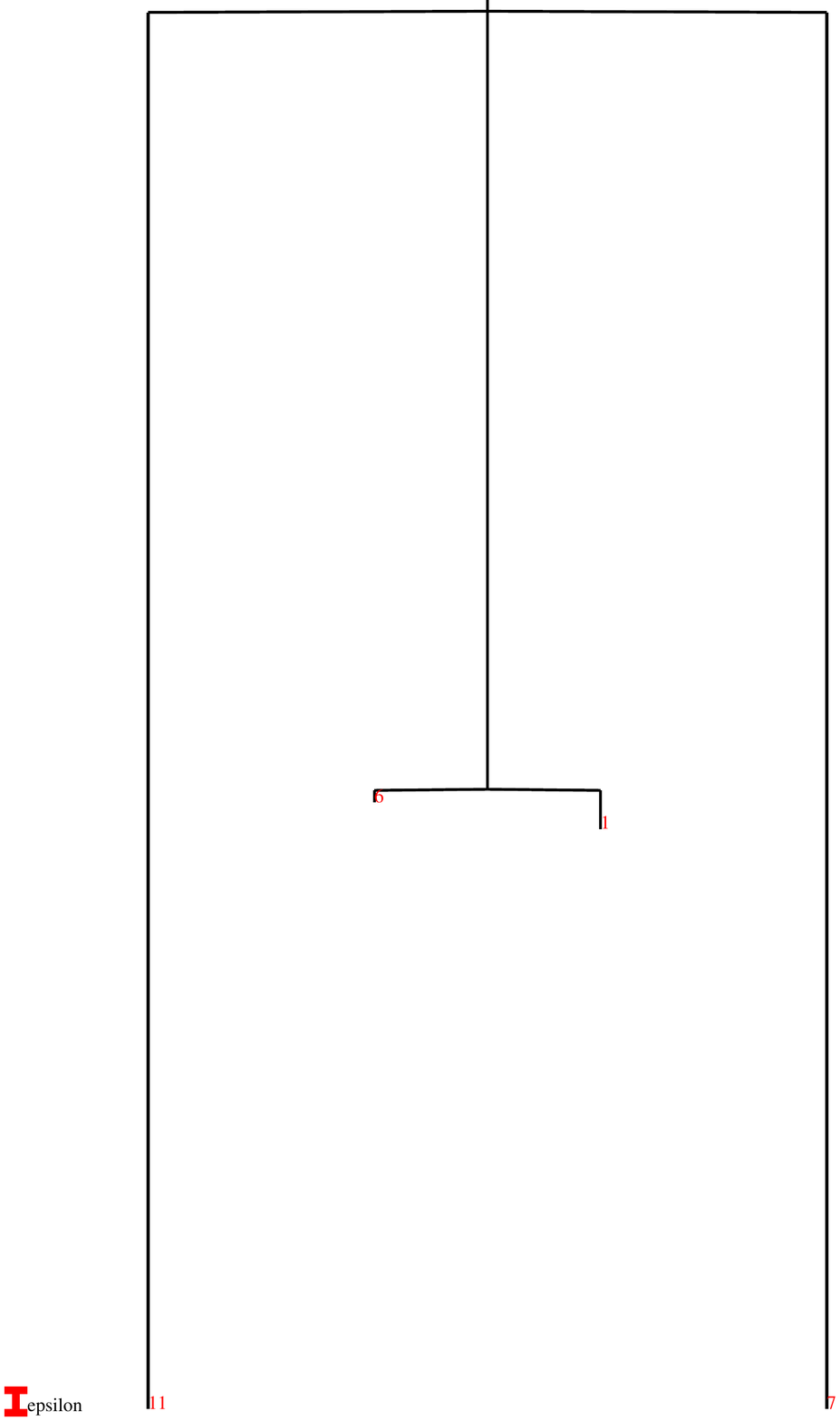}}
	\hspace{0.05\textwidth}
	\subfloat[$4.8$]{\includegraphics[width=0.05\textwidth, trim=3.7cm 0 3.7cm 0, clip]{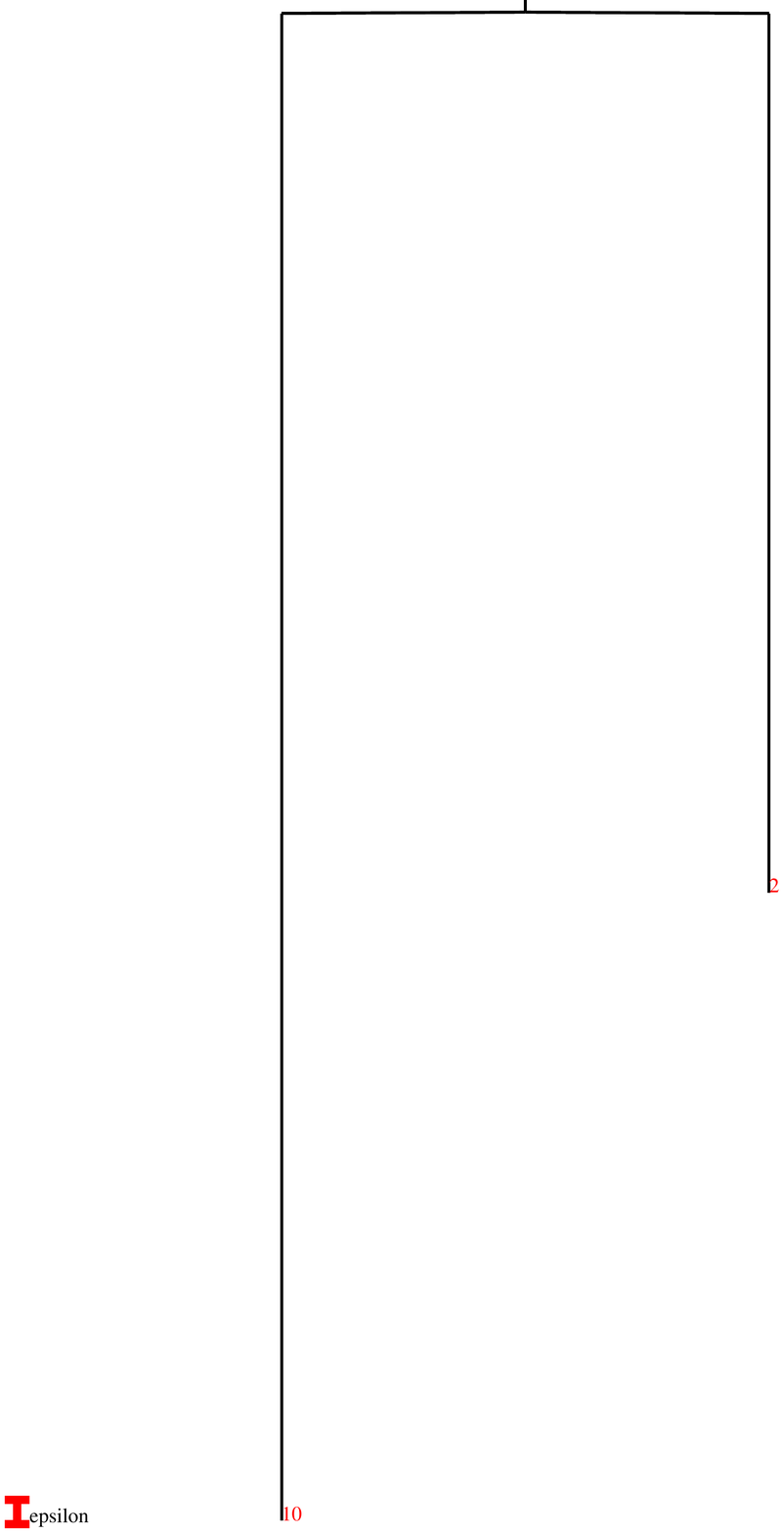}}
	\hspace{0.05\textwidth}
	\subfloat[$5.0$]{\includegraphics[width=0.05\textwidth, trim=3.7cm 0 3.7cm 0, clip]{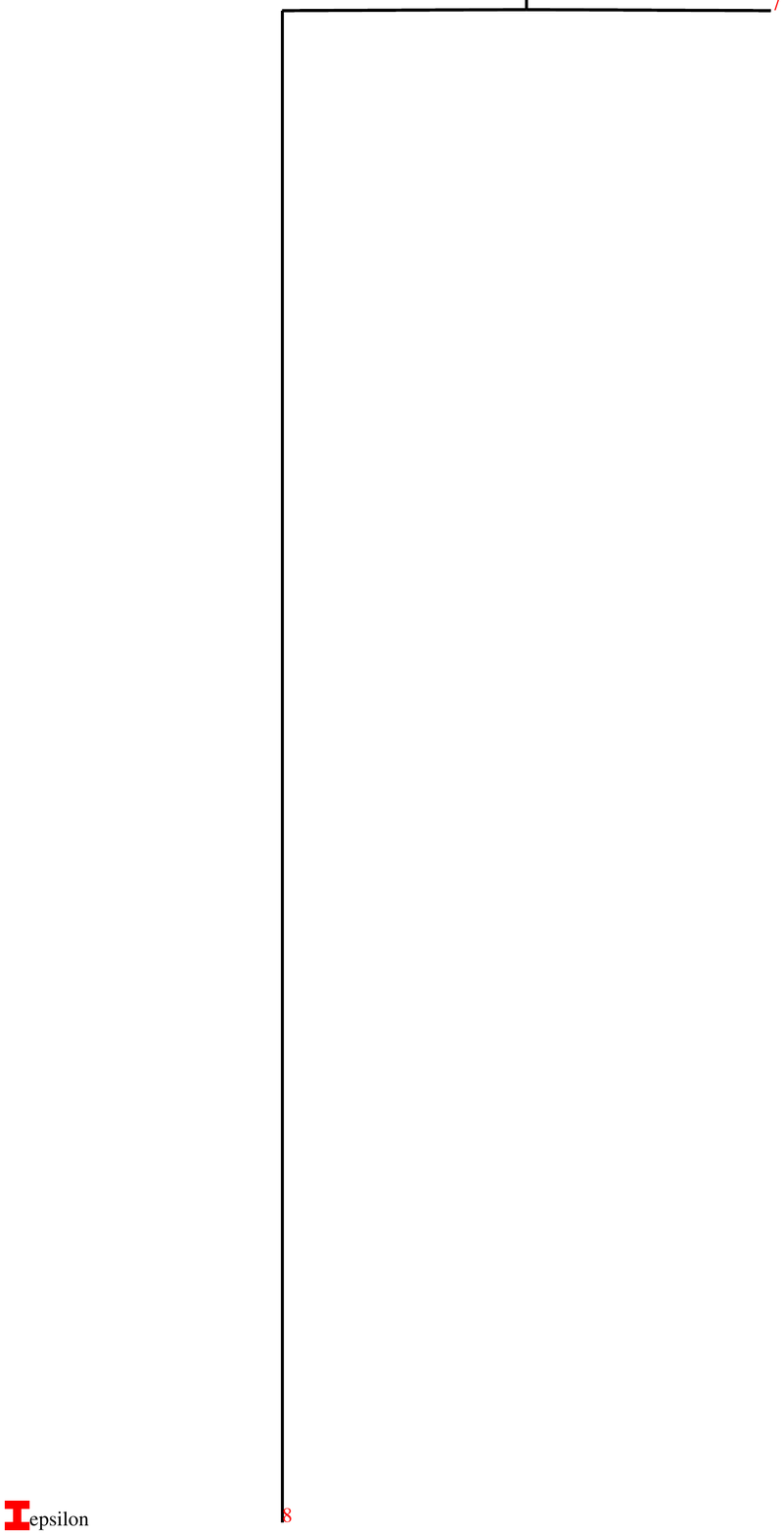}}
	\hspace{0.05\textwidth}
	\subfloat[$5.8$]{\includegraphics[width=0.05\textwidth, trim=3.7cm 0 3.7cm 0, clip]{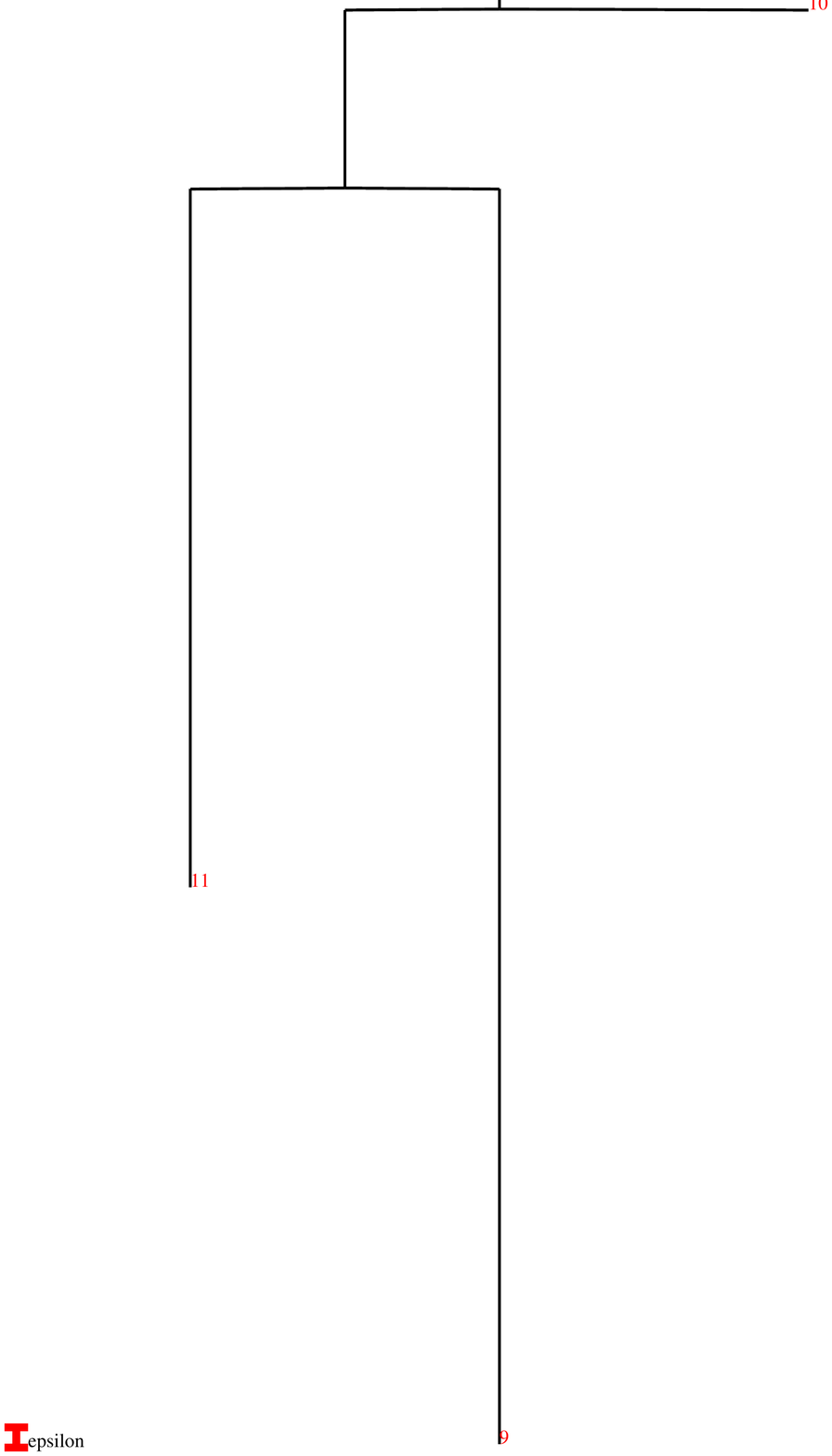}}
	\hspace{0.05\textwidth}
	\subfloat[$7.4$]{\includegraphics[width=0.05\textwidth, trim=3.7cm 0 3.7cm 0, clip]{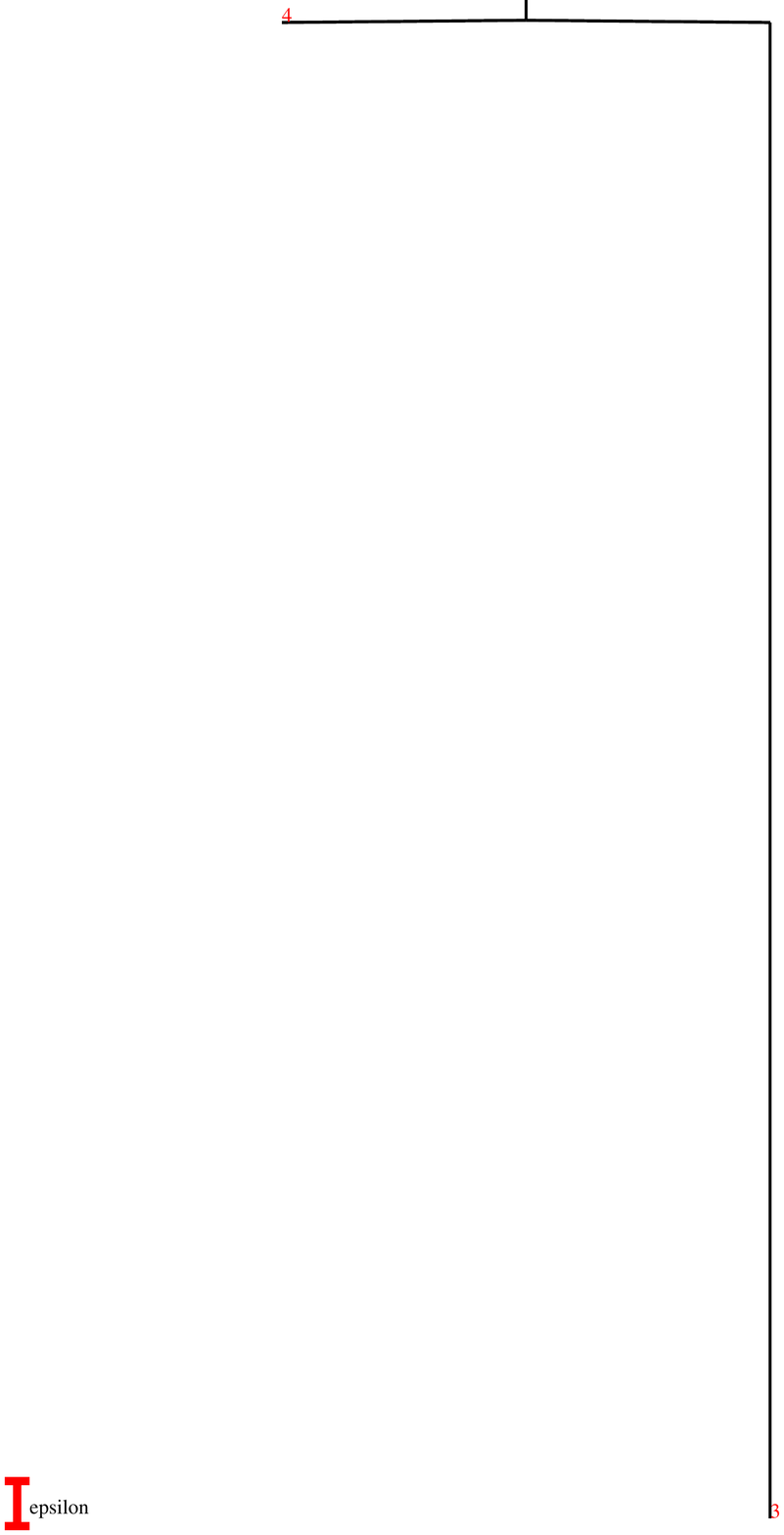}}
	\hspace{0.05\textwidth}
	\subfloat[$7.8$]{\includegraphics[width=0.05\textwidth, trim=3.7cm 0 3.7cm 0, clip]{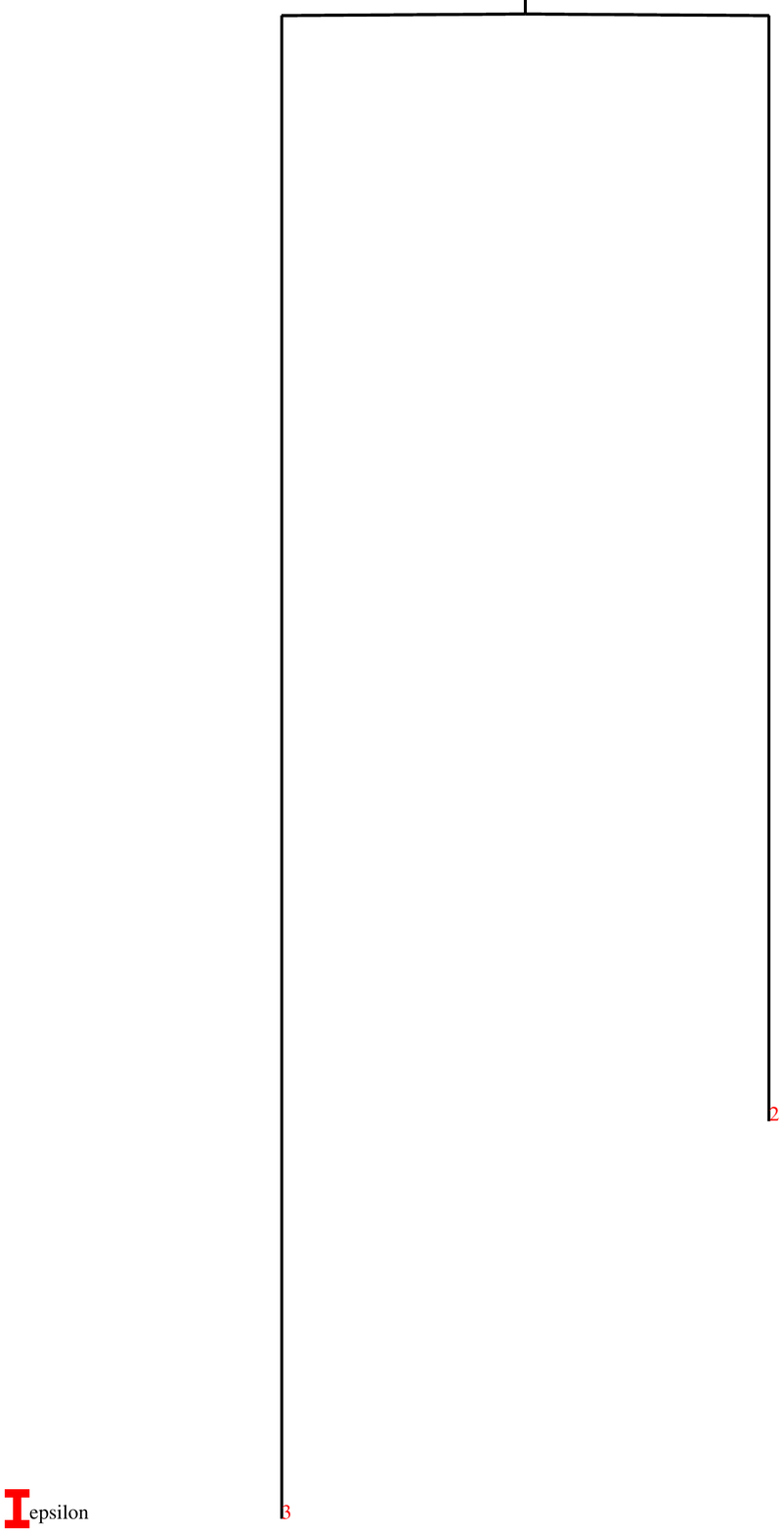}}
	\hspace{0.05\textwidth}
	\subfloat[$8.2$]{\includegraphics[width=0.05\textwidth, trim=3.7cm 0 3.7cm 0, clip]{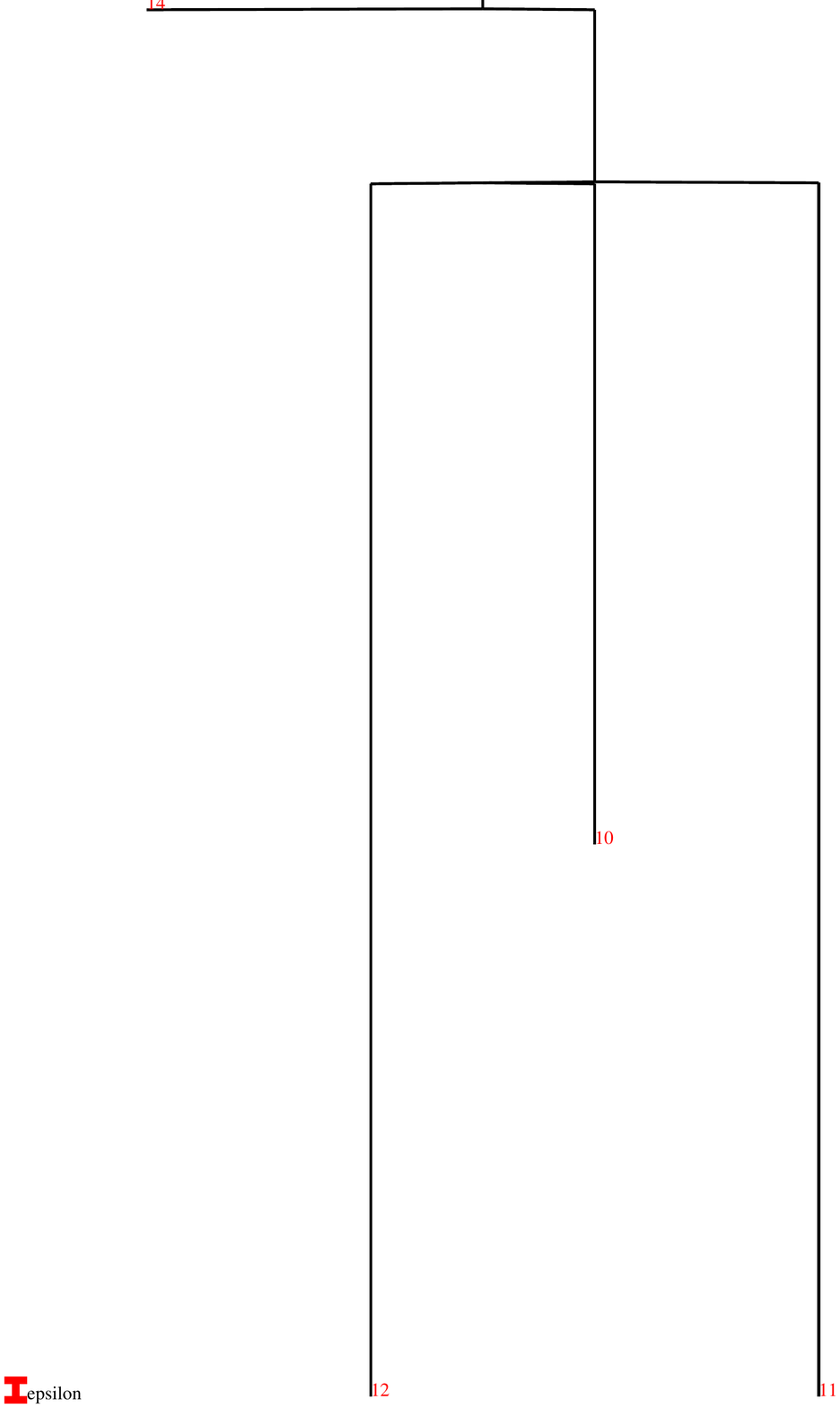}}
	\hspace{0.05\textwidth}
	\subfloat[$9.0$]{\includegraphics[width=0.05\textwidth, trim=3.7cm 0 3.7cm 0, clip]{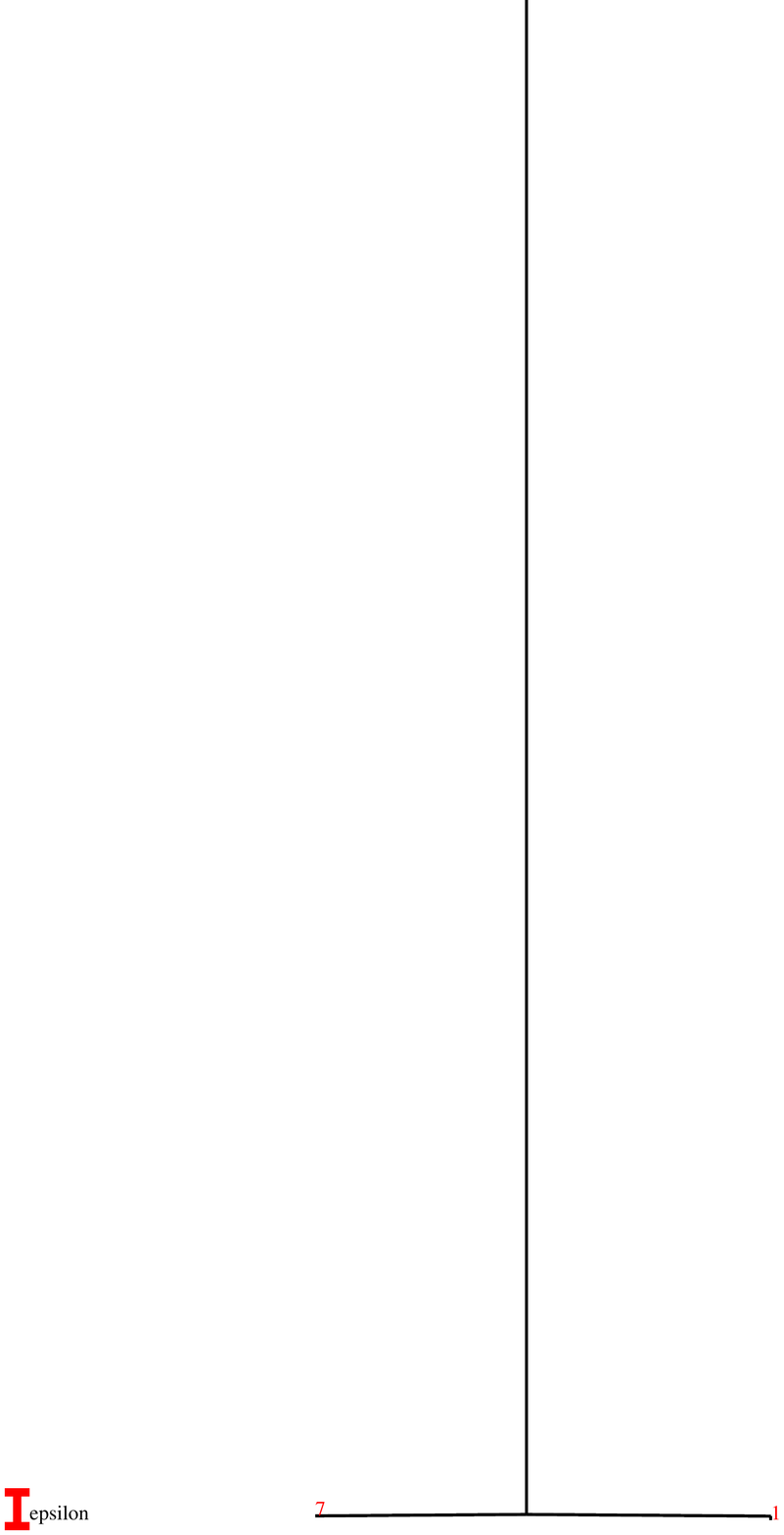}}
	\hspace{0.05\textwidth}
	\subfloat[$9.8$]{\includegraphics[width=0.05\textwidth, trim=3.7cm 0 3.7cm 0, clip]{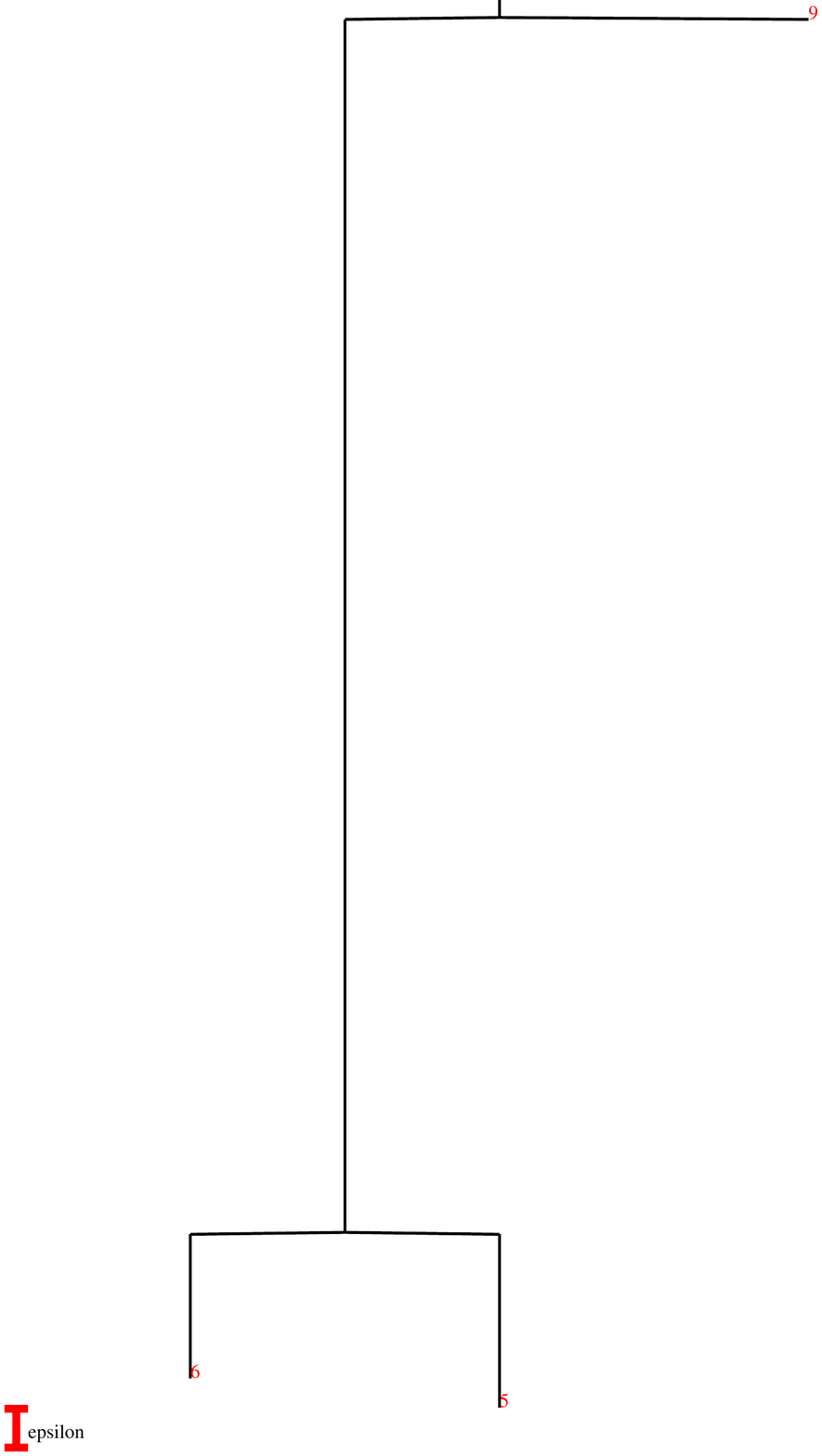}}
	\hspace{0.05\textwidth}
	\subfloat[$10.2$]{\includegraphics[width=0.05\textwidth, trim=3.7cm 0 3.7cm 0, clip]{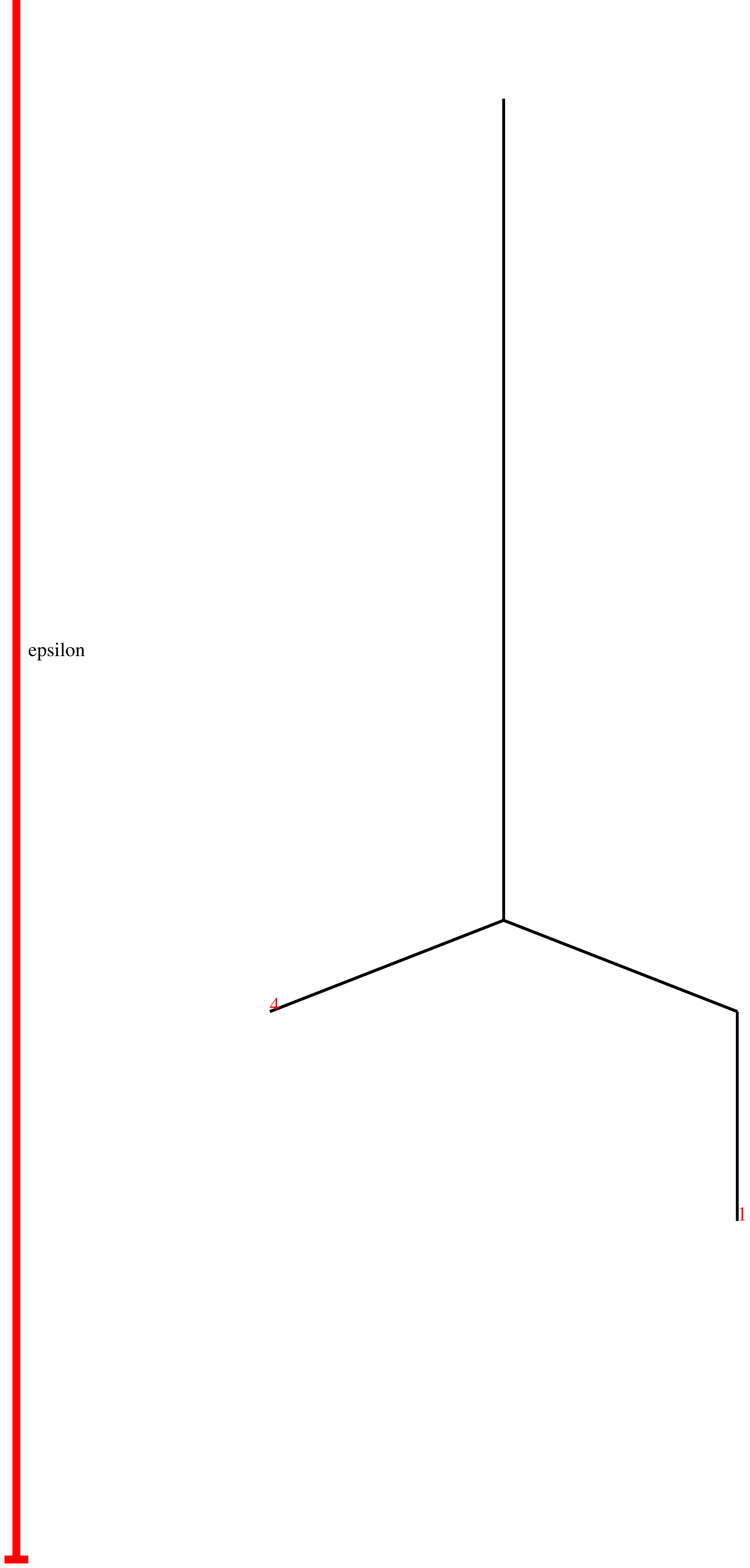}}
	\hspace{0.05\textwidth}
	\subfloat[$10.6$]{\includegraphics[width=0.05\textwidth, trim=3.7cm 0 3.7cm 0, clip]{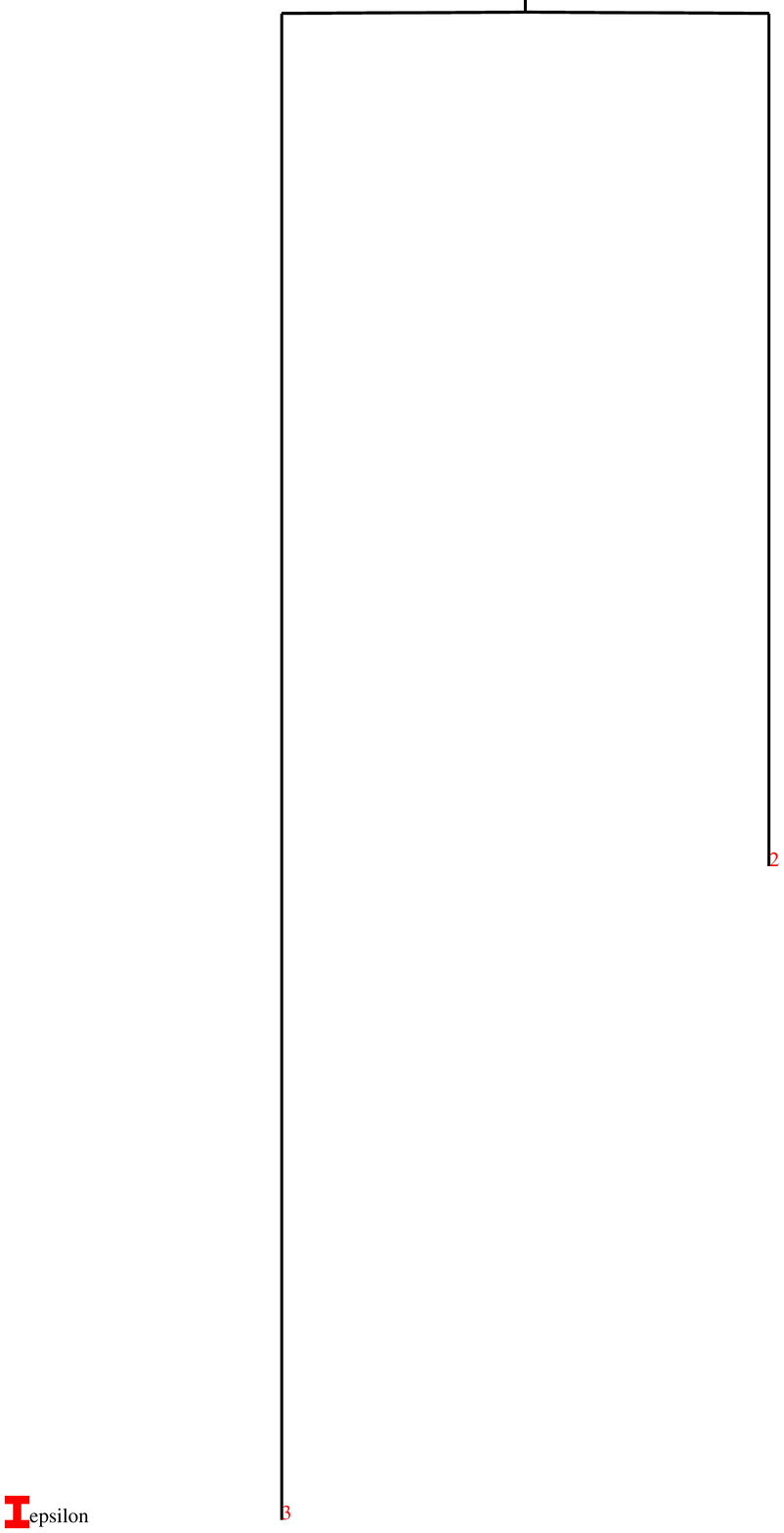}}
	\hspace{0.05\textwidth}

	\caption{LMLs disconnectivity graphs for the Mat\'{e}rn ($\nu=2.5$) kernels.  Labels underneath each graph denote the $\alpha$ parameter for the corresponding GP LML.}
	\label{fig:Mat\'{e}rn-disco}
\end{figure}

\par
\begin{figure}[H]
	\centering
	\captionsetup[subfigure]{labelformat=empty}
	\subfloat[$0.4$]{\includegraphics[width=0.05\textwidth, trim=3.7cm 0 3.7cm 0, clip]{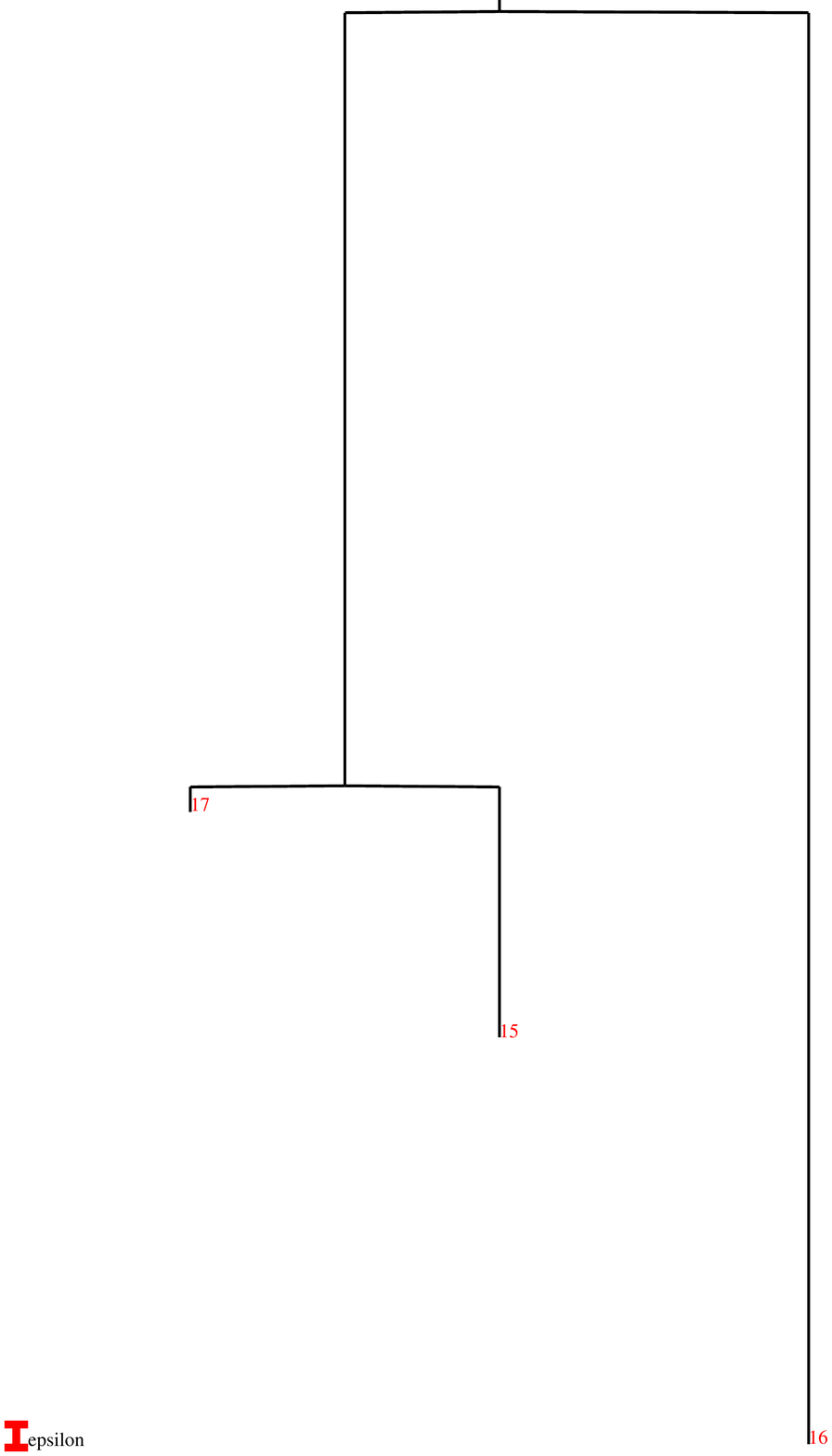}}
	\hspace{0.05\textwidth}
	\subfloat[$0.6$]{\includegraphics[width=0.05\textwidth, trim=3.7cm 0 3.7cm 0, clip]{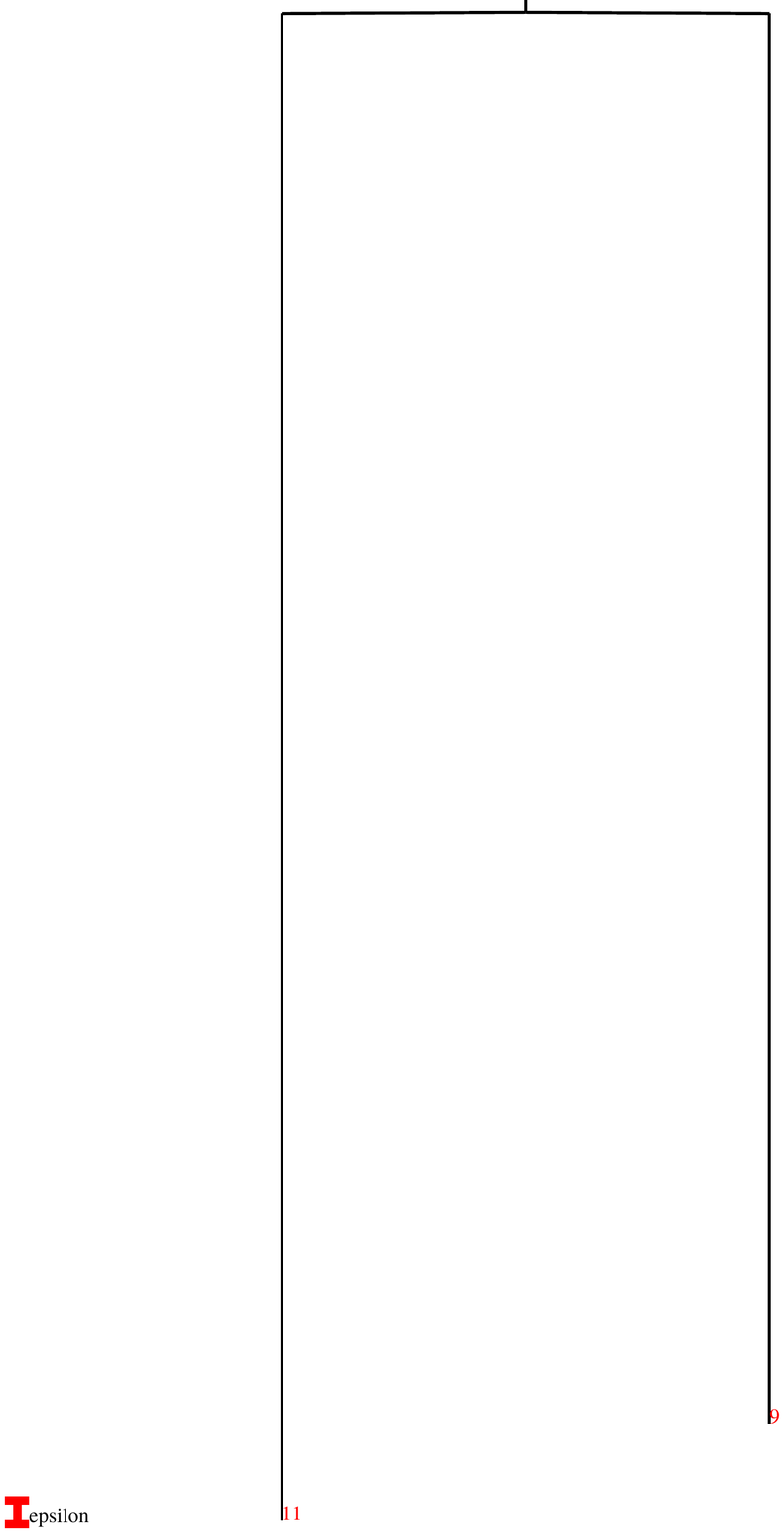}}
	\hspace{0.05\textwidth}
	\subfloat[$0.8$]{\includegraphics[width=0.05\textwidth, trim=3.7cm 0 3.7cm 0, clip]{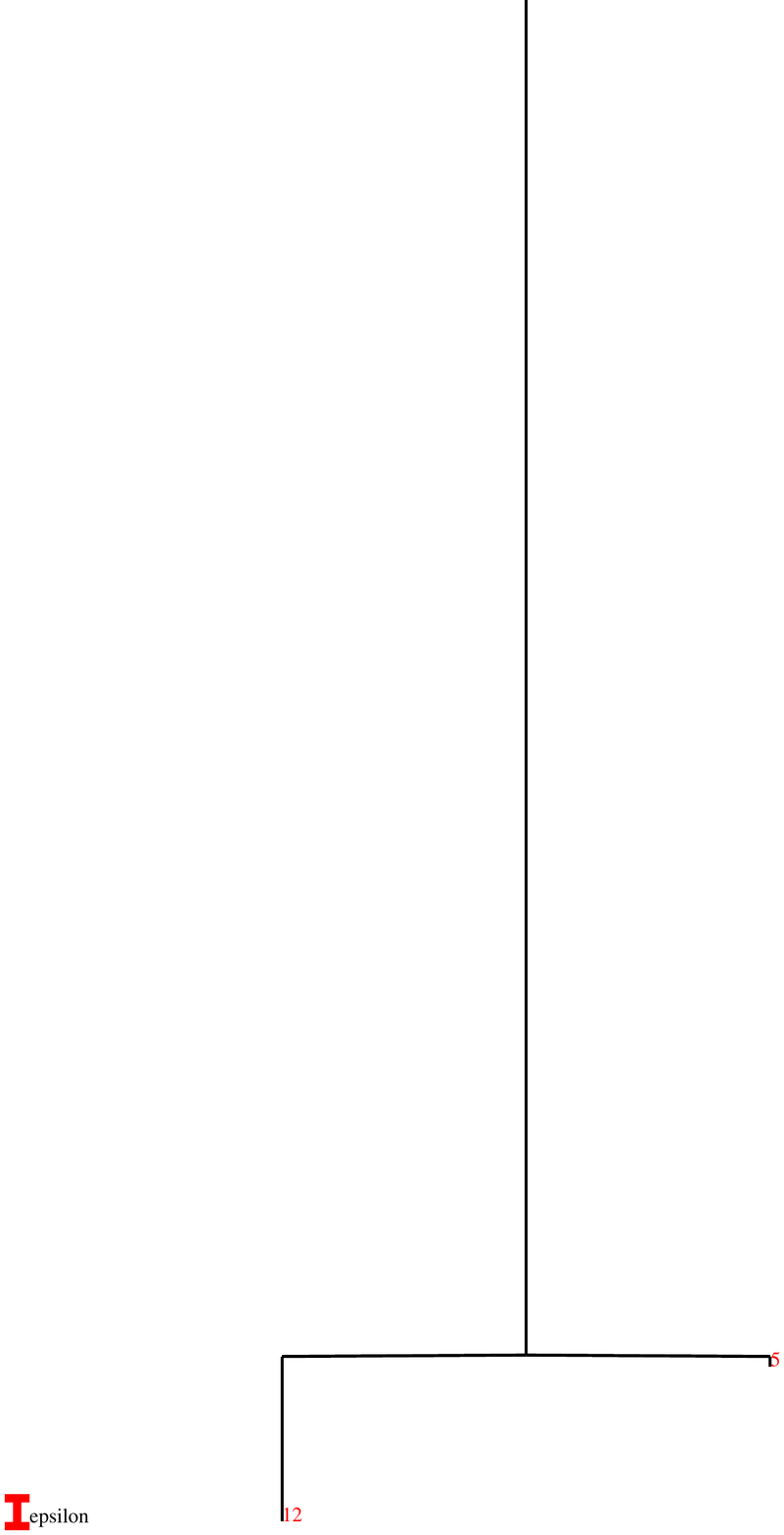}}
	\hspace{0.05\textwidth}
	\subfloat[$1.2$]{\includegraphics[width=0.05\textwidth, trim=3.7cm 0 3.7cm 0, clip]{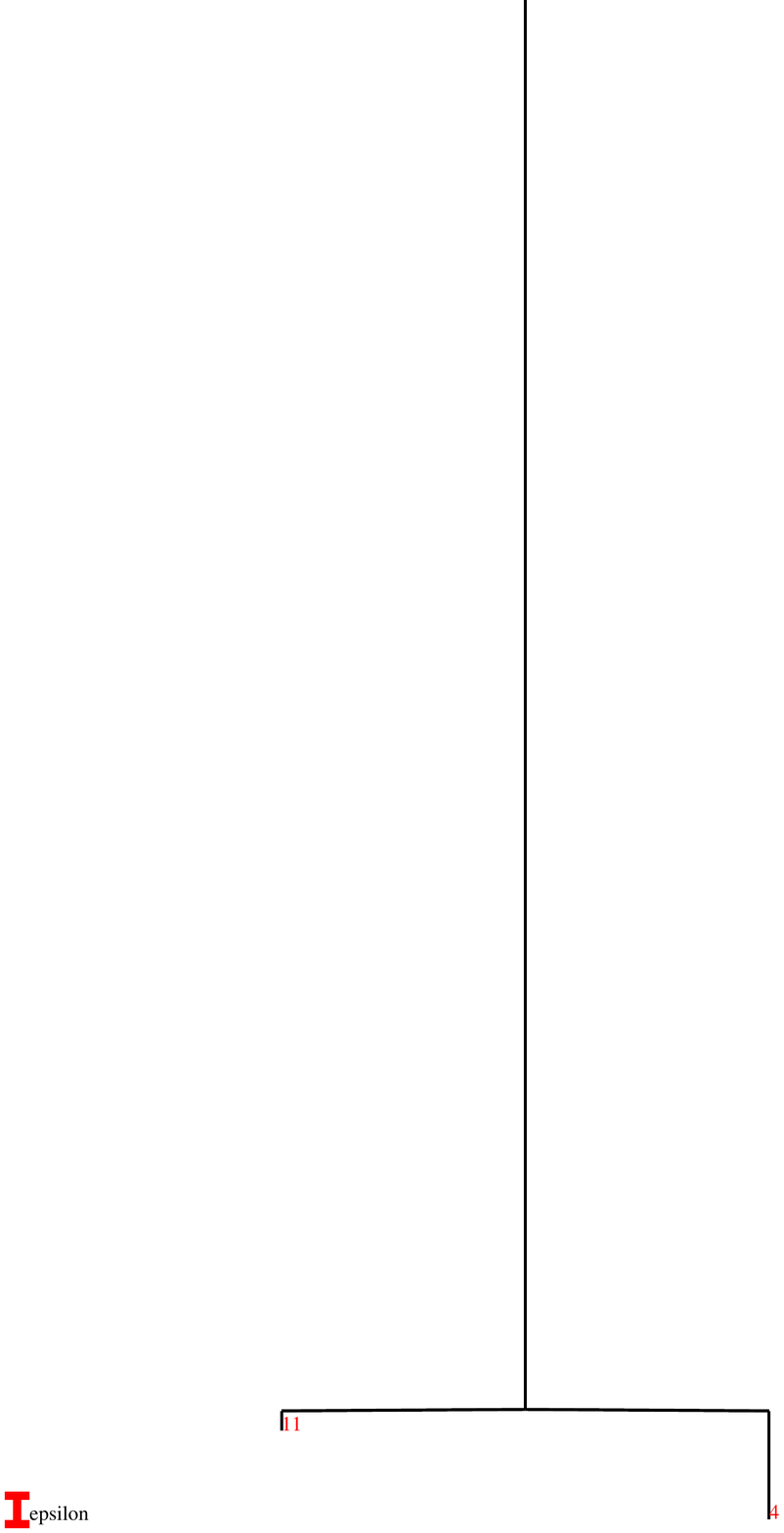}}
	\hspace{0.05\textwidth}
	\subfloat[$1.4$]{\includegraphics[width=0.05\textwidth, trim=3.7cm 0 3.7cm 0, clip]{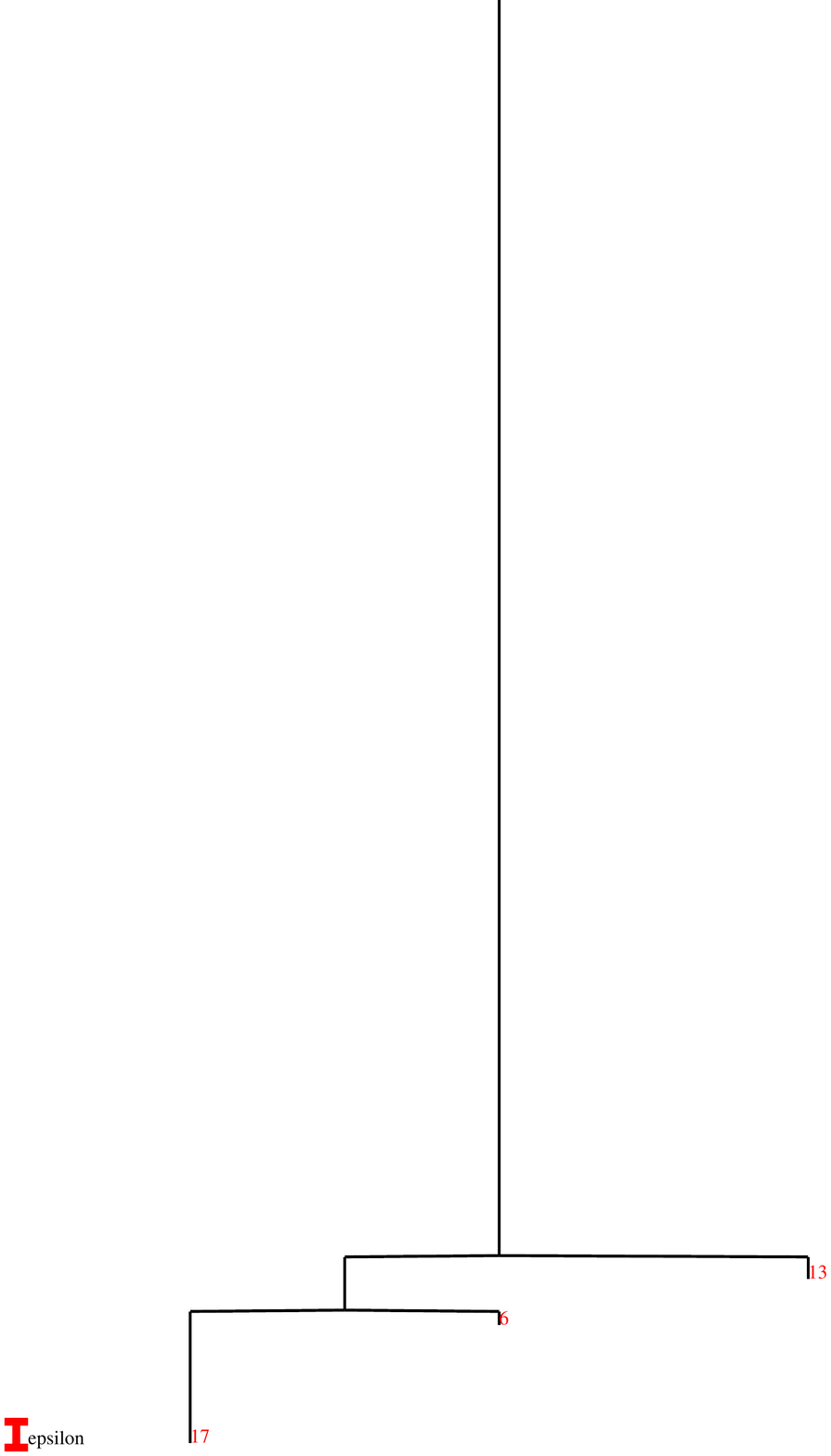}}
	\hspace{0.05\textwidth}
	\subfloat[$1.6$]{\includegraphics[width=0.05\textwidth, trim=3.7cm 0 3.7cm 0, clip]{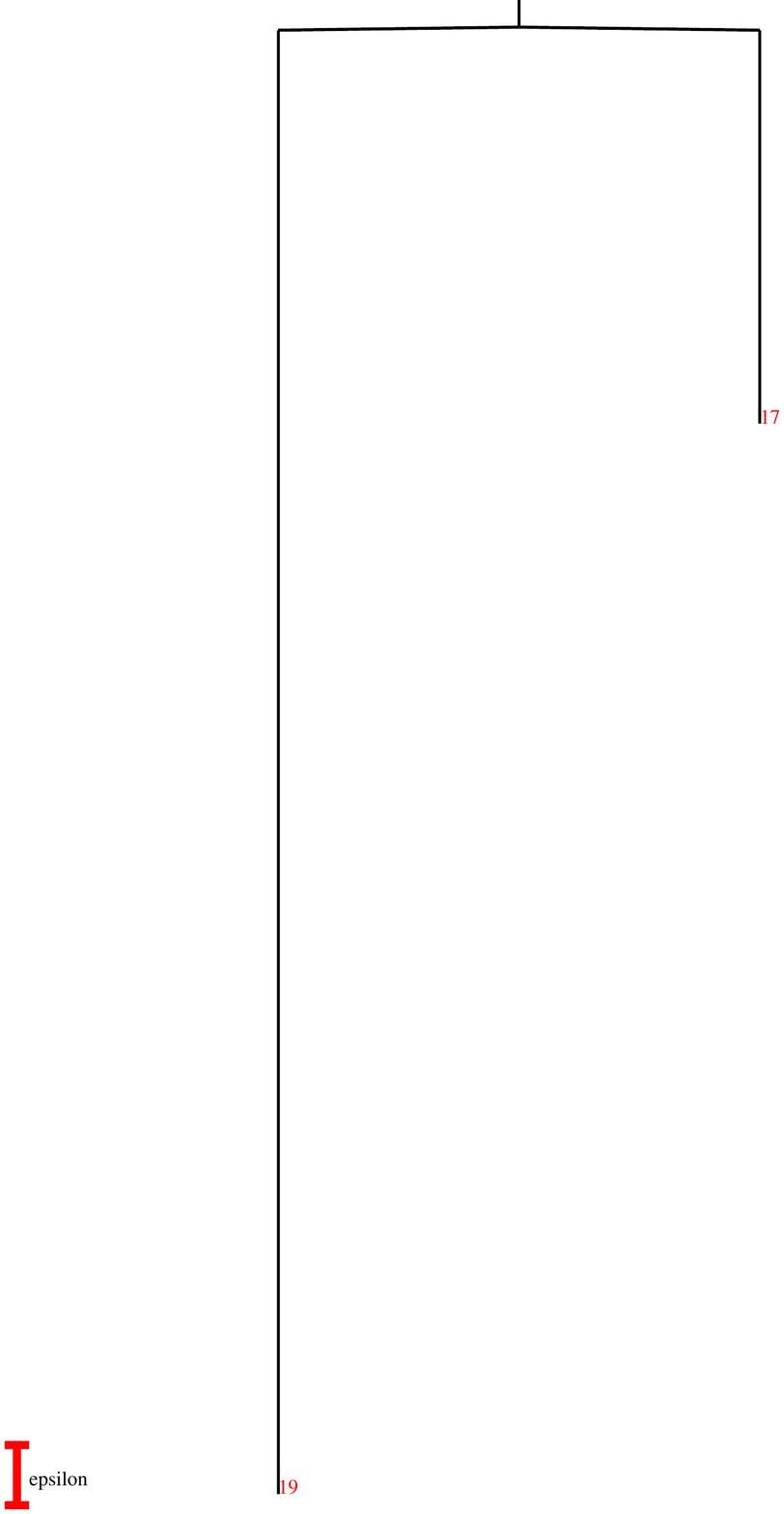}}
	\hspace{0.05\textwidth}
	\subfloat[$1.8$]{\includegraphics[width=0.05\textwidth, trim=3.7cm 0 3.7cm 0, clip]{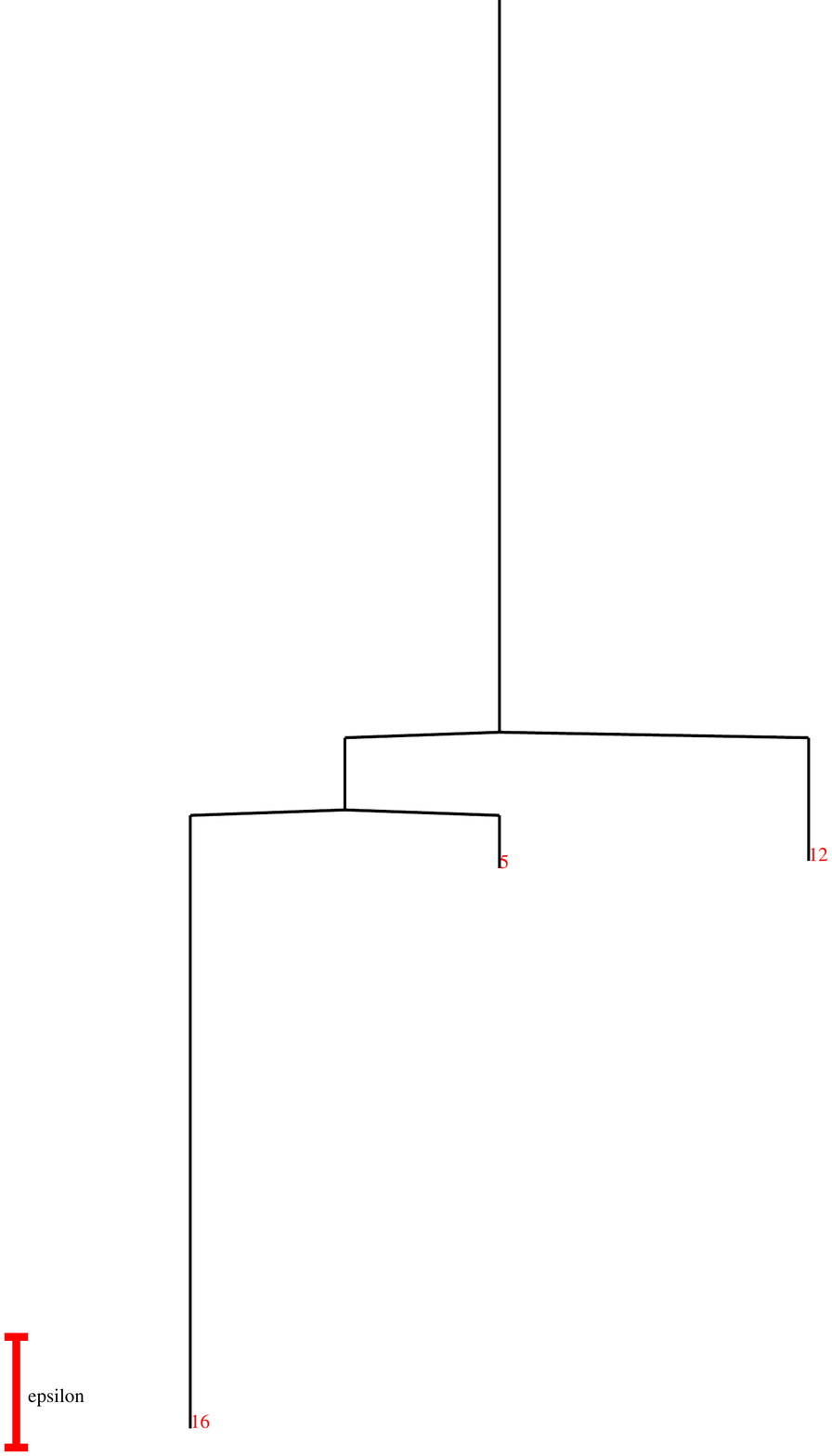}}
	\hspace{0.05\textwidth}
	\subfloat[$2.2$]{\includegraphics[width=0.05\textwidth, trim=3.7cm 0 3.7cm 0, clip]{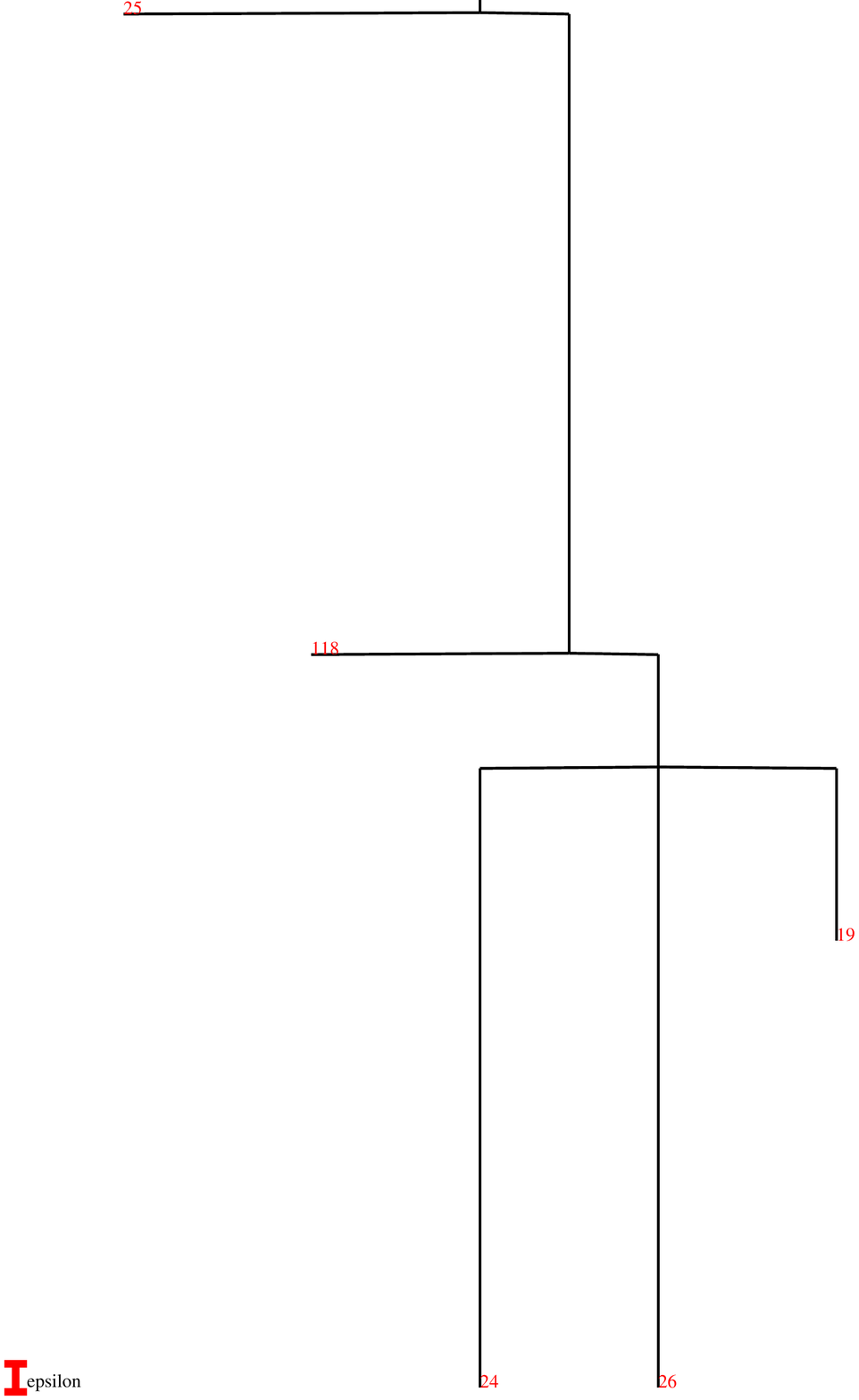}}
	\hspace{0.05\textwidth}
	\subfloat[$2.4$]{\includegraphics[width=0.05\textwidth, trim=3.7cm 0 3.7cm 0, clip]{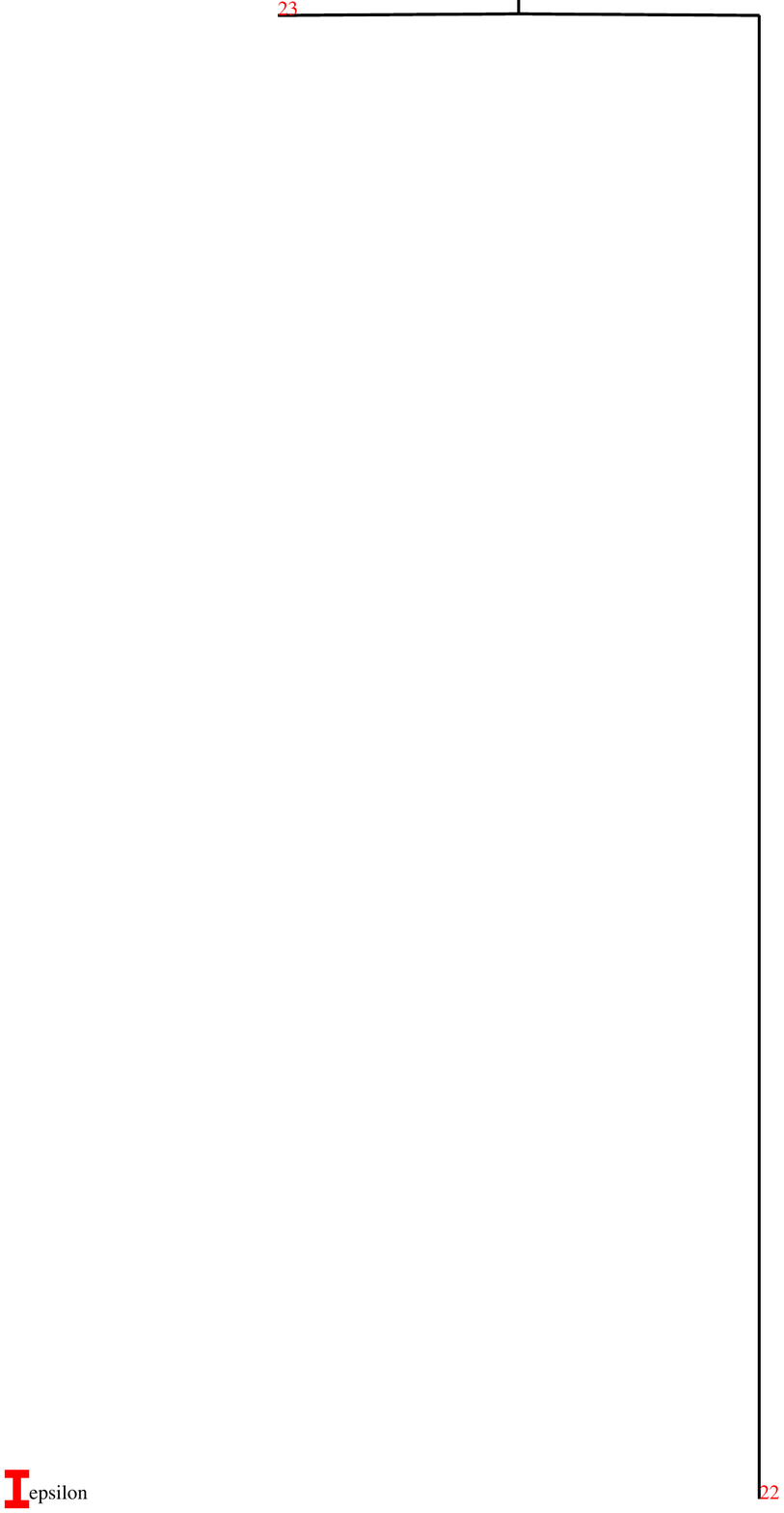}}
	\hspace{0.05\textwidth}
	\subfloat[$2.6$]{\includegraphics[width=0.05\textwidth, trim=3.7cm 0 3.7cm 0, clip]{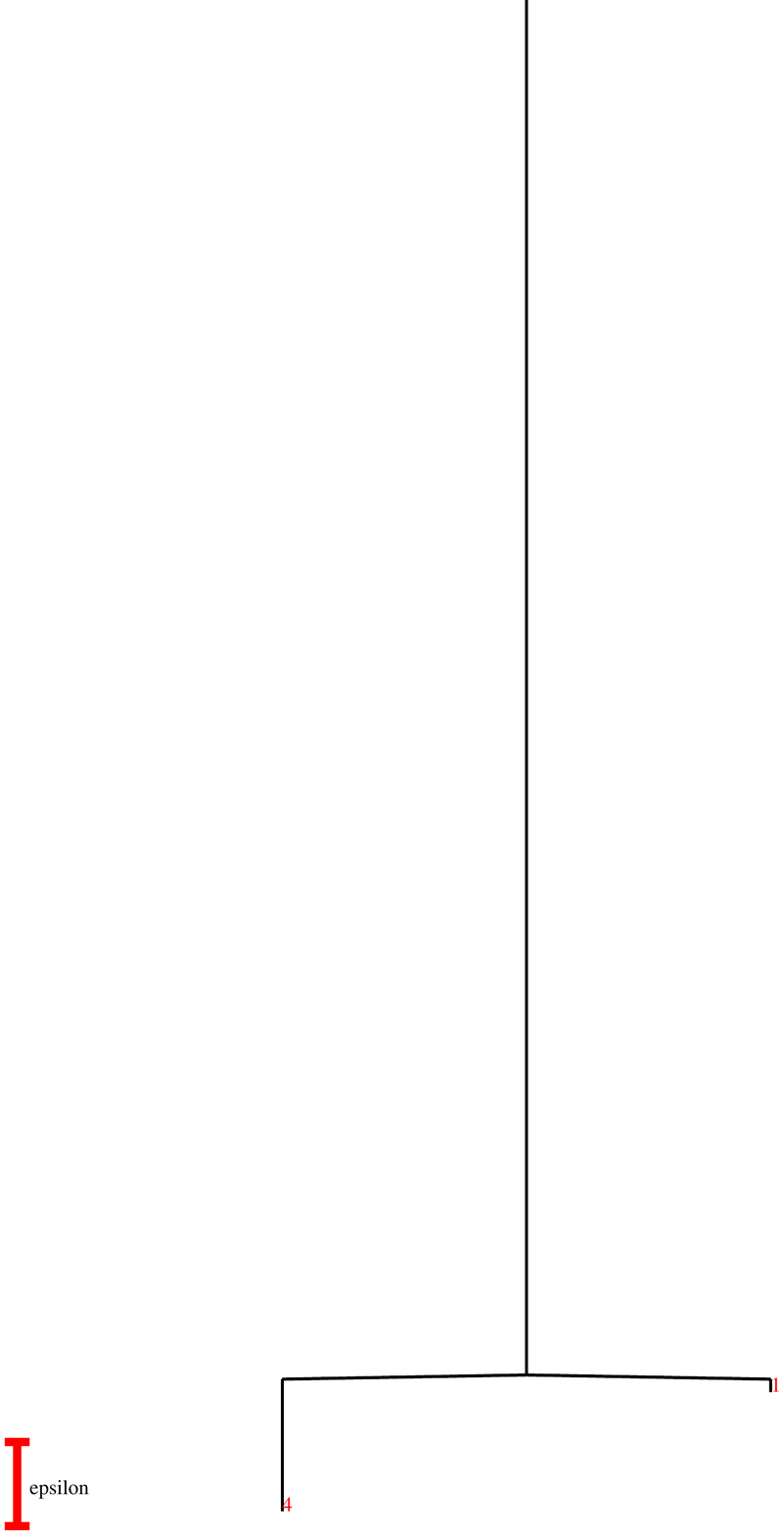}}
	\hspace{0.05\textwidth}
	\subfloat[$2.8$]{\includegraphics[width=0.05\textwidth, trim=3.7cm 0 3.7cm 0, clip]{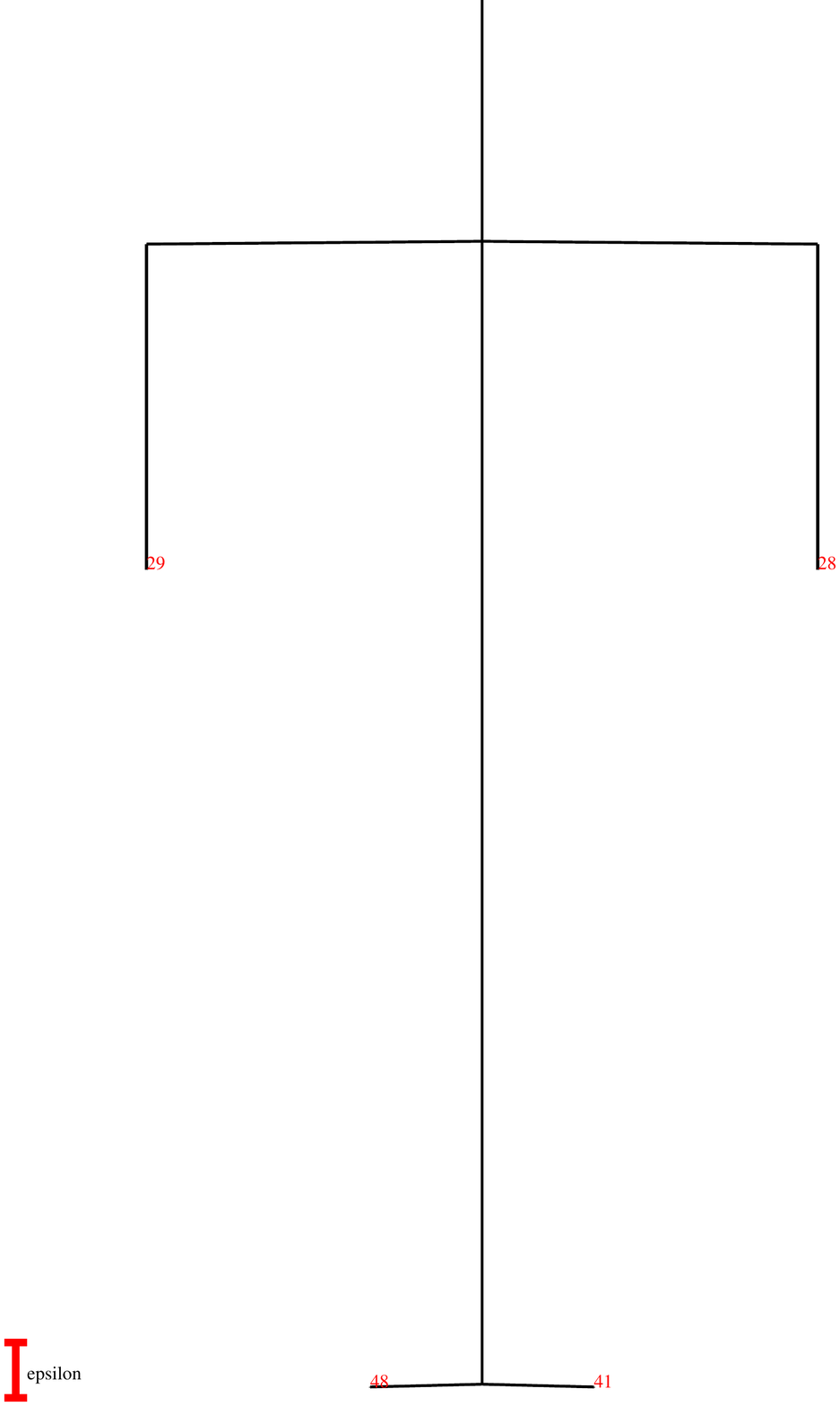}}
	\hspace{0.05\textwidth}
	\subfloat[$3.0$]{\includegraphics[width=0.05\textwidth, trim=3.7cm 0 3.7cm 0, clip]{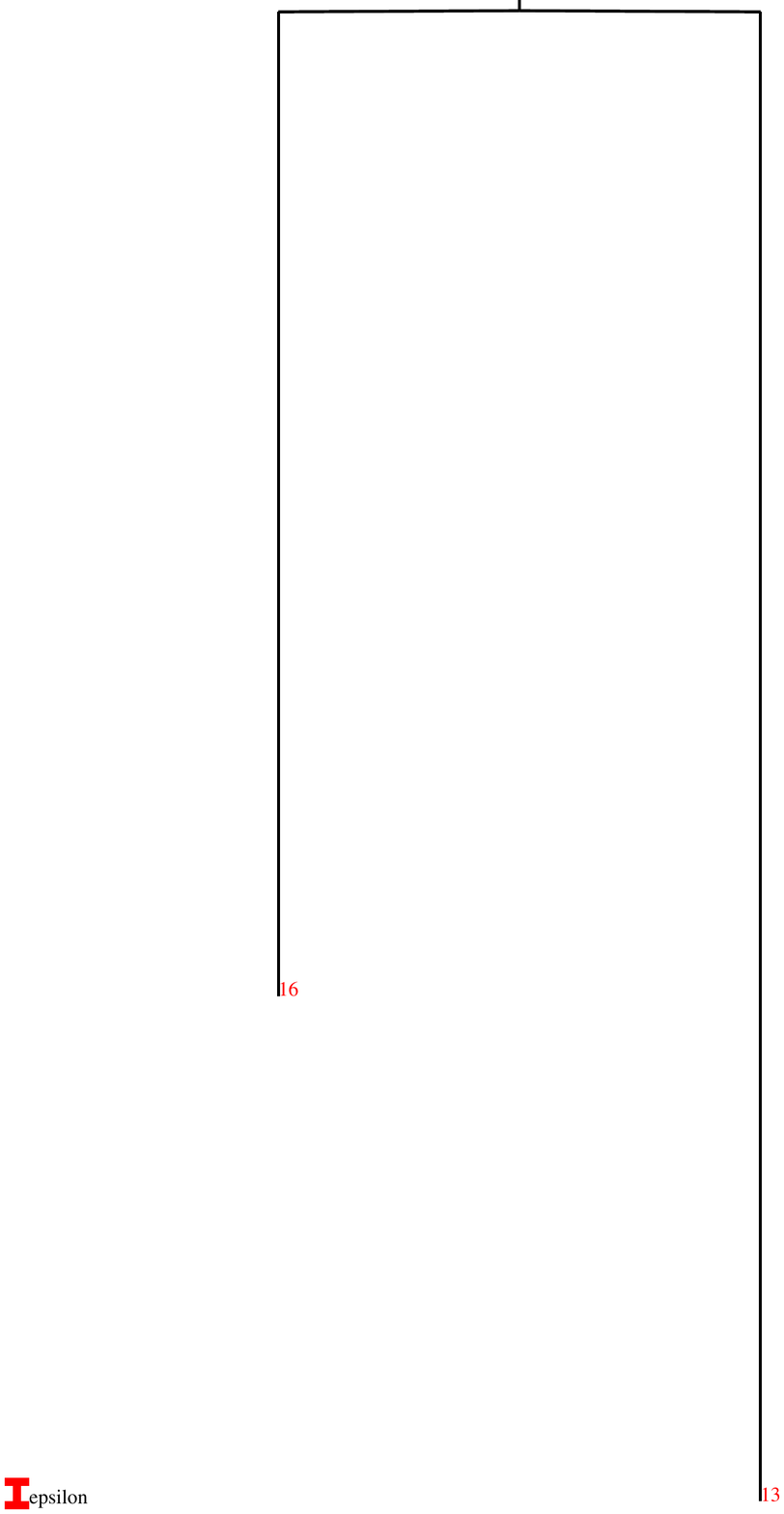}}
	\hspace{0.05\textwidth}
	\subfloat[$3.2$]{\includegraphics[width=0.05\textwidth, trim=3.7cm 0 3.7cm 0, clip]{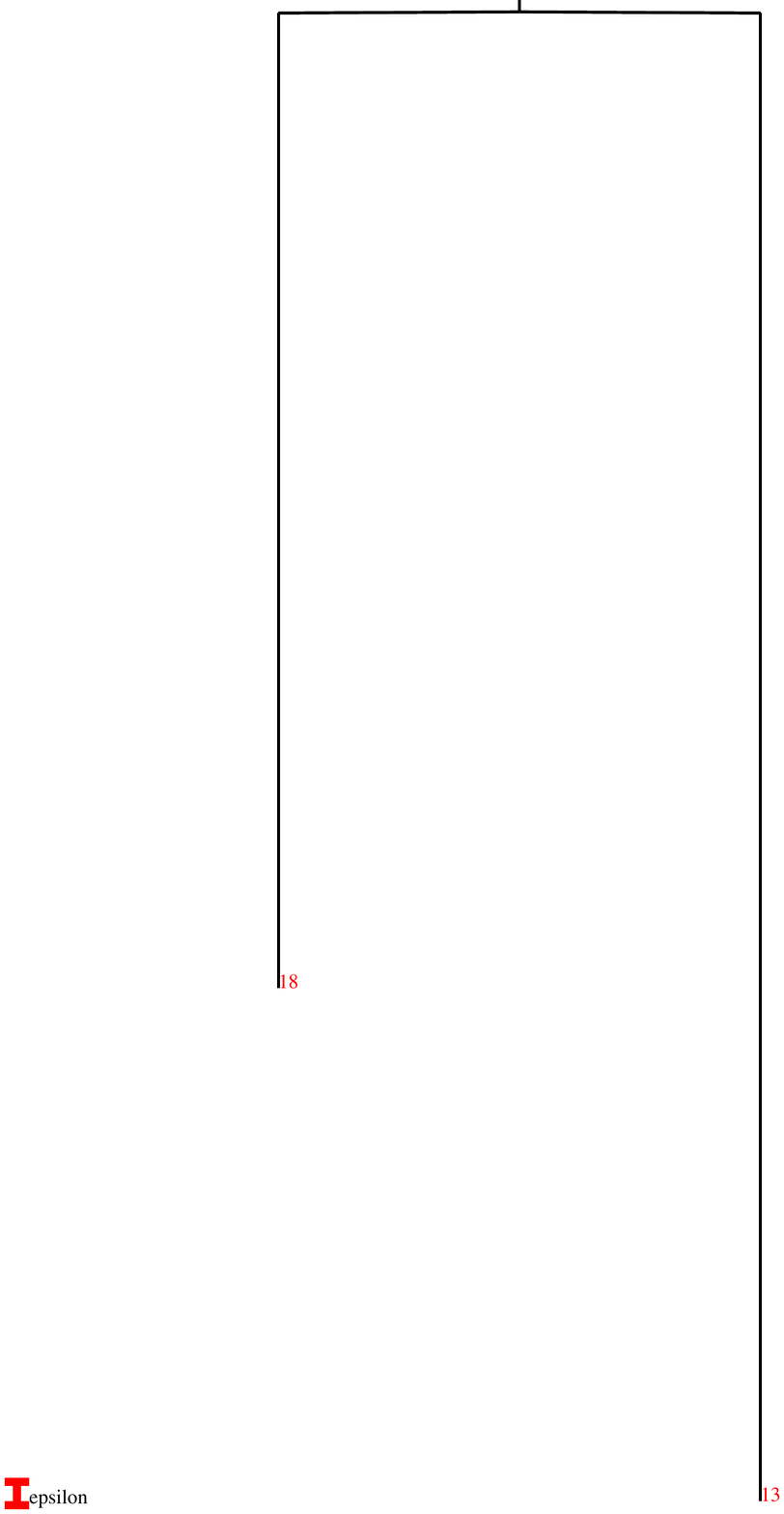}}
	\hspace{0.05\textwidth}
	\subfloat[$3.6$]{\includegraphics[width=0.05\textwidth, trim=3.7cm 0 3.7cm 0, clip]{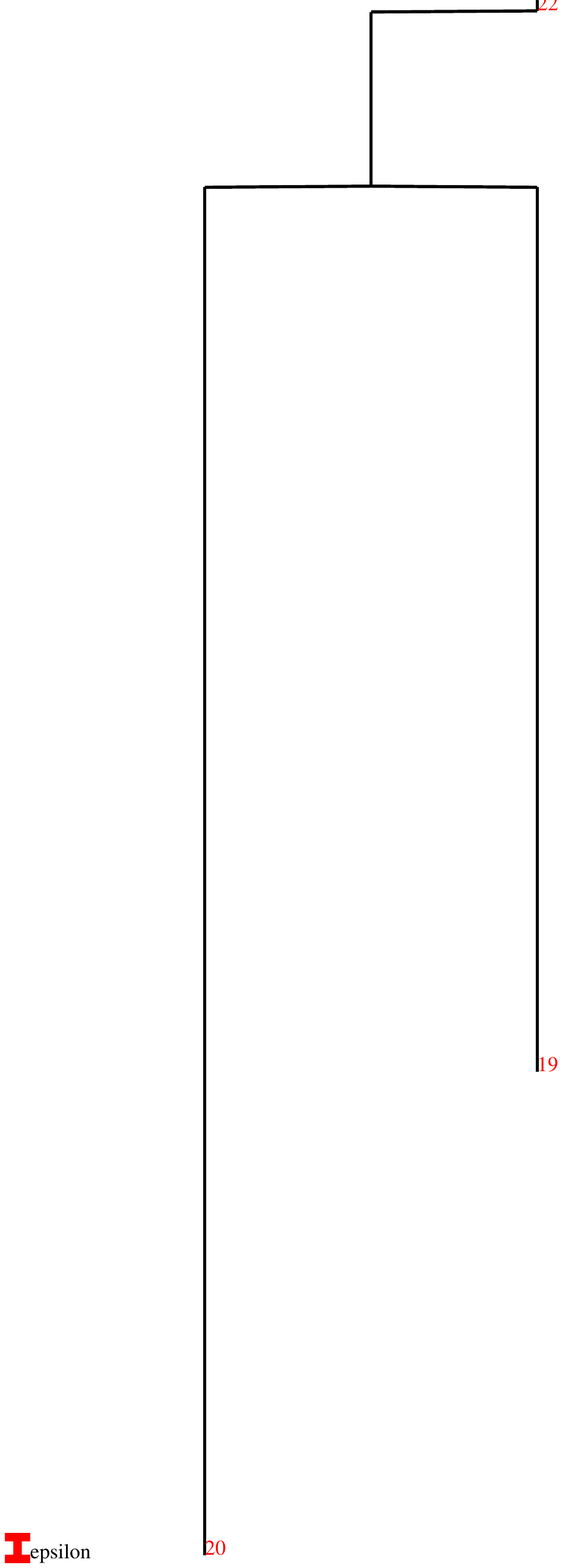}}
	\hspace{0.05\textwidth}
	\subfloat[$3.8$]{\includegraphics[width=0.05\textwidth, trim=3.7cm 0 3.7cm 0, clip]{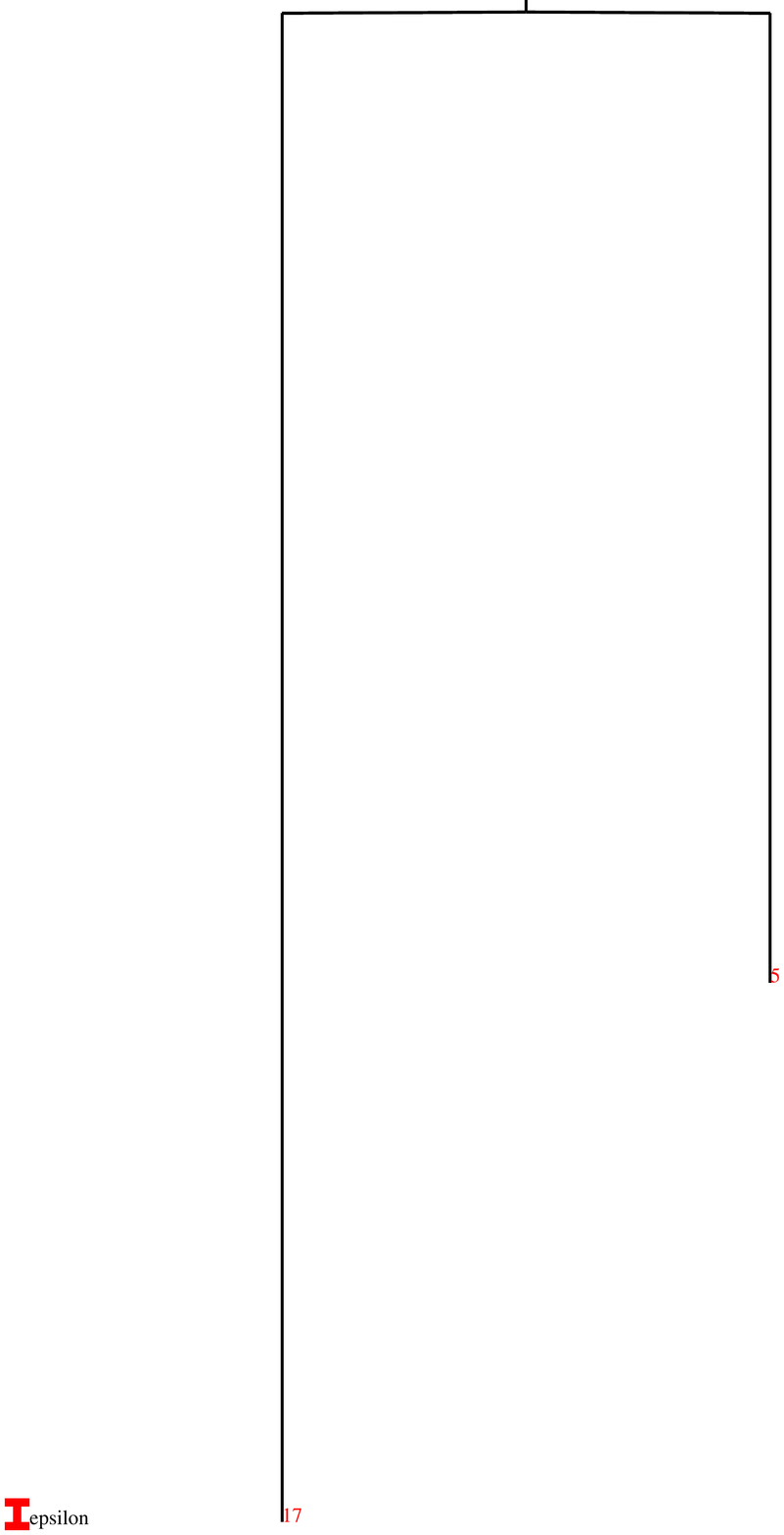}}
	\hspace{0.05\textwidth}
	\subfloat[$4.0$]{\includegraphics[width=0.05\textwidth, trim=3.7cm 0 3.7cm 0, clip]{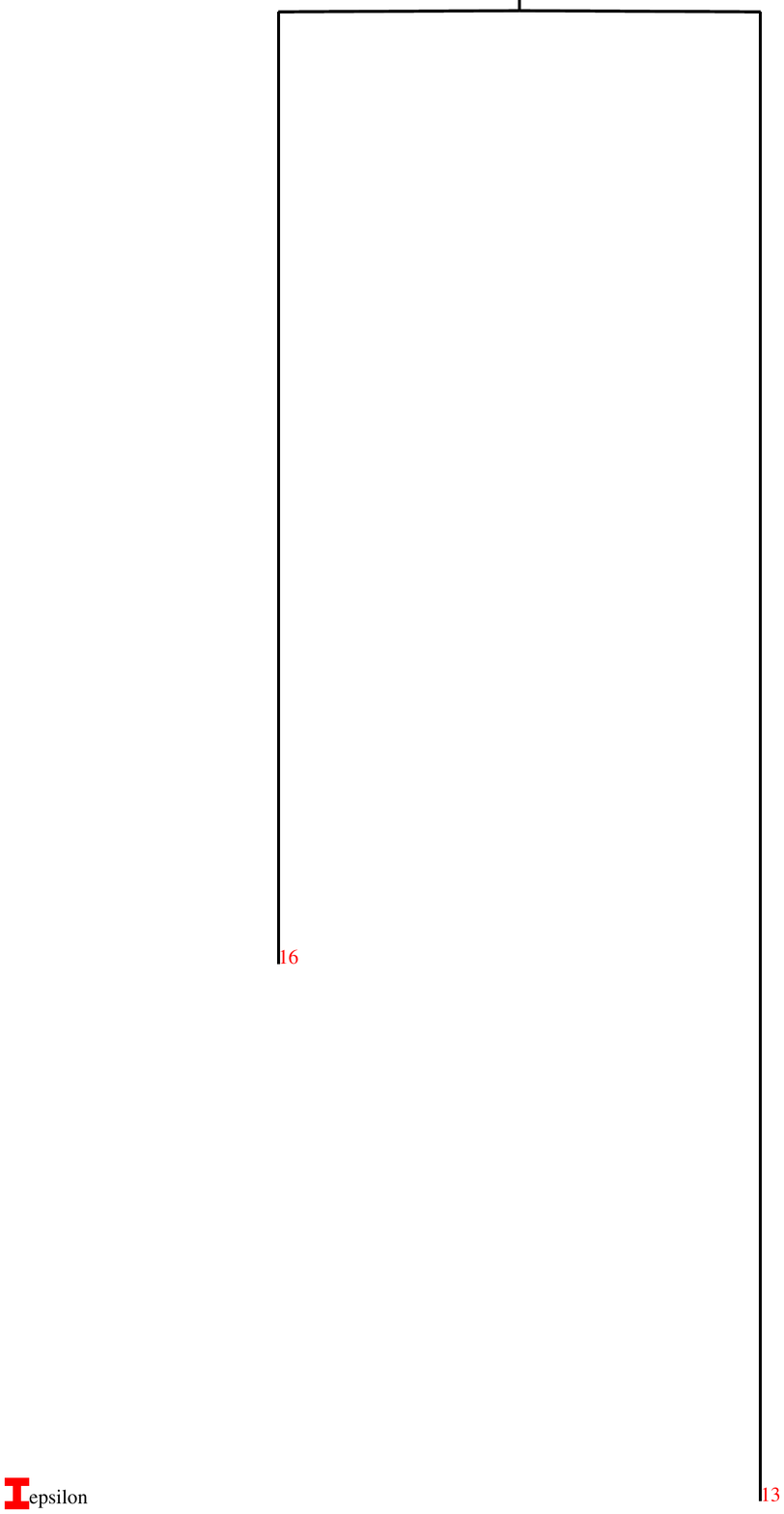}}
	\hspace{0.05\textwidth}
	\subfloat[$4.4$]{\includegraphics[width=0.05\textwidth, trim=3.7cm 0 3.7cm 0, clip]{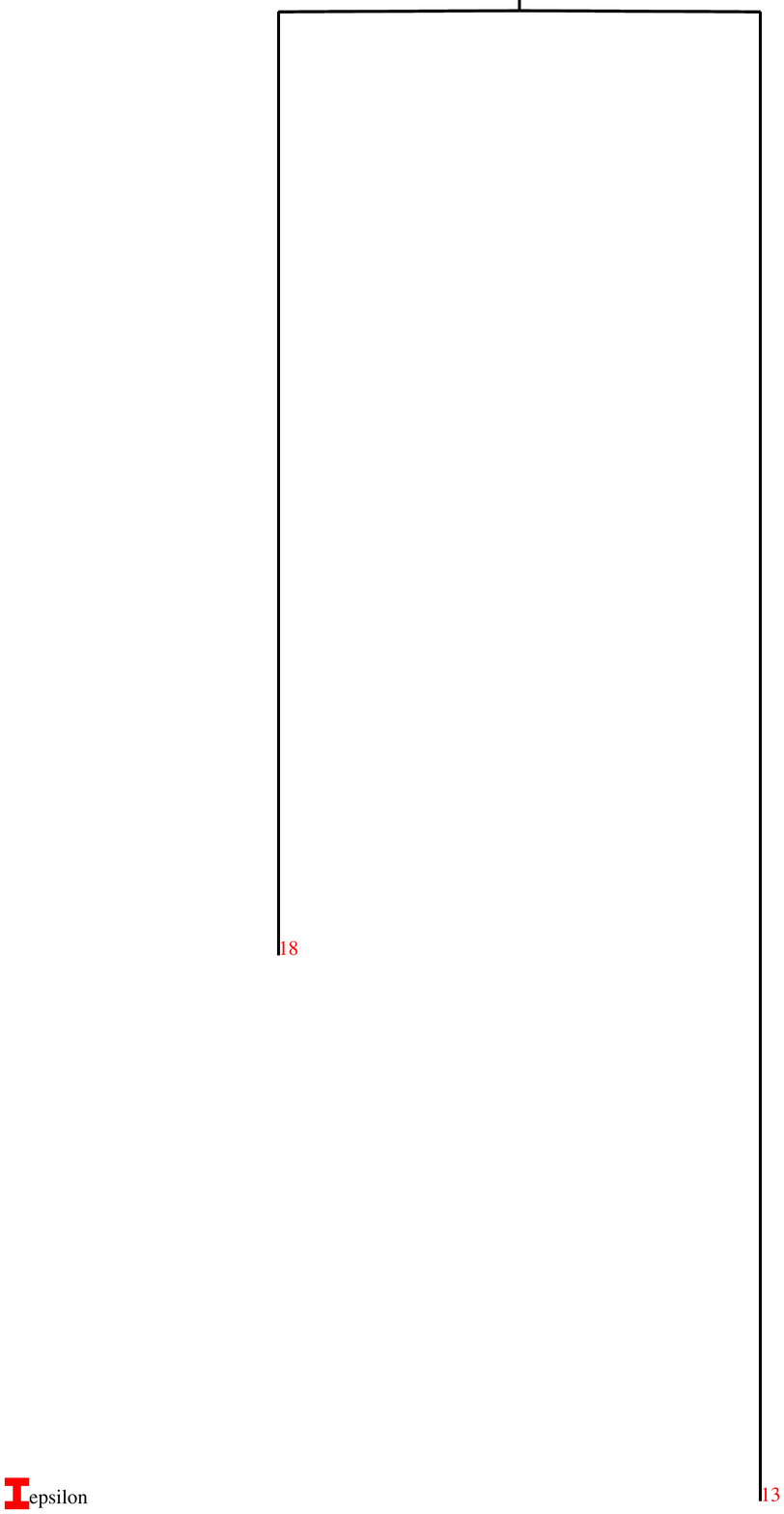}}
	\hspace{0.05\textwidth}
	\subfloat[$4.6$]{\includegraphics[width=0.05\textwidth, trim=3.7cm 0 3.7cm 0, clip]{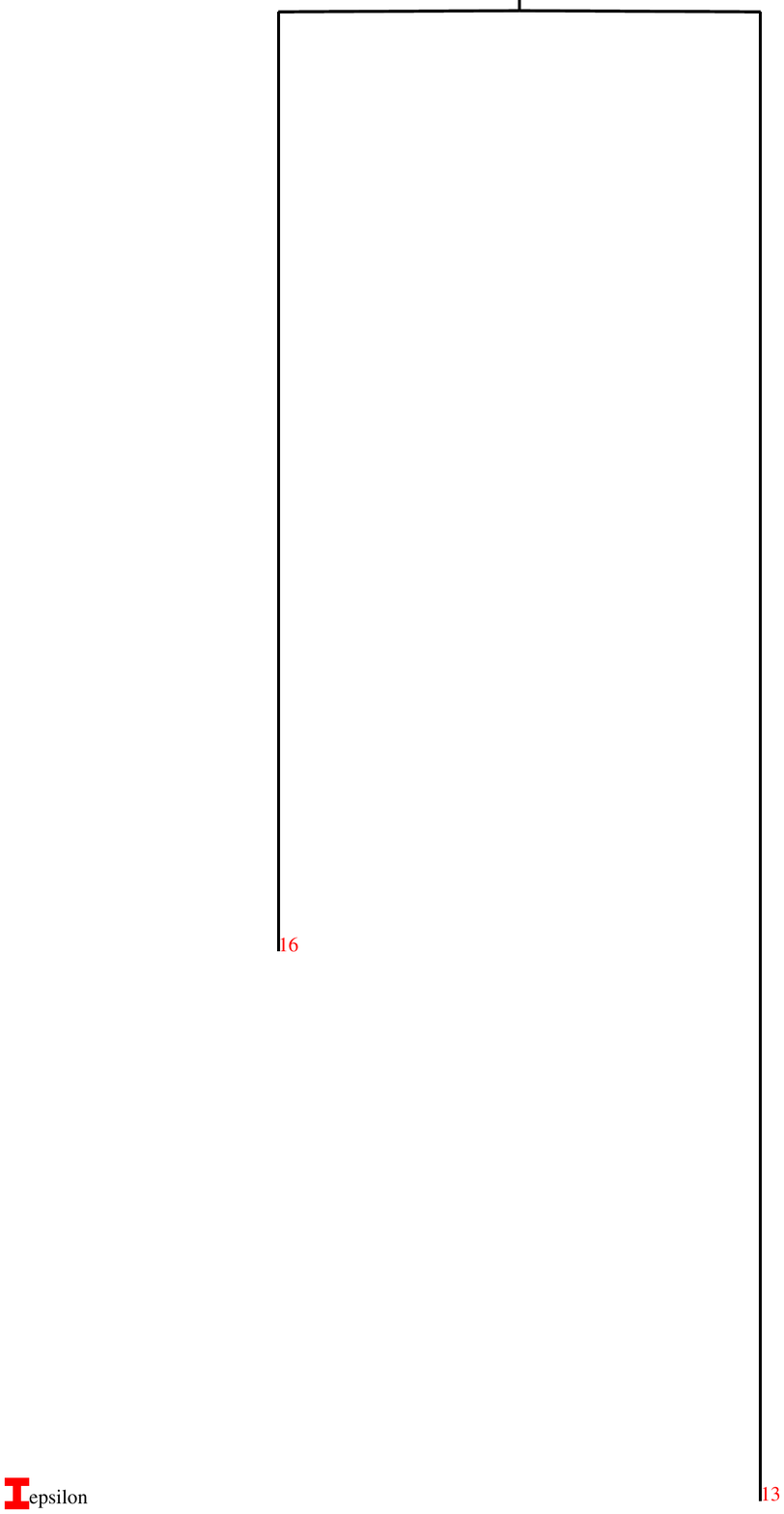}}
	\hspace{0.05\textwidth}
	\subfloat[$5.0$]{\includegraphics[width=0.05\textwidth, trim=3.7cm 0 3.7cm 0, clip]{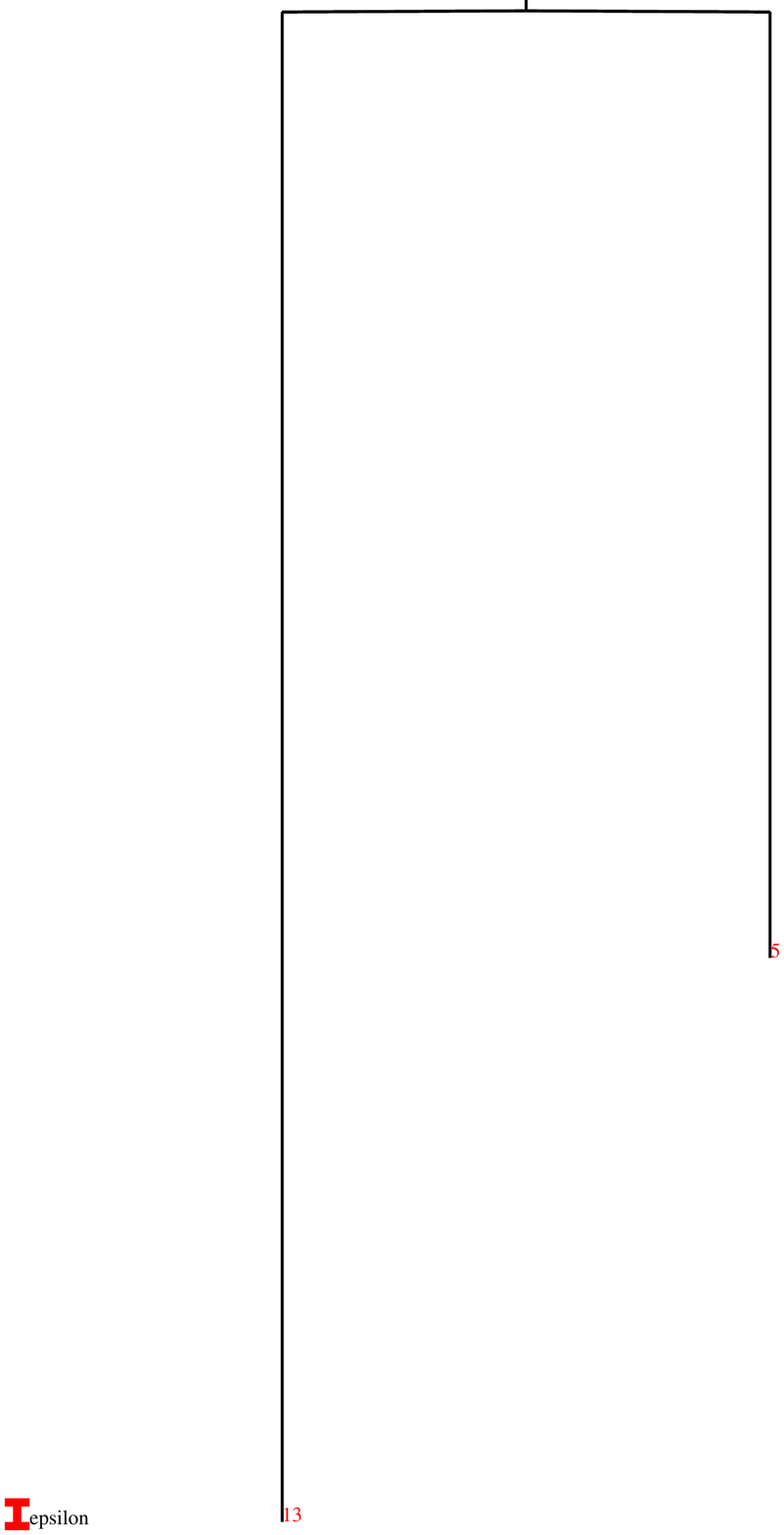}}
	\hspace{0.05\textwidth}
	\subfloat[$6.2$]{\includegraphics[width=0.05\textwidth, trim=3.7cm 0 3.7cm 0, clip]{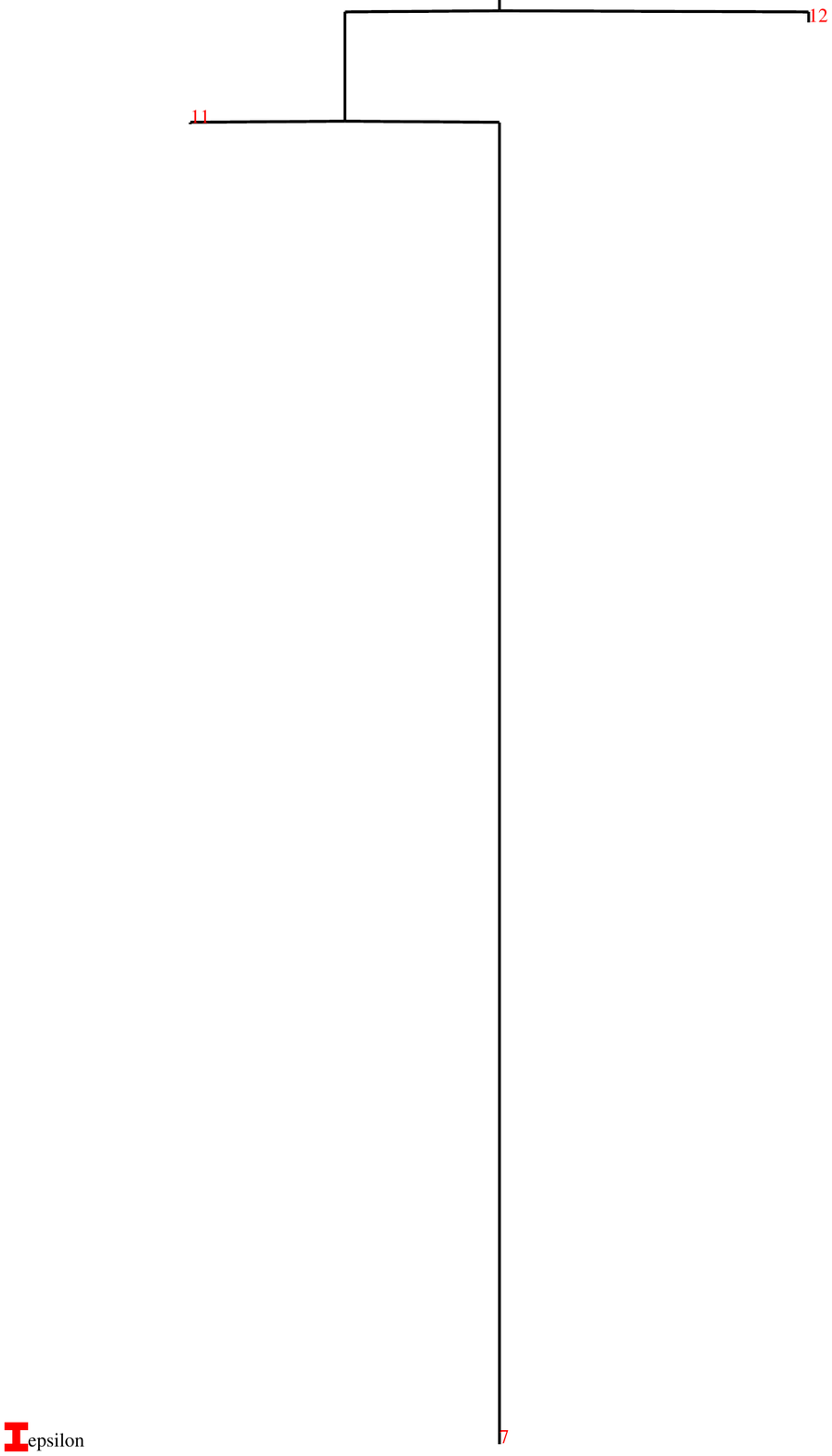}}
	\hspace{0.05\textwidth}
	\subfloat[$7.0$]{\includegraphics[width=0.05\textwidth, trim=3.7cm 0 3.7cm 0, clip]{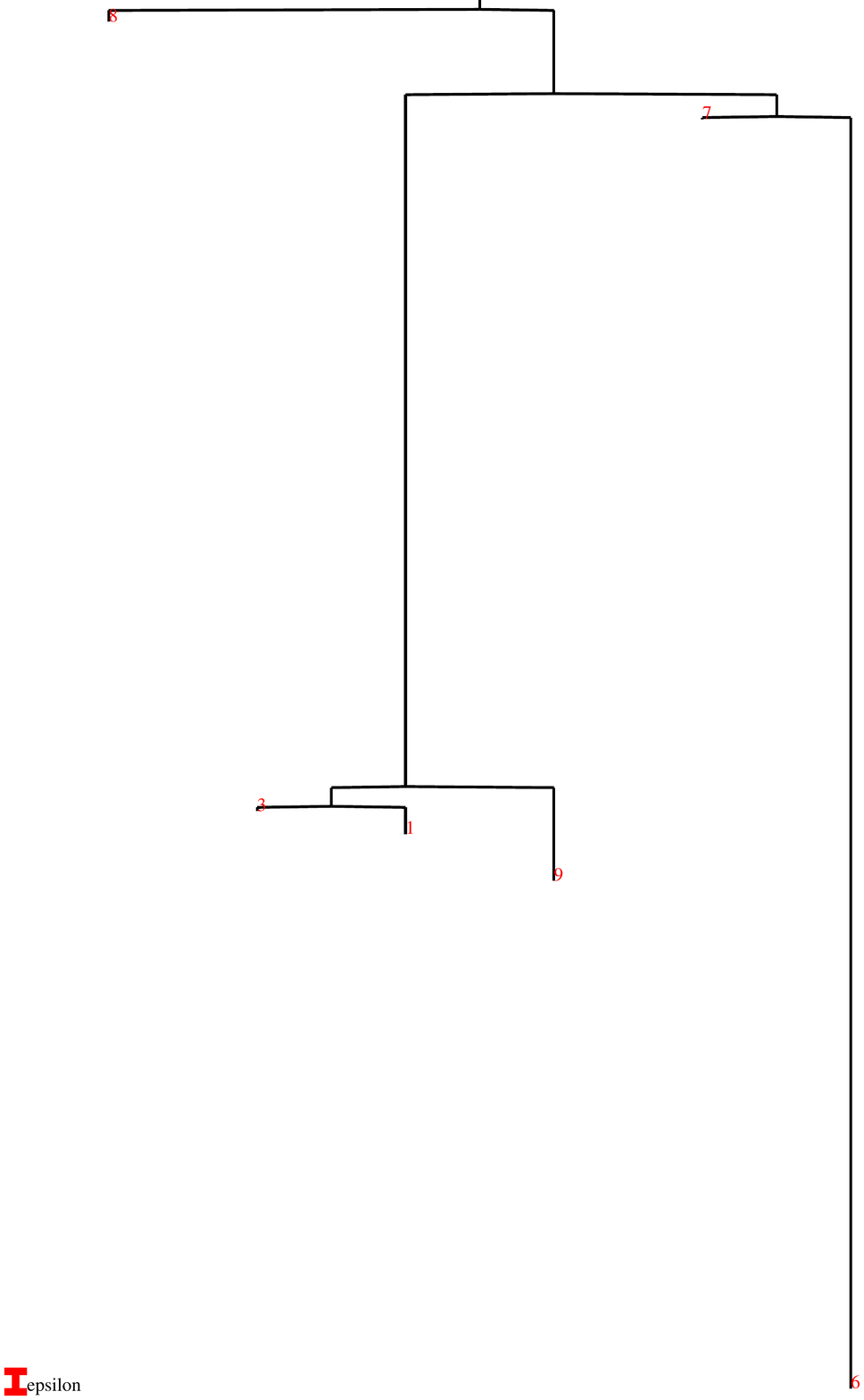}}
	\hspace{0.05\textwidth}
	\subfloat[$7.4$]{\includegraphics[width=0.05\textwidth, trim=3.7cm 0 3.7cm 0, clip]{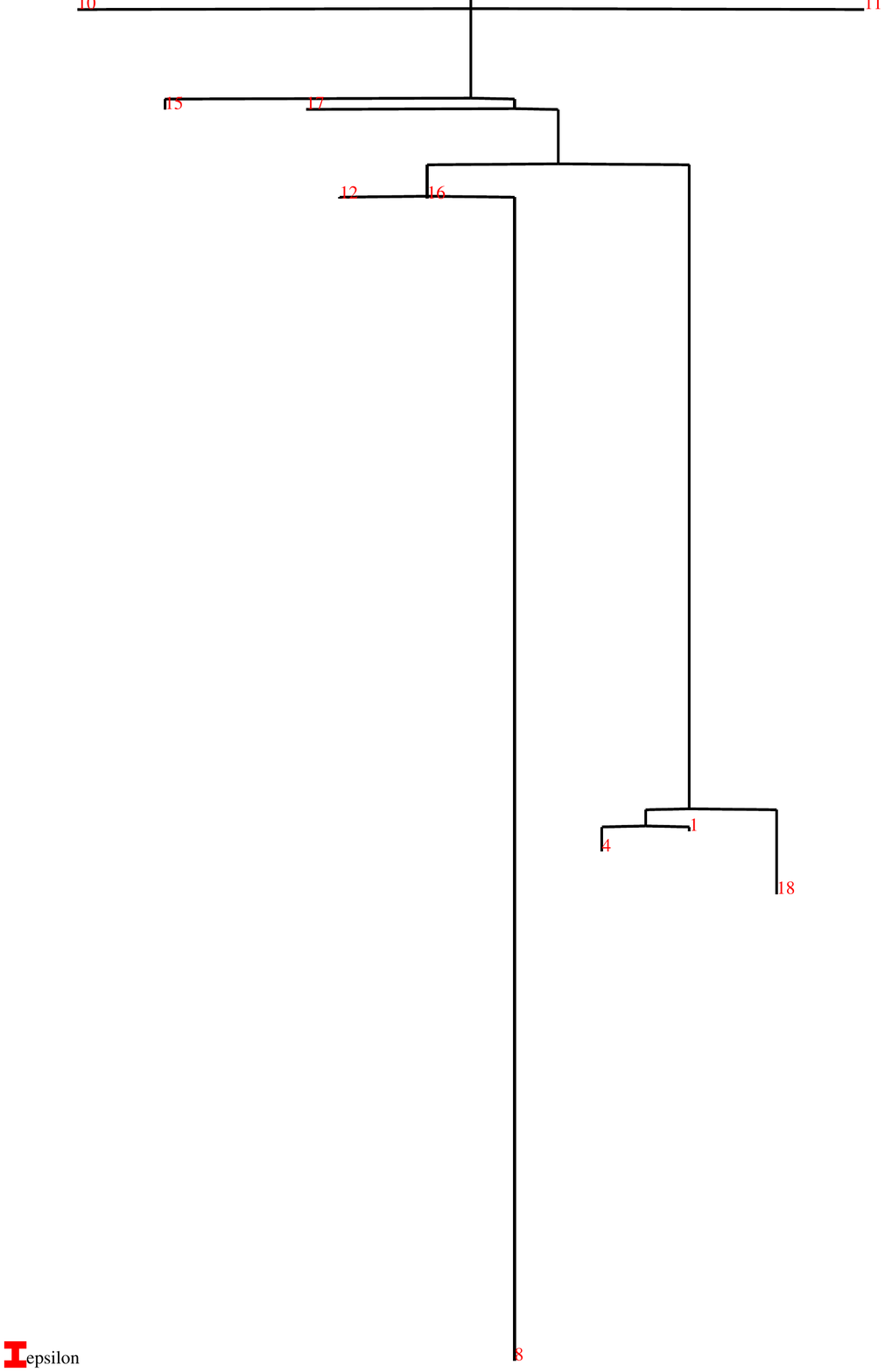}}
	\hspace{0.05\textwidth}
	\subfloat[$7.8$]{\includegraphics[width=0.05\textwidth, trim=3.7cm 0 3.7cm 0, clip]{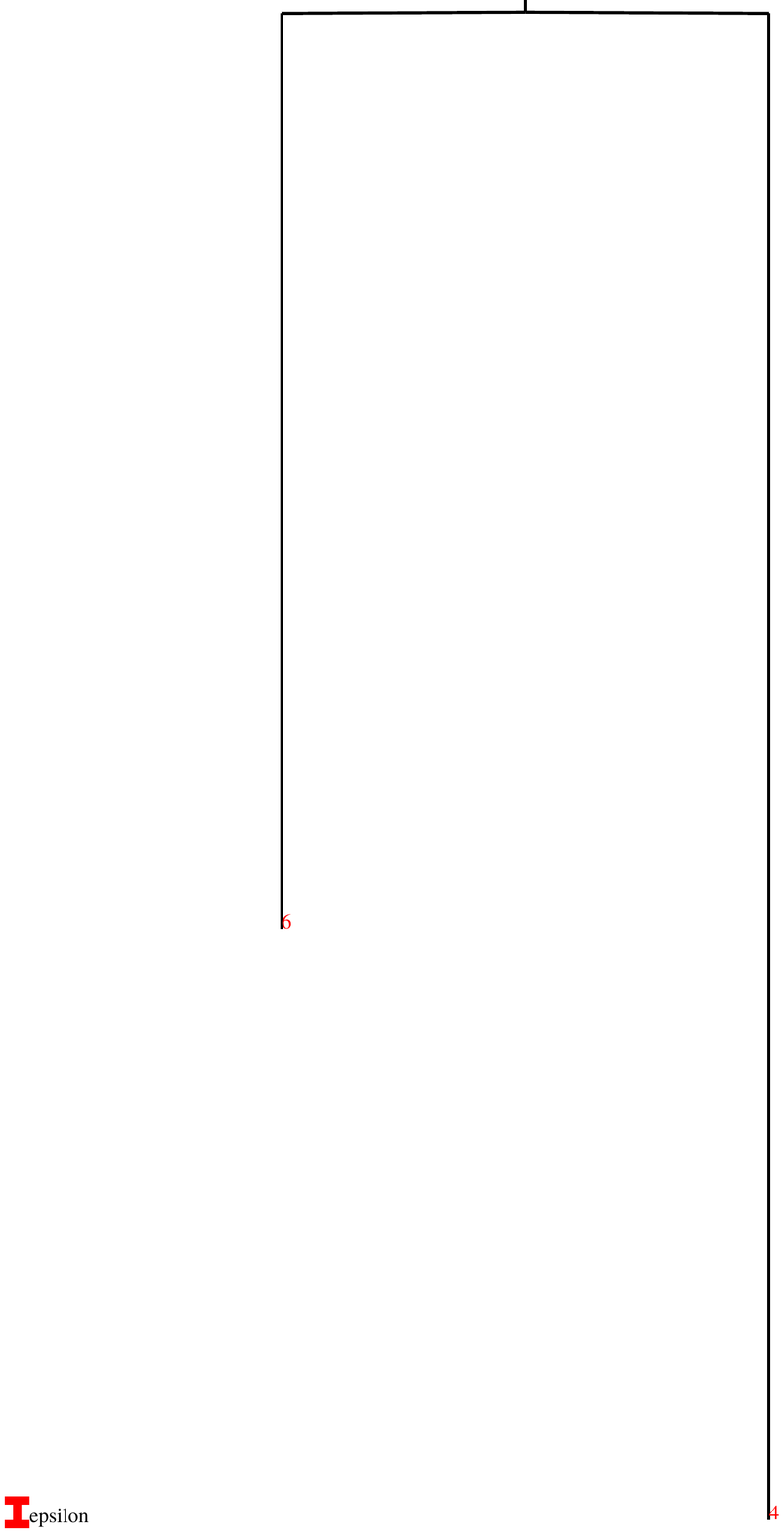}}
	\hspace{0.05\textwidth}
	\subfloat[$8.2$]{\includegraphics[width=0.05\textwidth, trim=3.7cm 0 3.7cm 0, clip]{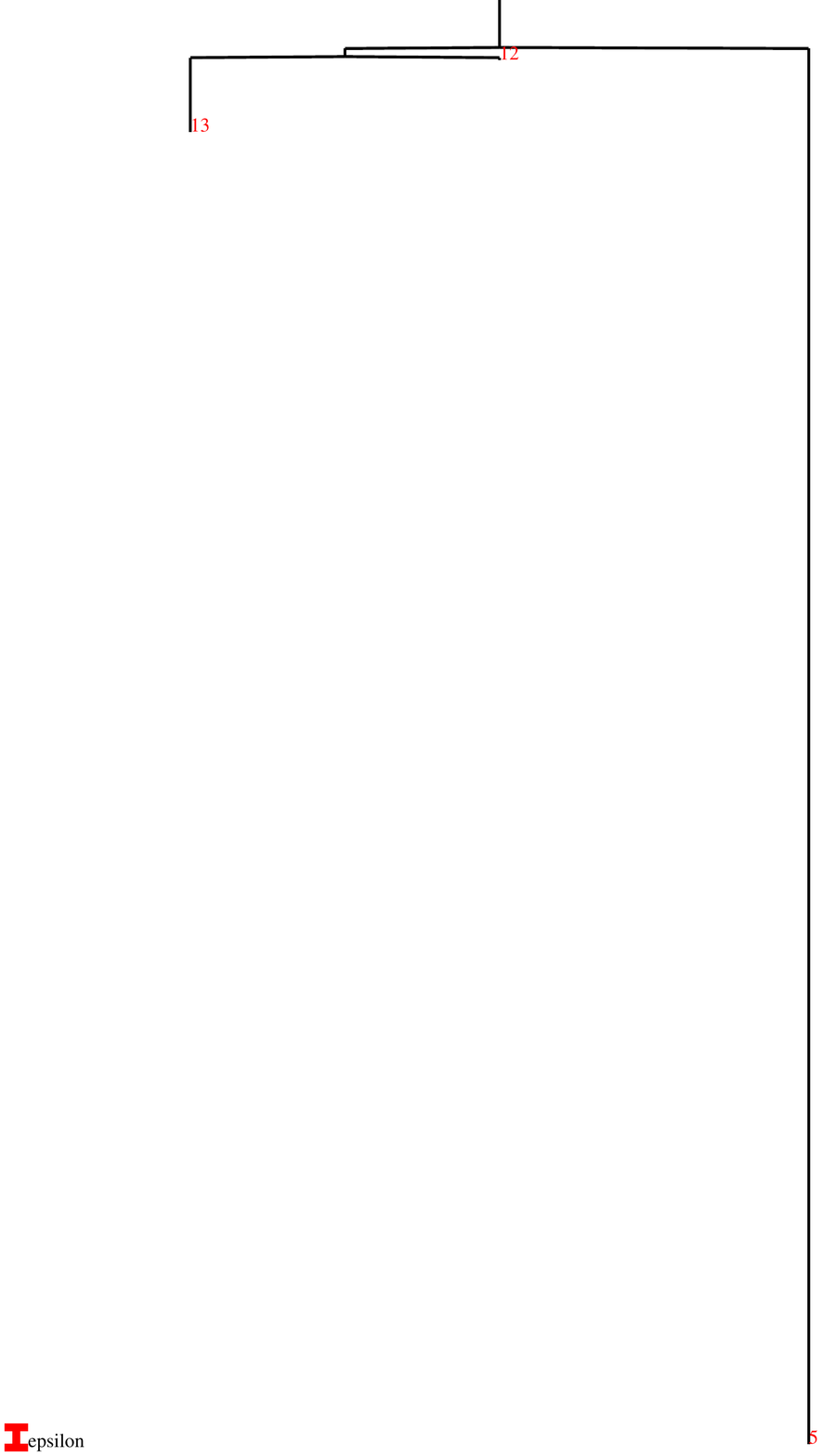}}
	\hspace{0.05\textwidth}
	\subfloat[$8.6$]{\includegraphics[width=0.05\textwidth, trim=3.7cm 0 3.7cm 0, clip]{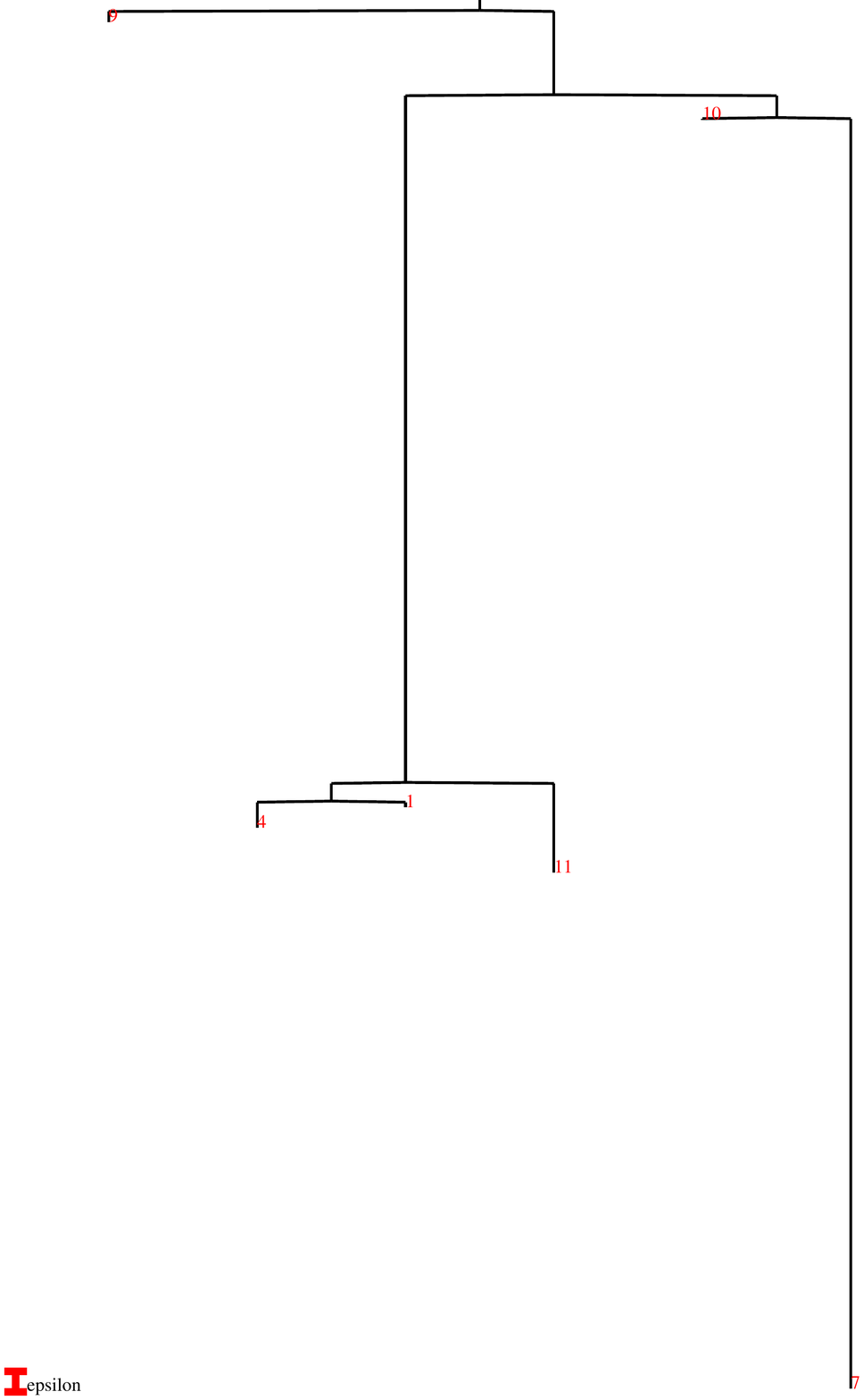}}
	\hspace{0.05\textwidth}
	\subfloat[$9.4$]{\includegraphics[width=0.05\textwidth, trim=3.7cm 0 3.7cm 0, clip]{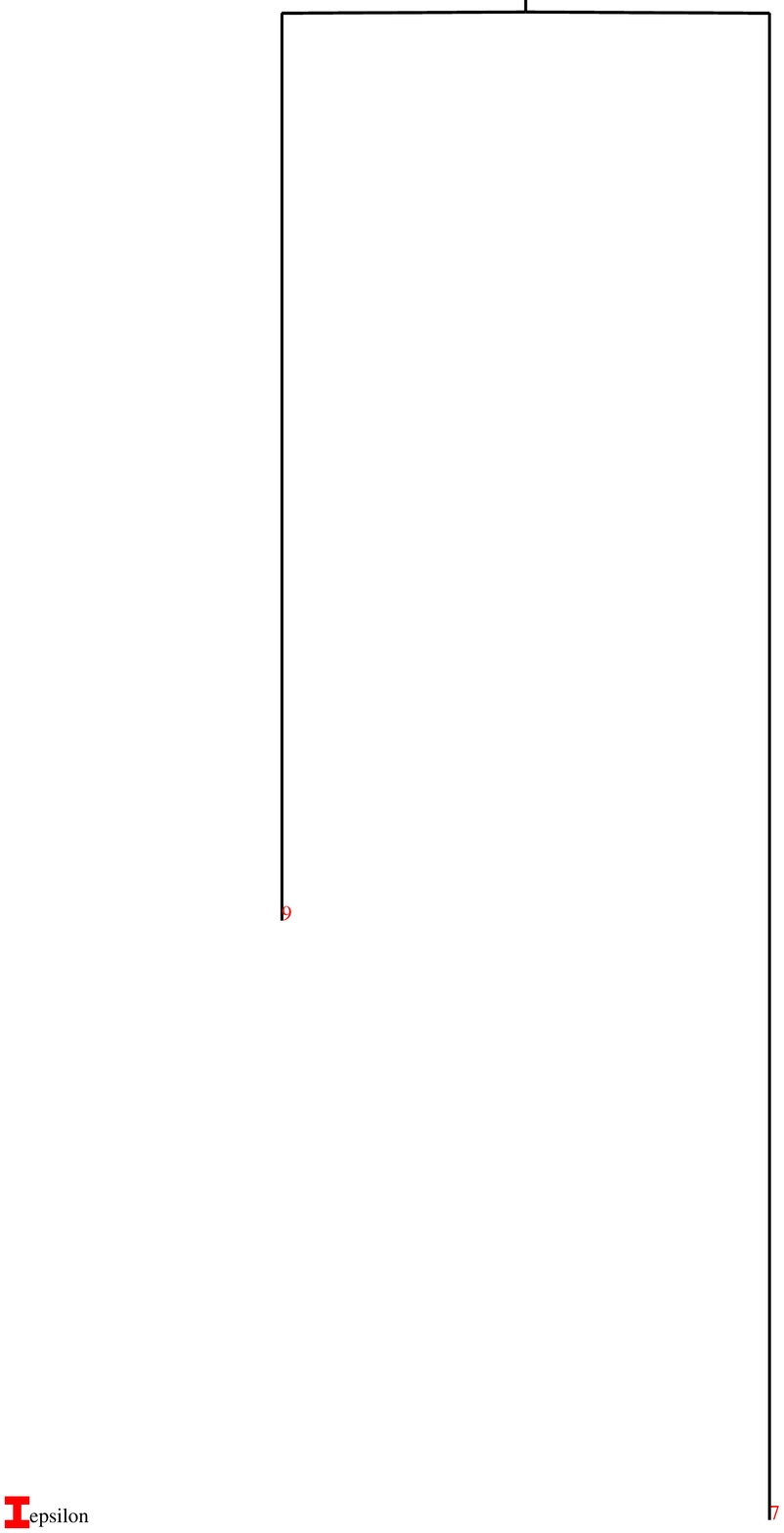}}
	\hspace{0.05\textwidth}
	\subfloat[$9.8$]{\includegraphics[width=0.05\textwidth, trim=3.7cm 0 3.7cm 0, clip]{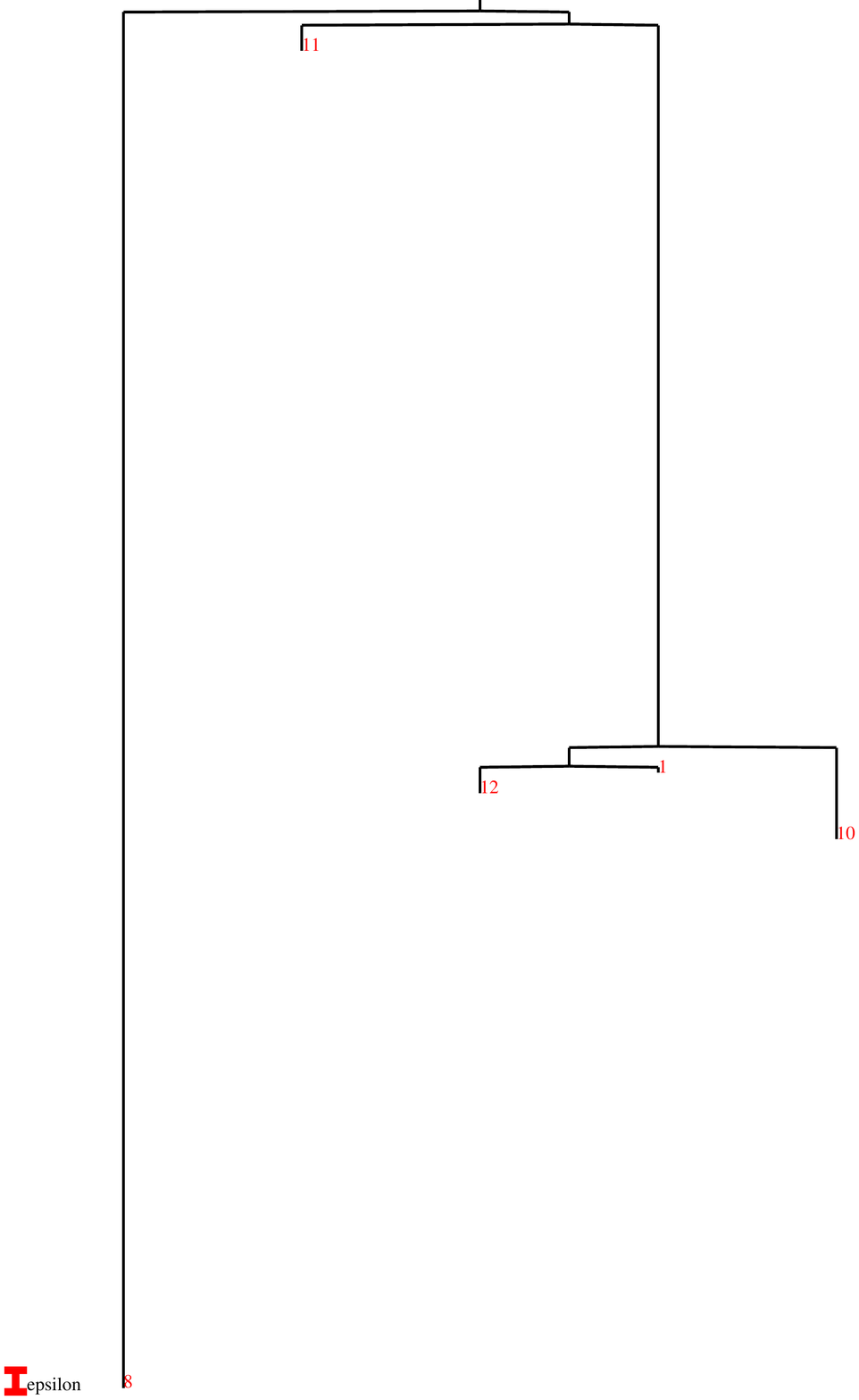}}
	\hspace{0.05\textwidth}
	\subfloat[$10.2$]{\includegraphics[width=0.05\textwidth, trim=3.7cm 0 3.7cm 0, clip]{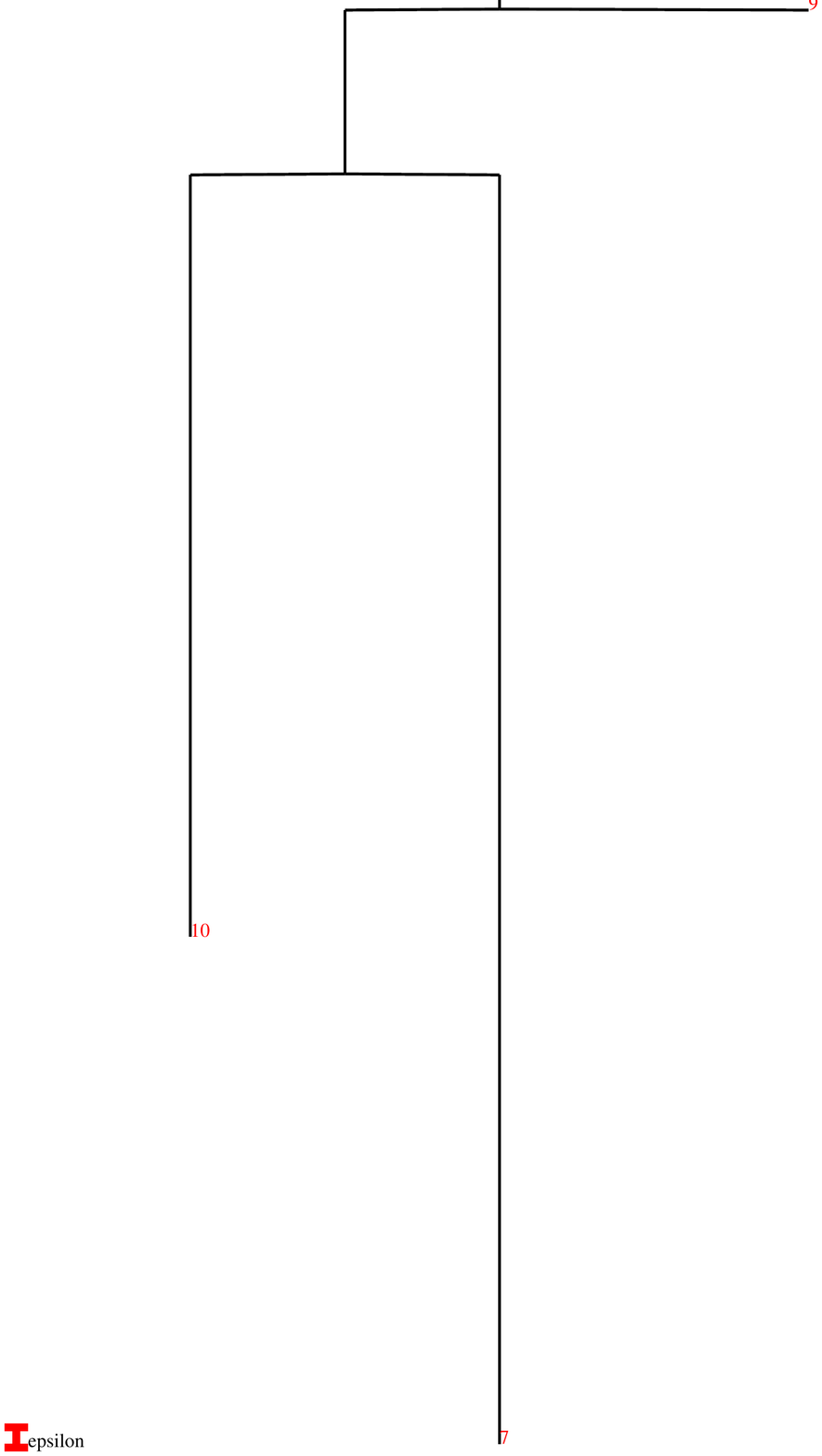}}
	\hspace{0.05\textwidth}
	\subfloat[$10.6$]{\includegraphics[width=0.05\textwidth, trim=3.7cm 0 3.7cm 0, clip]{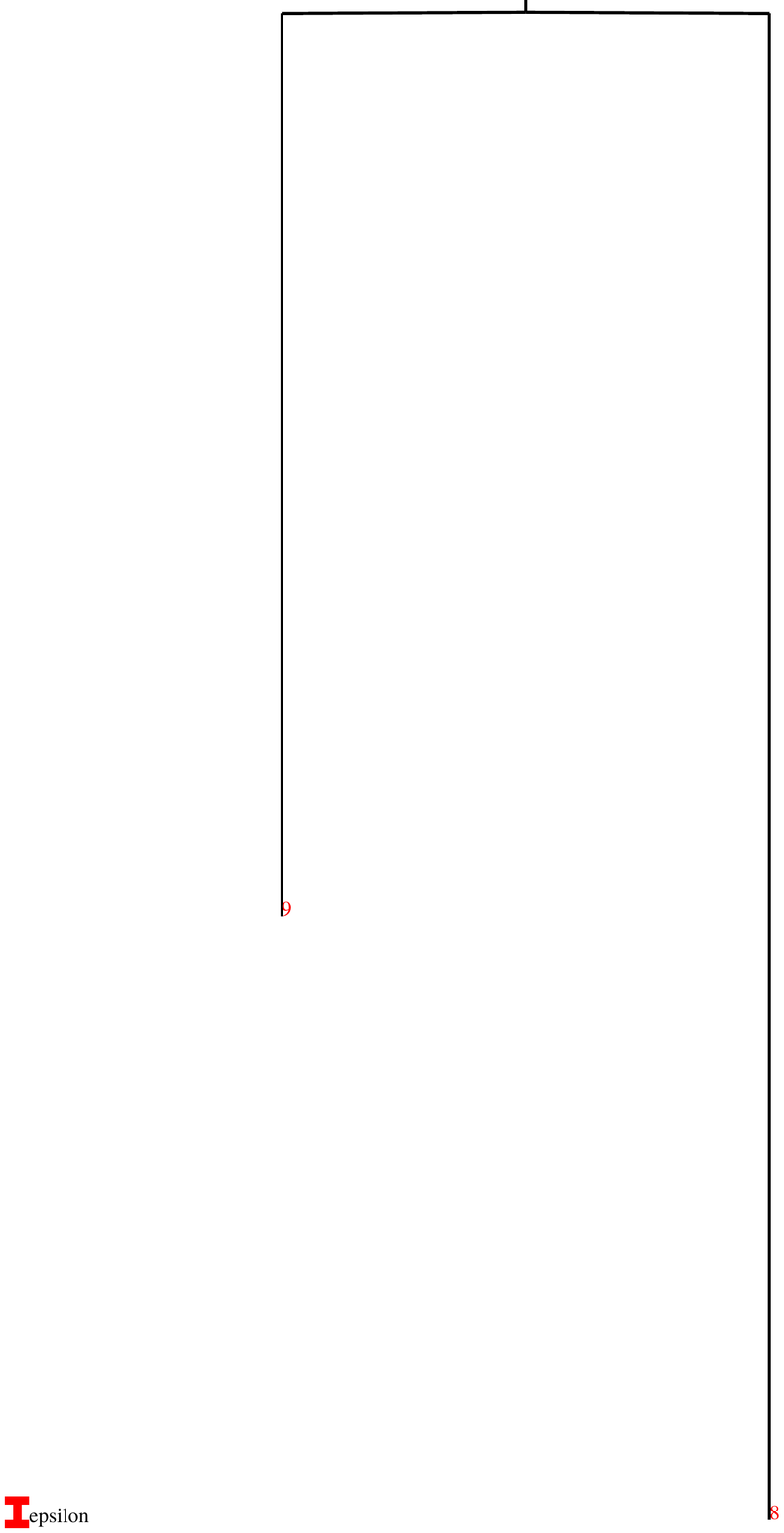}}
	\hspace{0.05\textwidth}

	\caption{LMLs disconnectivity graphs for the RBF kernels.  Again,  the labels underneath each graph denote the $\alpha$ parameter for the corresponding GP LML.}
	\label{fig:RBF-disco}
\end{figure}
\begin{figure}[H]
	\centering
	\includegraphics[width=0.3\textwidth]{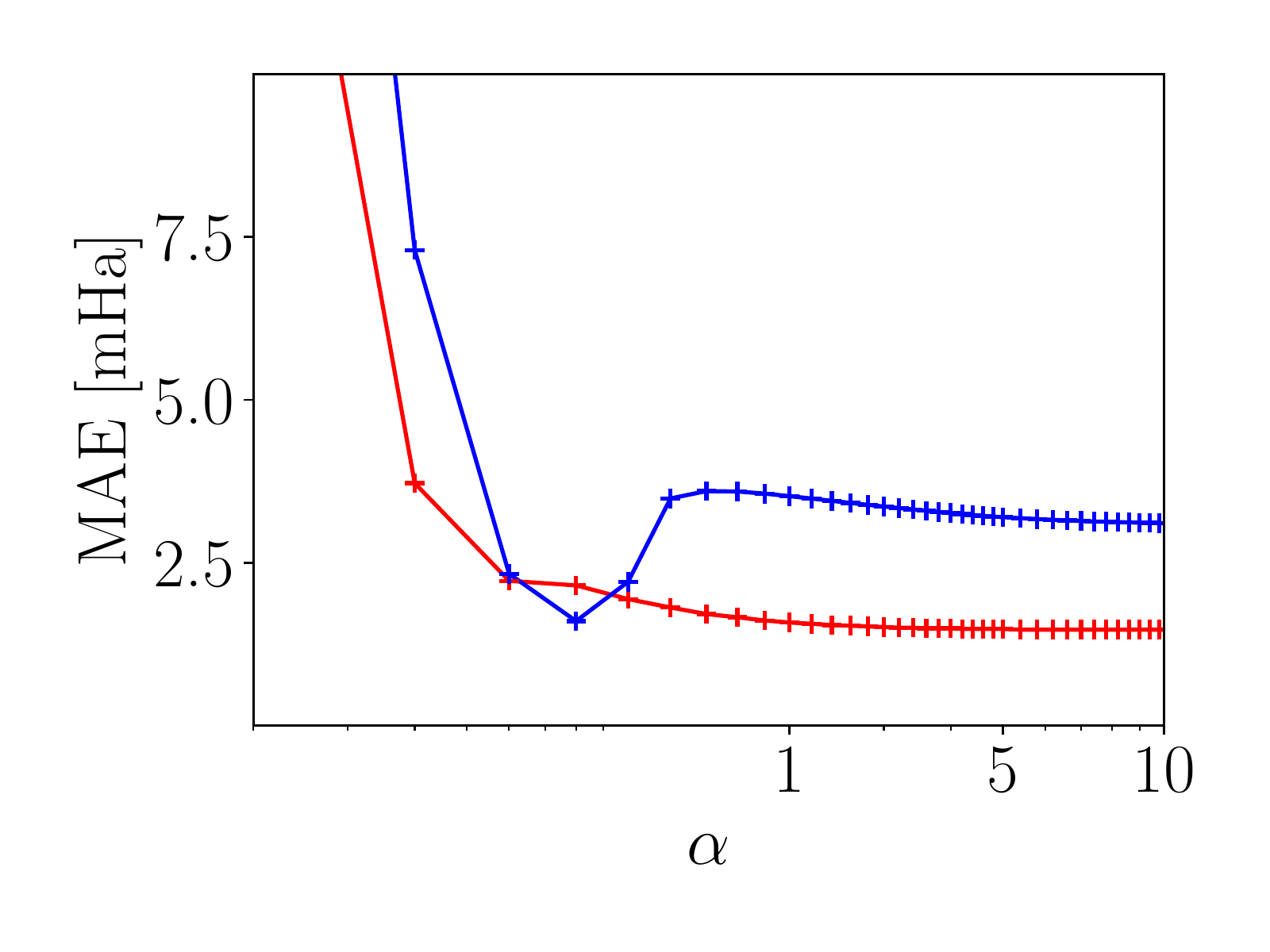}
	\caption[Summary of the optimisation through the ``best-fit'' approach.]{MAE curves for the RBF (blue) and Mat\'{e}rn (red) kernels against the Morse parameter.  One can see that the Mat\'{e}rn curve does not show a minimum and thus indicates the best transform is the linear transform,  \textit{i.e.} the limit of the Morse transform when $\alpha \to \infty$.}
	\label{fig:MAE-vs-none}
\end{figure}

The minimum on the LML with the lowest MAE is always shown on the graphs and,  despite its connectivity to other LML minima changing,  is easy to follow.  GPs converge towards a ``good'' model when the Morse parameter is large enough and all latent function for GPs with $\alpha>1$ perform similarly,  as shown by the plateau in figure \ref{fig:hyp-MAE-morse}.  The latent functions,  given in figure \ref{fig:Mat\'{e}rn-models},  are also similar.  The PES models are shown for the GP with a Morse parameter of $\alpha=2.0$ (as it has been used extensively in the previous chapter) and for comparison purposes for the GP with a Morse parameter of $\alpha=5.0$ where the MAE of the best model is reaching a ``plateau''.
\begin{figure}[H]
	\centering
	MAE: 1.58 mHa \par
	\includegraphics[width=0.4\textwidth]{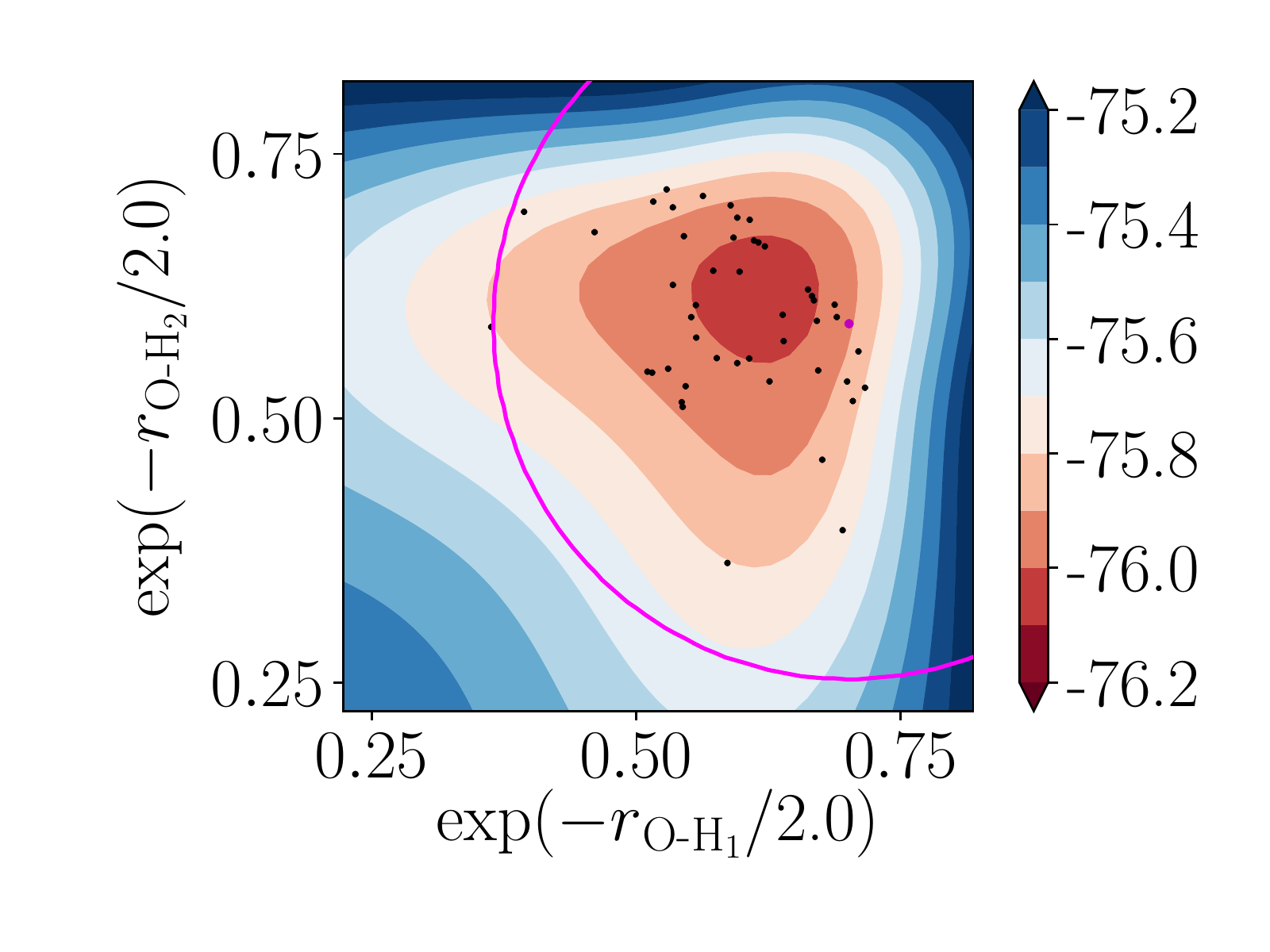}\par
	MAE: 1.47 mHa \par
	\includegraphics[width=0.4\textwidth]{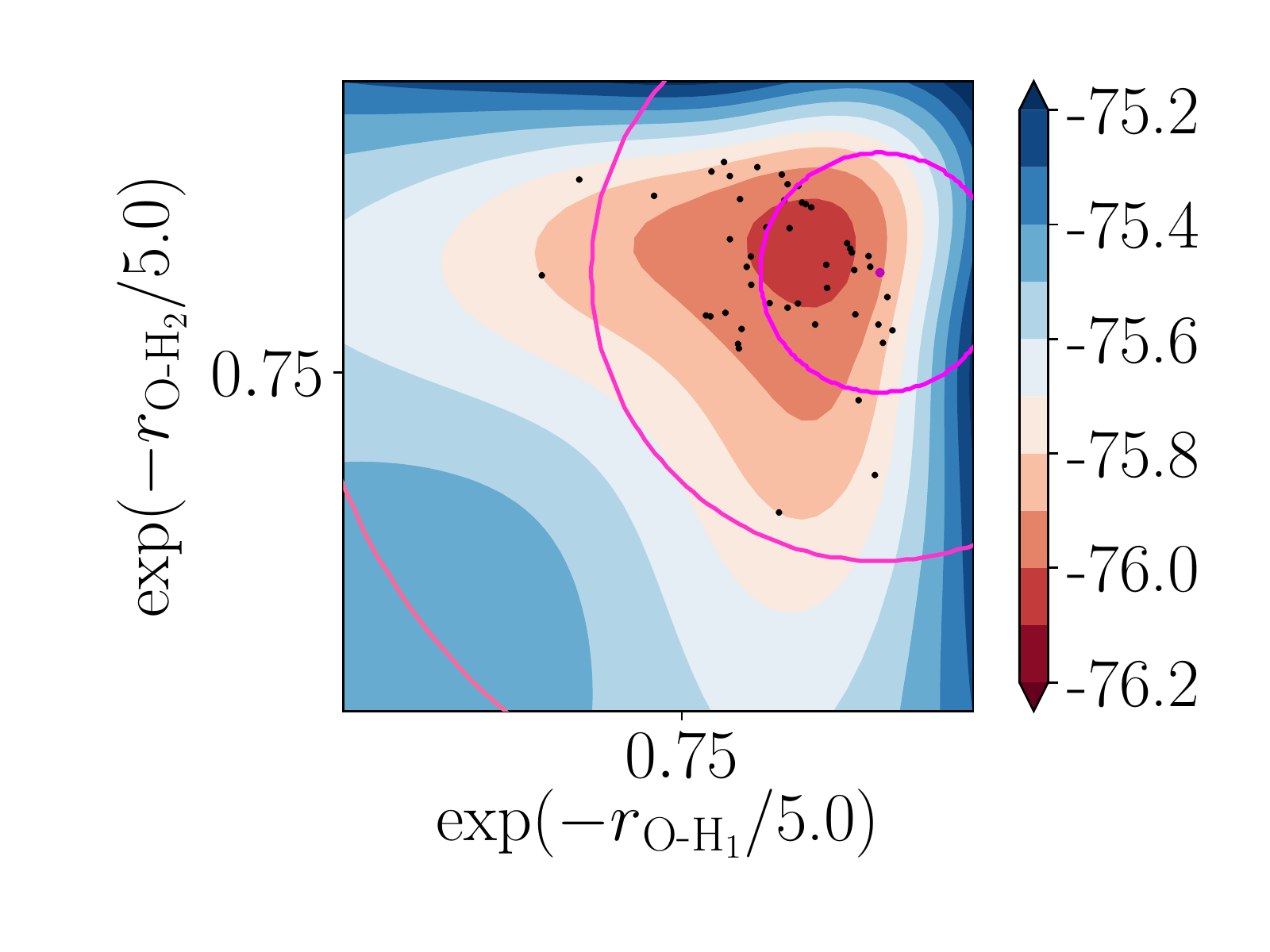}
	\hfill
	\caption[Latent function of Gaussian processes with Mat\'{e}rn kernel projected on different representations.]{Resulting PES,  projected on the Morse transformed O-H nuclear distances,  for Mat\'{e}rn kernels trained on Morse transformed spaces with parameters $\alpha=$2.0 (higher graph) and $\alpha=5.0$ (lower graph).  The magenta lines are isovalue contours of the kernel function where the covariance to the highlighted point is equal $n \sigma^2 / 4$ for $n=3,2,1$ where $\sigma$ is the amplitude hyperparameter of equation \ref{eq:covariance-Mat\'{e}rn}.}
	\label{fig:Mat\'{e}rn-models}
\end{figure}
The RBF kernel seem to have a much less stable LML landscapes (see figure \ref{fig:RBF-disco}). However,  the lowest MAE($\boldsymbol{\theta}$) models is always found to be the lowest minimum on the LML($\boldsymbol{\theta}$).  As opposed to the Mat\'{e}rn kernel,   the optimal Morse parameter value to minimise the MAE is more distinct (see figure \ref{fig:MAE-vs-none}).
\par
\begin{figure}[H]
	\centering
	MAE: 1.60 mHa\par
	\includegraphics[width=0.4\textwidth]{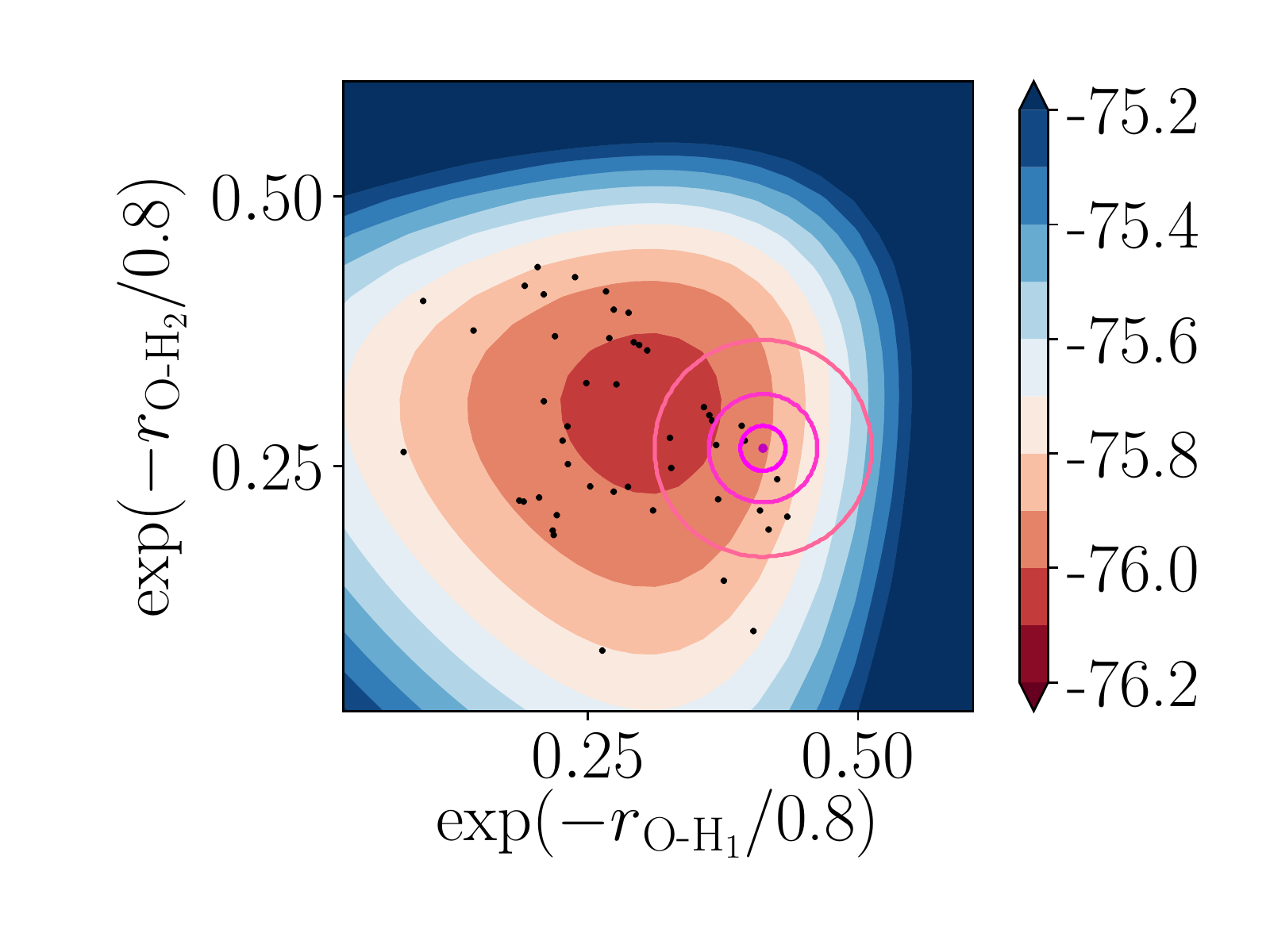} \par
	MAE: 3.52 mHa \par
	\includegraphics[width=0.4\textwidth]{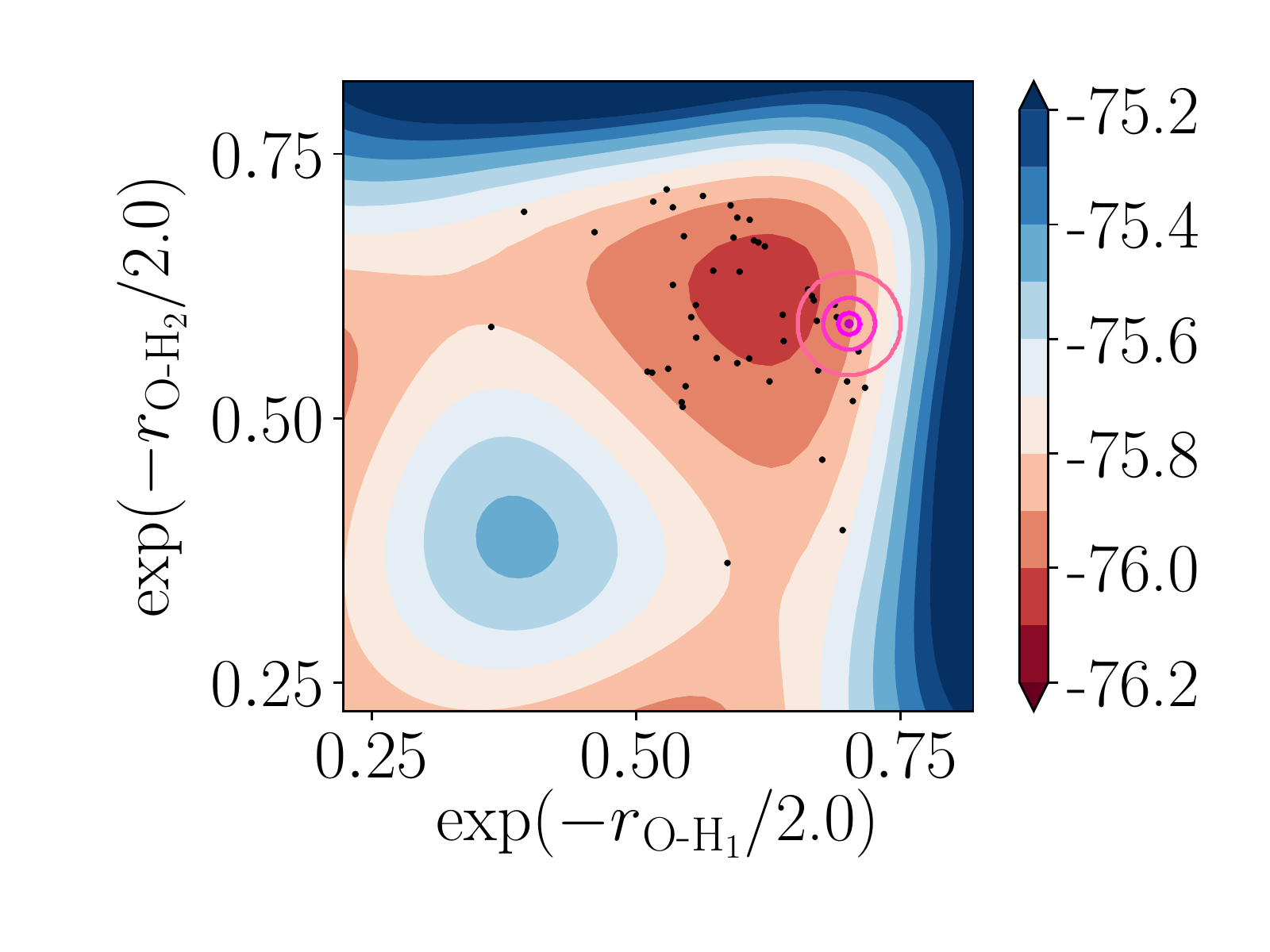}
	\hfill
	\caption[Latent function of Gaussian processes with RBF kernels and different Morse projections.]{Resulting PES,  projected on the Morse transformed O-H nuclear distances,  for RBF kernels trained on Morse transformed spaces with parameters $\alpha=0.8$ (higher graph) and $\alpha=2.0$ (lower graph) respectively.  These correspond to the minimum along the MAE plot in figure \ref{fig:hyp-MAE-morse} for the RBF and the Mat\'{e}rn kernels.  The magenta lines are isovalue contours of the kernel function.}
	\label{fig:rbf-models}
\end{figure}

The Gaussian process for the optimal value seem to correspond to a value where the training data is not too compressed and does not allow the RBF kernel to over fit.  Despite the MAE being similar to the Mat\'{e}rn kernel best performing models,  the length scales are shorter and the latent function,  displayed on figure \ref{fig:rbf-models},  shows that the RBF model predictions are only reliable close to the training data and do not ``carry'' any of the information to longer bond lengths,  like the Mat\'{e}rn kernel does.

Compared to the graphs of the Mat\'{e}rn kernel in figure \ref{fig:Mat\'{e}rn-disco},  the graphs for the RBF kernel in figure \ref{fig:RBF-disco} show more minima and show stronger changes.  This is due to the tendency for the RBF kernel to over fit data giving a more complex LML landscape in the region with short length scales.  The TSs are also harder to optimise,  in these short length scale regions, making the graphs rapidly changing. 
\par
The most performant GP is found for $\alpha=0.8$ and one can see, in figure \ref{fig:MAE-vs-none},  that its MAE is similar to the best Mat\'{e}rn GPs.  The resulting latent function is shown in figure \ref{fig:rbf-models} alongside the latent function for the GP trained with the Morse parameter set to $\alpha=2.0$ to compare with the model of the Mat\'{e}rn kernel in figure \ref{fig:Mat\'{e}rn-models}.  One can see that,  for $\alpha=2.0$,  despite similar MAEs,  the RBF kernel is more ``local'' and does not predict a meaningful PES at longer bond lengths.
\vspace{0.5cm}
In figure \ref{fig:rbf-models},  as before,  the contours represent an isocontour of the kernel from a given sample\footnote{The first contour is where the covariance function evaluates to $0.75\sigma^2$,  where $\sigma$ is the amplitude hyperparameter,  while the second one correspond to $0.5\sigma^2$.}.  One can see that despite the surfaces covering the same geometry stretches,  for the larger $\alpha$,  the optimised length scales is much shorter and only allows a sample to span `` influence'' over a small part of the considered space.  This leads to partial over fitting of the training data and a more complicated PES model.
\par
To summarise both the MAE optimisation of the GPs trained with RBF and Mat\'{e}rn kernels,  we plot the MAE curves against the Morse parameter.  This is the curve that one minimises in the ``best-fit'' approach and leads to selecting $\alpha=0.8$ for the RBF kernel and a larger $\alpha>2.0$ for the Mat\'{e}rn kernel.  The latter produces a monotonically decreasing line which indicates that the optimal Morse transform is a linear transform (since the limit of $\alpha \to \infty$ reduces the Morse transform to the latter,  as explained in equation \ref{eq:morse-trans-limit}).  This is an indication that it is not optimal , for the Mat\'{e}rn kernel,  to do the transformation and that the initial internuclear distances produce a better feature space to learn on.

\section{Optimisable Morse Kernels}

We now explore the optimisation of GP with the same training data projected on the internuclear distances that are Morse transformed in the kernel,  for example as given by equation \ref{eq:cov-morse-rbf} for the MorseRBF kernel.  As usual the kernels are scaled by an optimisable CK and have an added optimisable noise given by a WK.  The additional hyperparameter,  $\alpha$,  means we approach the Morse parameter optimisation in a fully Bayesian manner through the LML minimisation.  As mentioned before this does remove the testing set in the optimisation and only the training data affects its optimisation.
\par
\begin{figure}[H]
	\centering
	\includegraphics[width=0.22\textwidth]{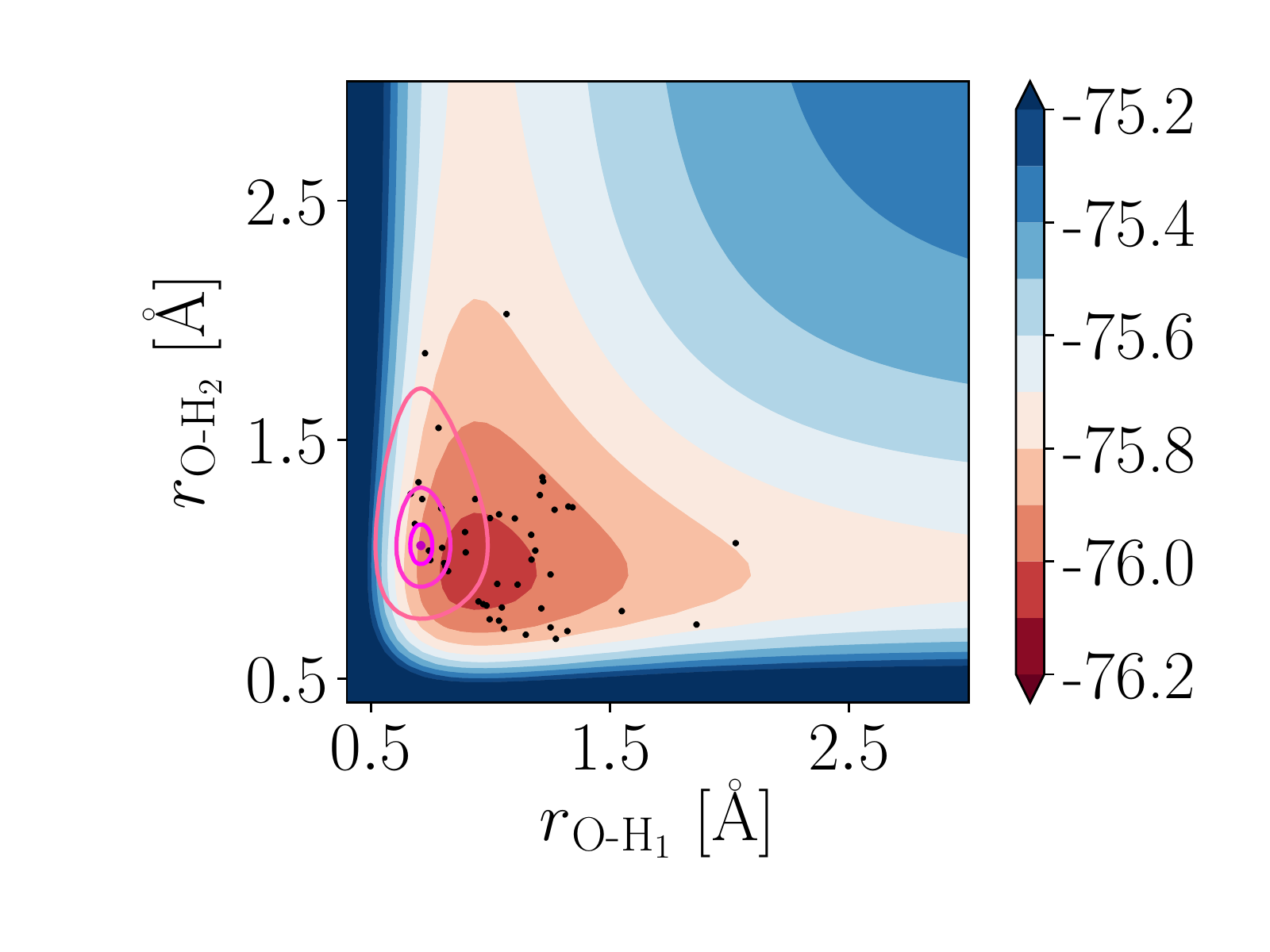}
	\includegraphics[width=0.22\textwidth]{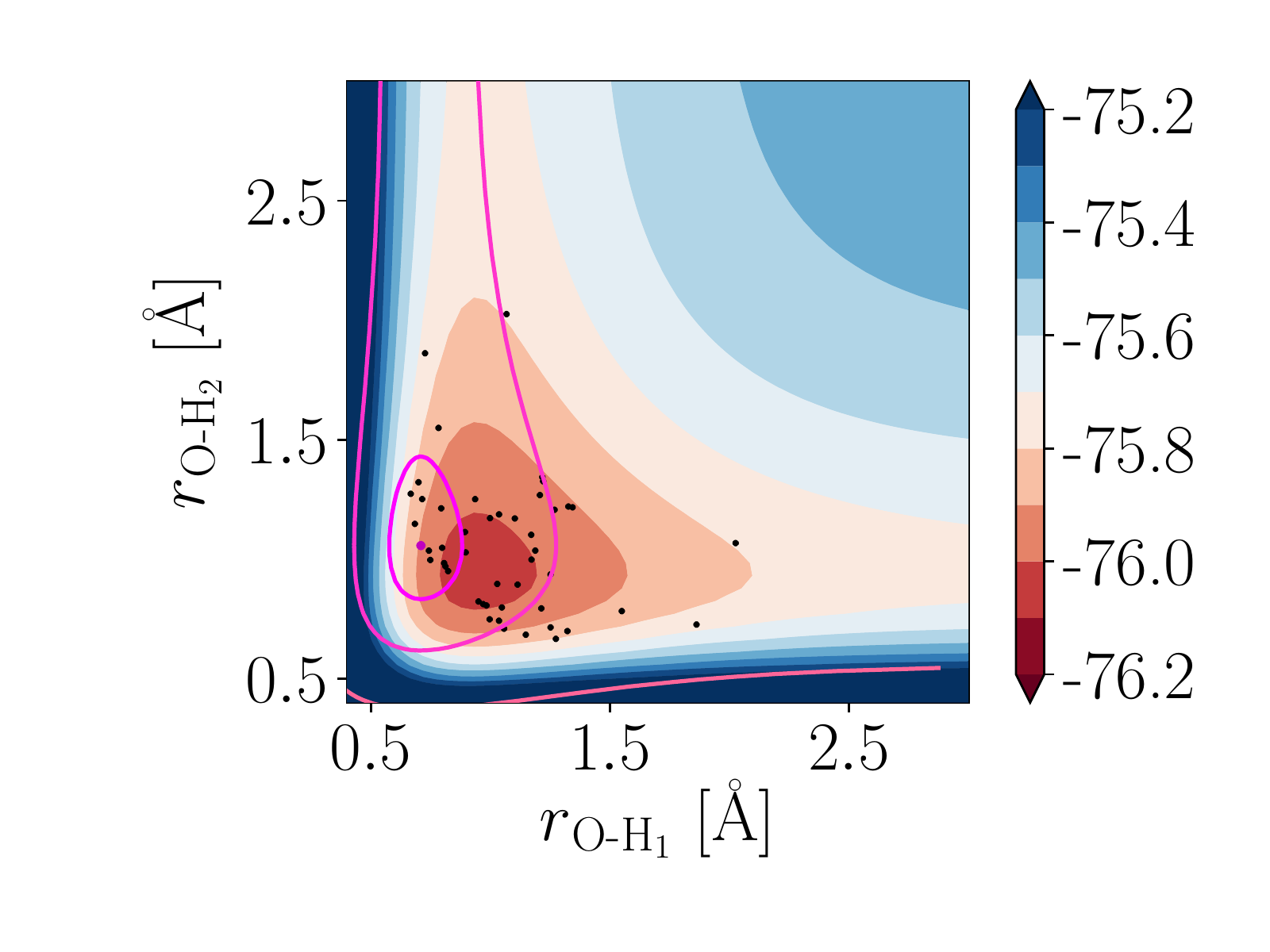}
	\hfill
	\caption[Latent function of Gaussian processes with MorseRBF kernel compared to RBF model.]{Latent functions of GPs trained with a standard RBF kernel (higher graph) and a fixed Morse parameter close to the one exhibited by the lowest LML minimum of the other GP,  trained with a MorseRBF kernel (lower graph).  The change in length scale (shown by the kernel isovalue contour extending further out) is simply a consequence of the small difference in $\alpha$ value and the models are essentially the same.}
	\label{fig:morserbf-model0}
\end{figure}

For the MorseRBF,  multiple minima on the LML are obtained and,  when ranked with their respective MAEs,  the best GP models are found to be in the region of $0.5<\alpha<1.0$.  This is in accordance with the MAE curve,  as seen in figure \ref{fig:MAE-vs-none},  of the GPs trained with standard RBF kernels and fixed Morse transforms.  As expected,  the MorseRBF GP best models latent function are very similar to the RBF GP latent function with small $\alpha$ parameters\footnote{The MorseRBF kernel is equal to a RBF kernel and a fixed Morse projection with the optimal Morse hyperparameter.}: the lowest minima on the LML for the MorseRBF is shown in figure \ref{fig:morserbf-model0}. 
%
Obtaining MorseRBF models that resembles the RBF ones is important as it tells us that the added hyperparameter dimension creates a convex LML hypersurface\footnote{It should be made clear again that we are technically talking about the $-$LML surface which we are optimising.  On the true LML surface this would be concave.} that can be optimised.  The other hyperparameters of the kernel are quite close to the ones of the kernel that does not include the transformation when one fixes the latter with the parameters found by the Morse kernel.
\par

\begin{figure}[H]
	\centering
	\includegraphics[width=0.3\textwidth]{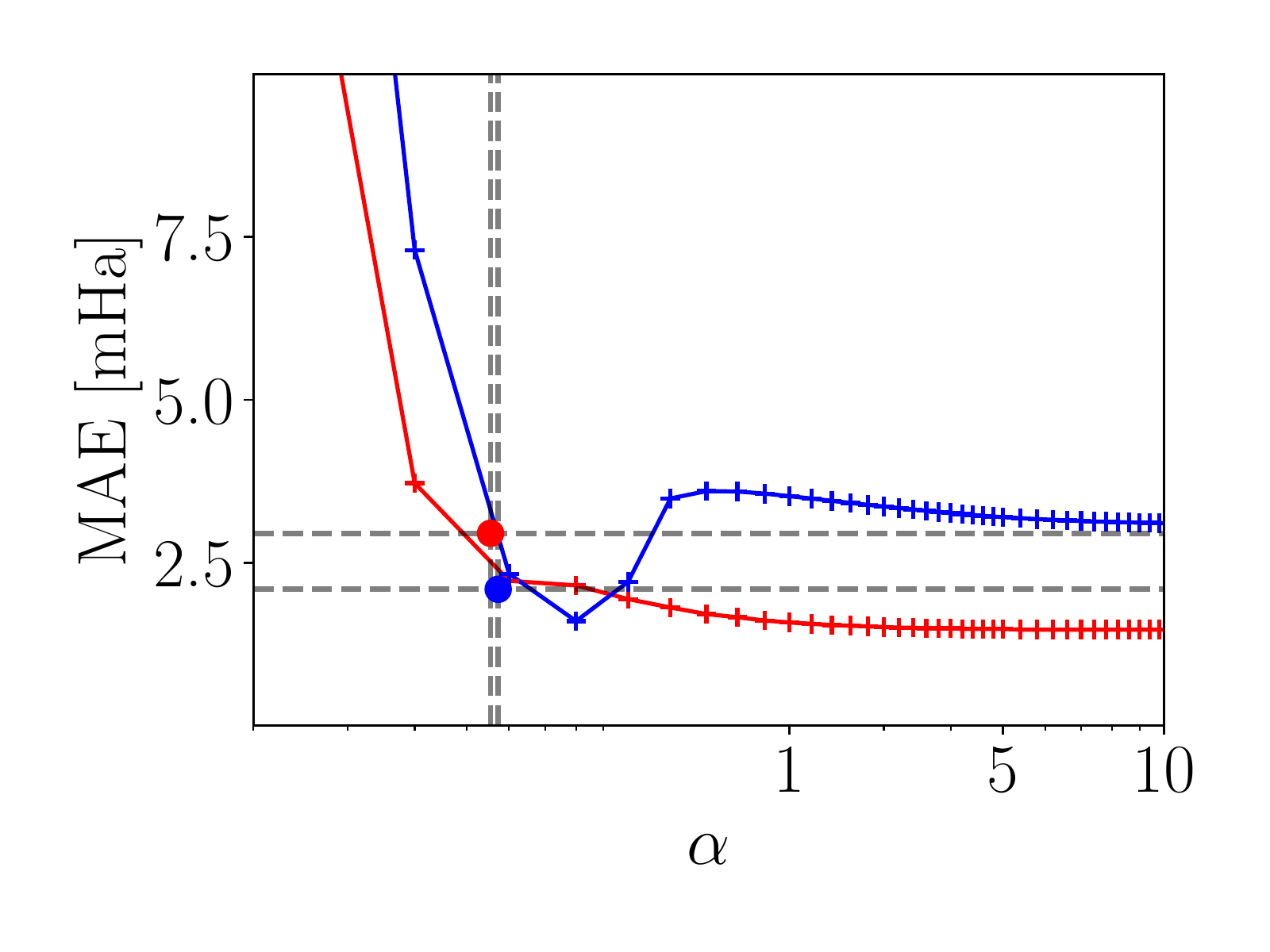}
	\caption[Summary of the optimisation through two approaches.]{MAE curves for the RBF (blue) and Mat\'{e}rn (red) kernels against the Morse parameter.  The dots represent the optimised Morse hyperparameter of the Morse kernels (blue for MorseRBF and red for MorseMat\'{e}rn) with grey line to aid clarity.}
	\label{fig:MAE-vs-bayesian}
\end{figure}

To summarise,  the Morse kernels do optimise to lower values which agree better with the optimal MAE($\alpha$) for the RBF kernel but not for the Mat\'{e}rn kernel.  Figure \ref{fig:MAE-vs-bayesian} shows the disparity between the two Morse kernel ability to replicate the ``best-fit'' approach..

\section{Changing the Training Data}
Since the feature space is optimised differently with respect to the selected training data,  we will consider the effect of adding data to the previously discussed models.  We still use the MAE of the final GP model but we use two different testing sets.  The new set is also taken from a Boltzmann distibution but at a higher temperature which allows data to be sampled 0.4 Ha above the equilibrium energy.
\par
The training data is changed by adding data sampled from NM clusters\footnote{There is no overlap of the two training set.  The additional data is added incrementally to the original training data with batches of 5 random samples drawn from the NM clusters.}.  Two things are interesting to follow: the effect of increasing the size of the dataset on the MAE($\alpha$) curves as well as the progression of LML-optimised Morse hyperparameters.

\begin{figure}[H]
	\centering
	\subfloat[$N=29$]{\includegraphics[width=0.15\textwidth]{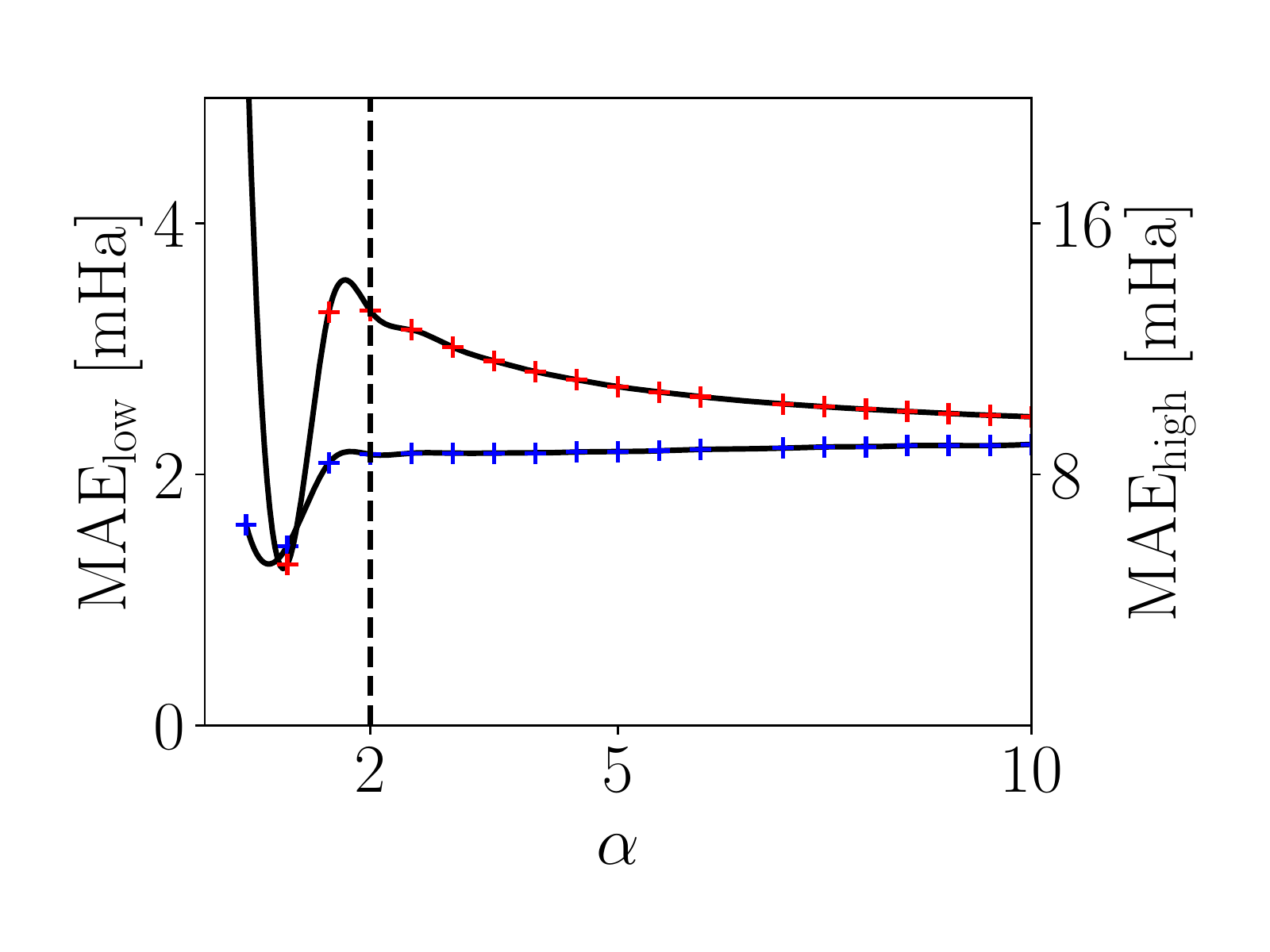}}
	\hfill
	\subfloat[$N=39$]{\includegraphics[width=0.15\textwidth]{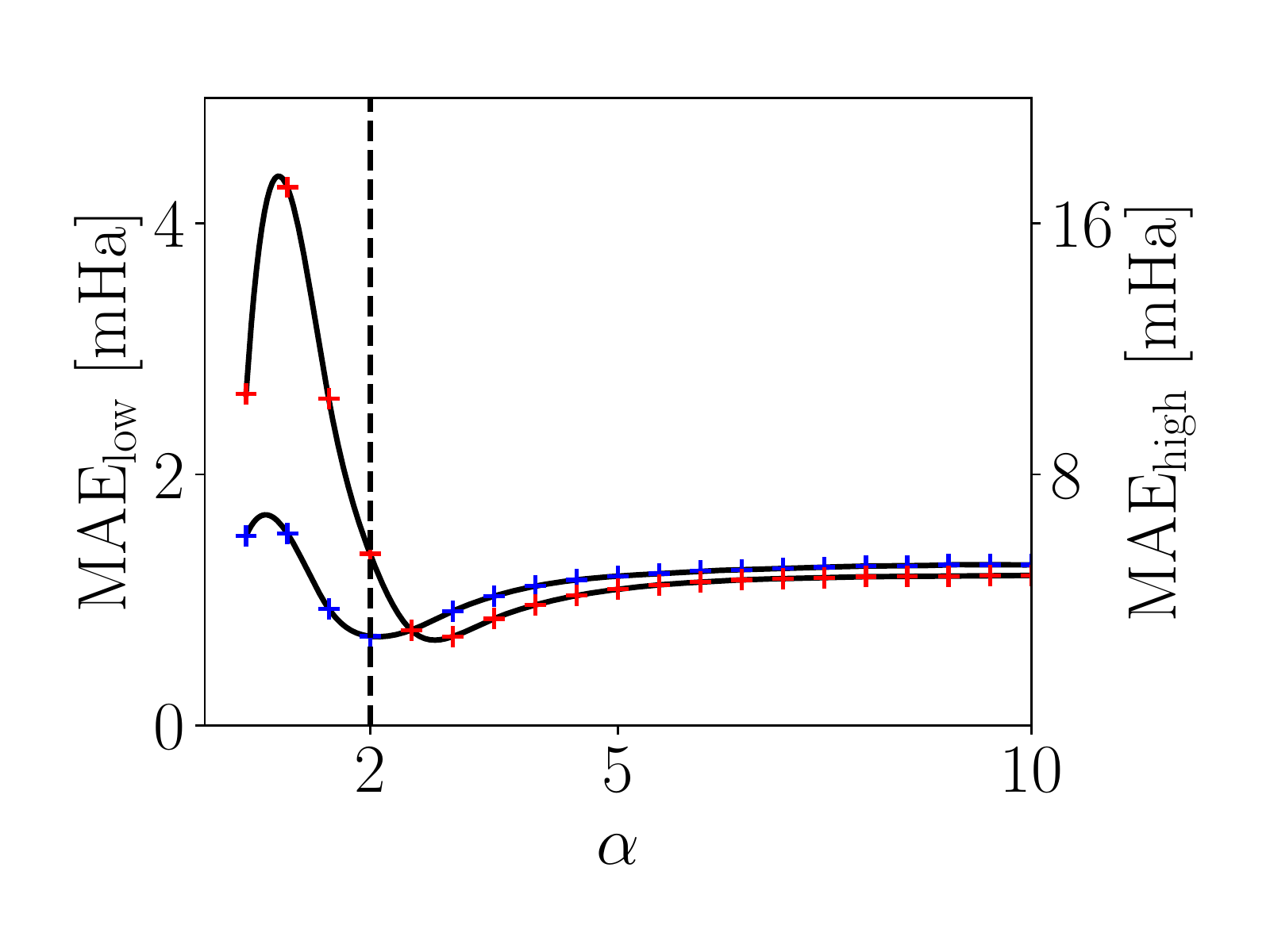}}
	\hfill
	\subfloat[$N=49$]{\includegraphics[width=0.15\textwidth]{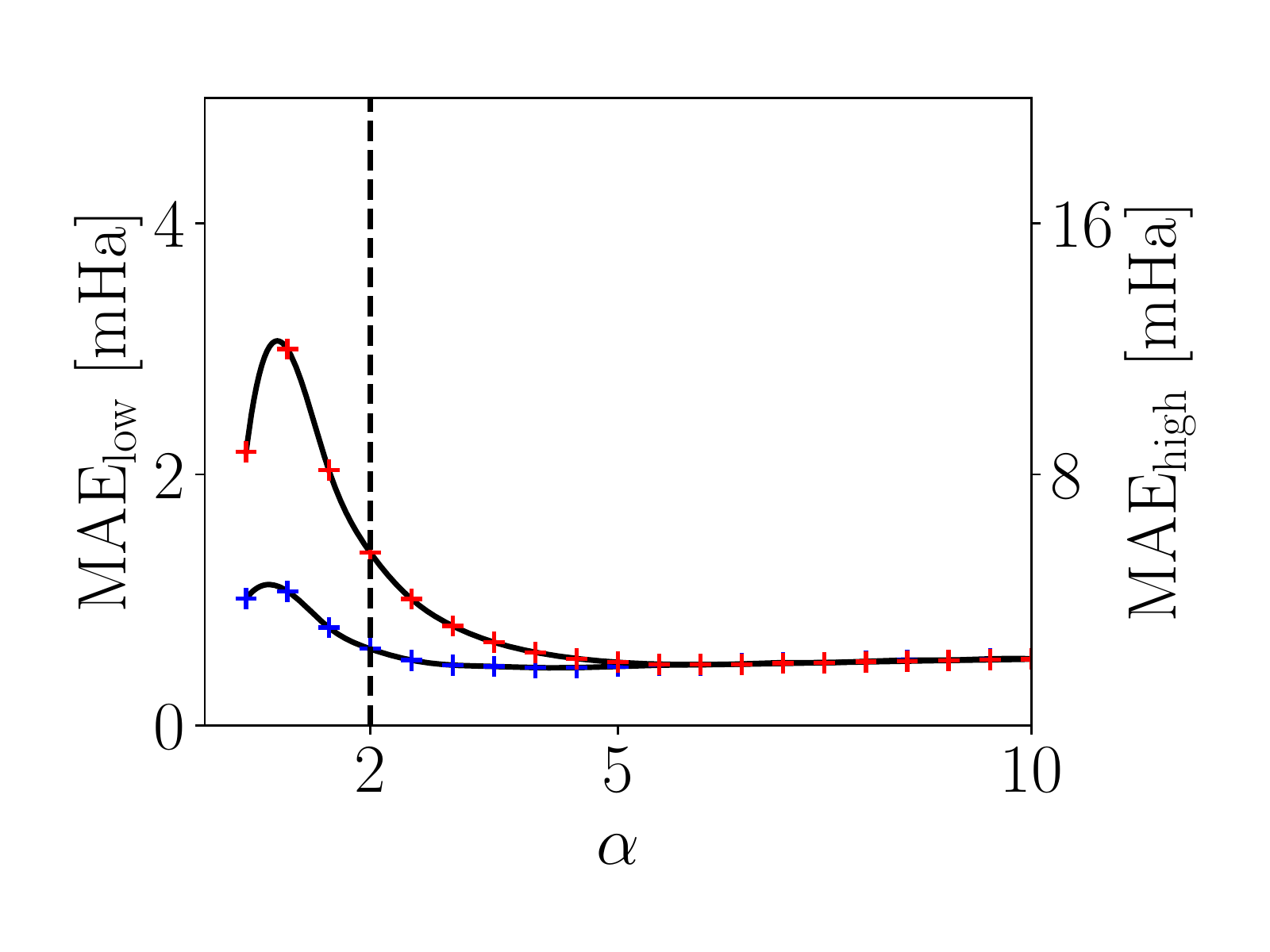}}
	\hfill
	\subfloat[$N=59$]{\includegraphics[width=0.15\textwidth]{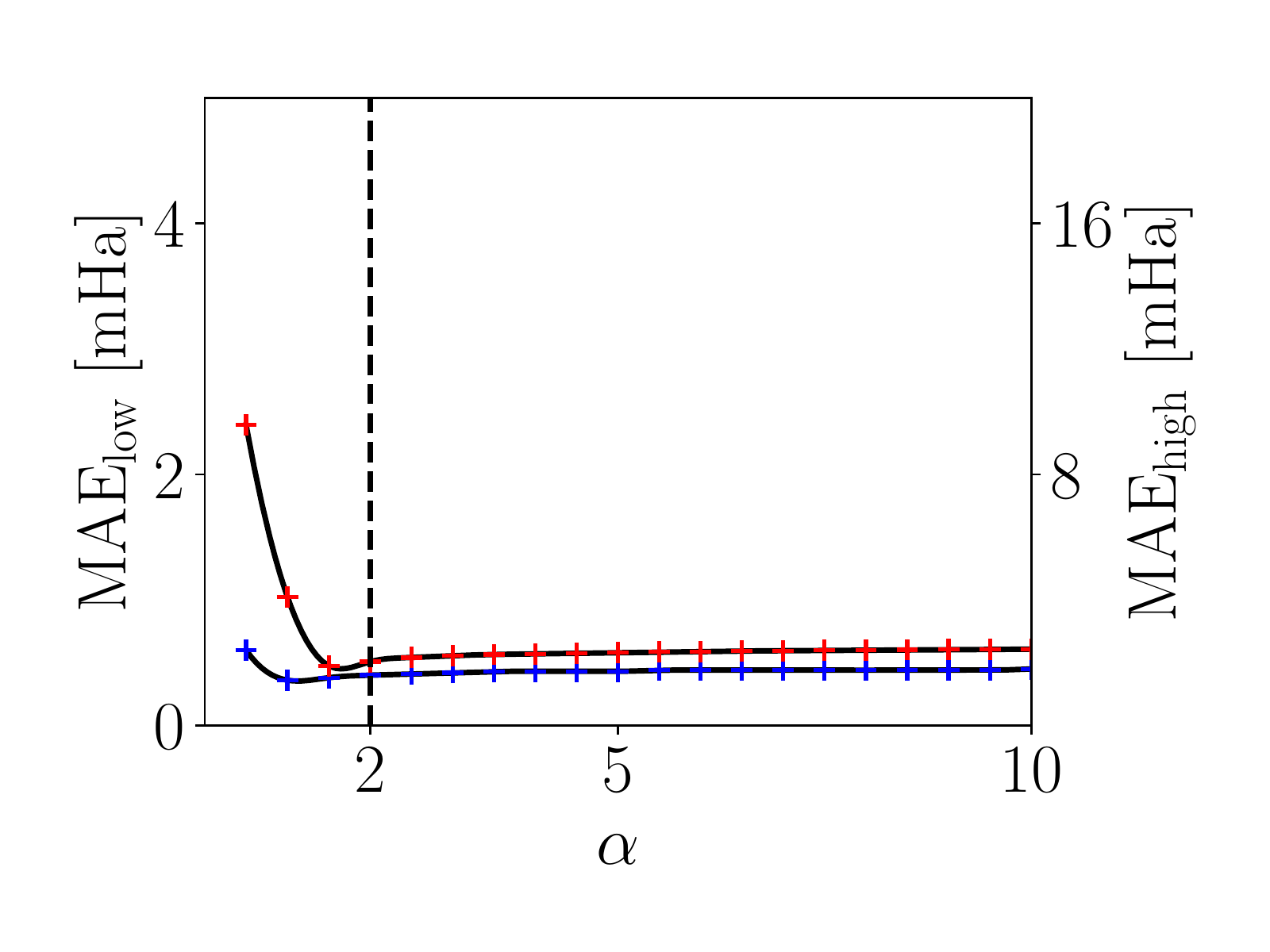}}
	\hfill
	\subfloat[$N=69$]{\includegraphics[width=0.15\textwidth]{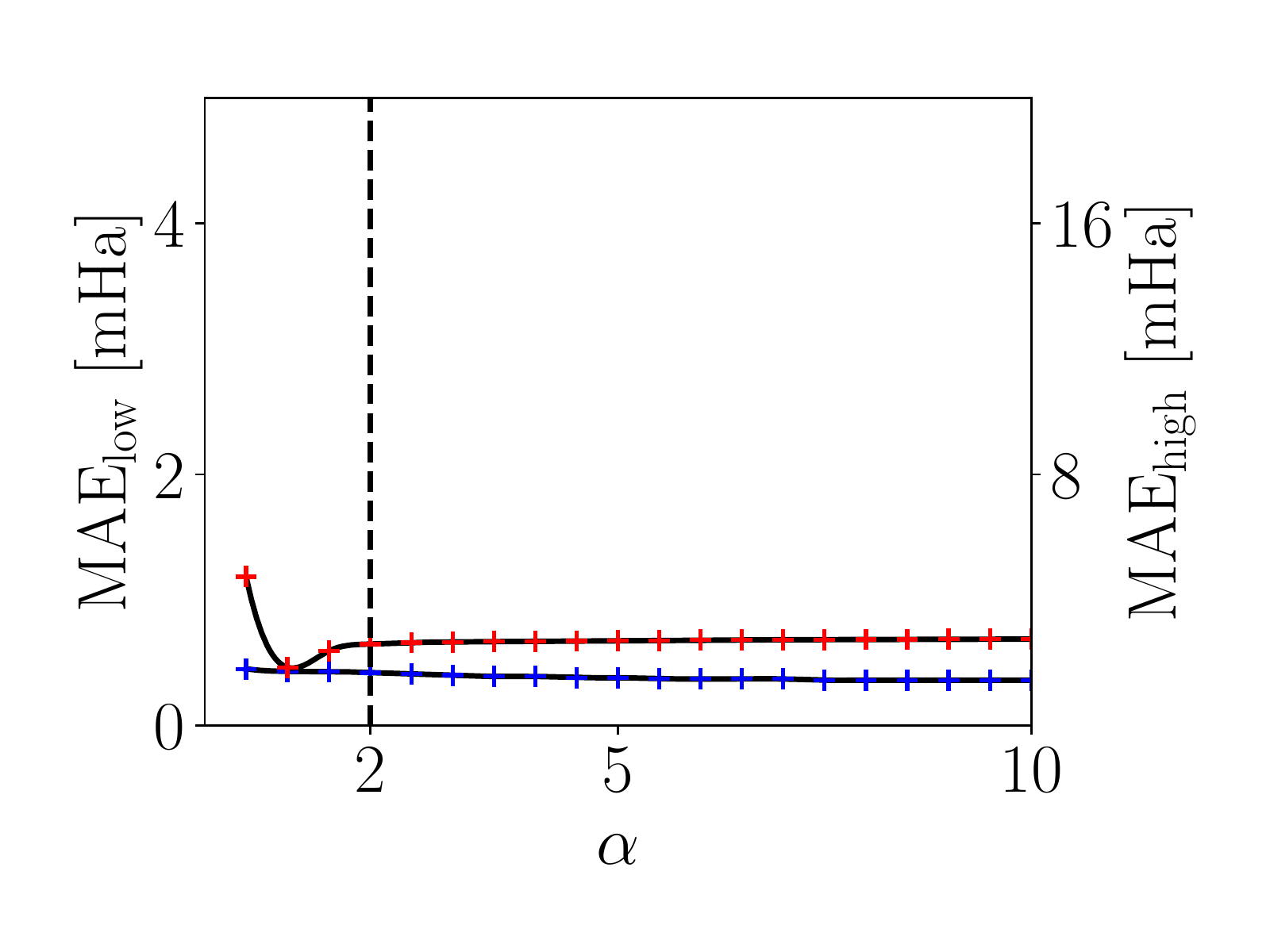}}
	\hfill
	\subfloat[$N=79$]{\includegraphics[width=0.15\textwidth]{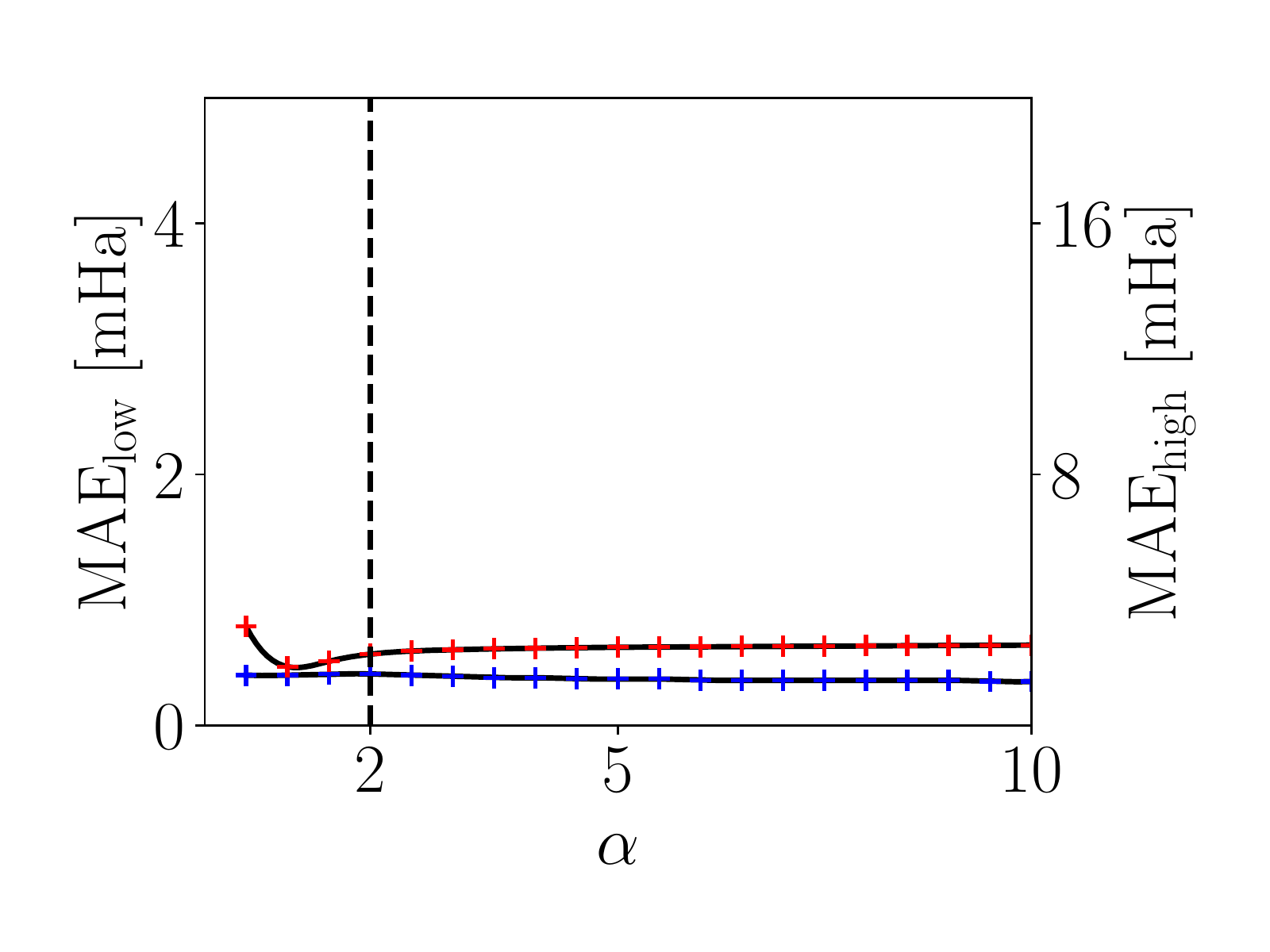}}
	\hfill
	\caption[MAEs for different sizes of training sets across the $\alpha$ Morse parameter.]{Different MAE($\alpha$) curves for different datasets.  The two colours represent the MAE on different testing sets\footnote{The two testing sets are the one used in chapter \ref{chap:feature} and represent two different Boltzmann population with the blue testing set having a smaller energy spread than the red testing set.} to see the dependency of the minimum of the MAE($\alpha$) with respect to the chosen testing set.  There is no clear choice for an $\alpha$,  although there seem to be a preference for small $\alpha$ values in the RBF kernel,  until training data becomes rather large and most Morse parameters perform equally.}
	\label{fig:morse-different-sets}
\end{figure}

A first observation that can be made from the curves, in figure \ref{fig:morse-different-sets}, is that a different testing set can lead to a different optimal Morse parameter.  A second important aspect is that small changes to the training set (in this case adding training data) can importantly alter the curves.  For the latter,  it is a surprising result since the new training data does not differ from the original training data in terms of what it describes.  The new sampled training data does not allow the GP to understand new patterns in the target function,  which were not seen in the original set.  One could expect that consequently the changes in the training data would not affect the relative performance of GPs with different Morse parameters.
\par
As data is added to the original training set,  the MAE curves tend to flatten and stop exhibiting a clear minimum.  The optimal Morse parameter is not well defined and GPs,  despite having different projections on their feature spaces,  perform similarly in terms of MAE.  The sparsity of training data is reduced,  which makes the feature space less relevant: dense training data is likely to perform well in any way it is projected. 
\par
Consequently,  instead of additional data making the minimum of the MAE($\alpha$) more and more distinct,  one sees the minimum disappearing.
\begin{figure}[H]
	\centering
	\captionsetup[subfigure]{labelformat=empty}
	\subfloat[RBF]{\includegraphics[height=2.5cm]{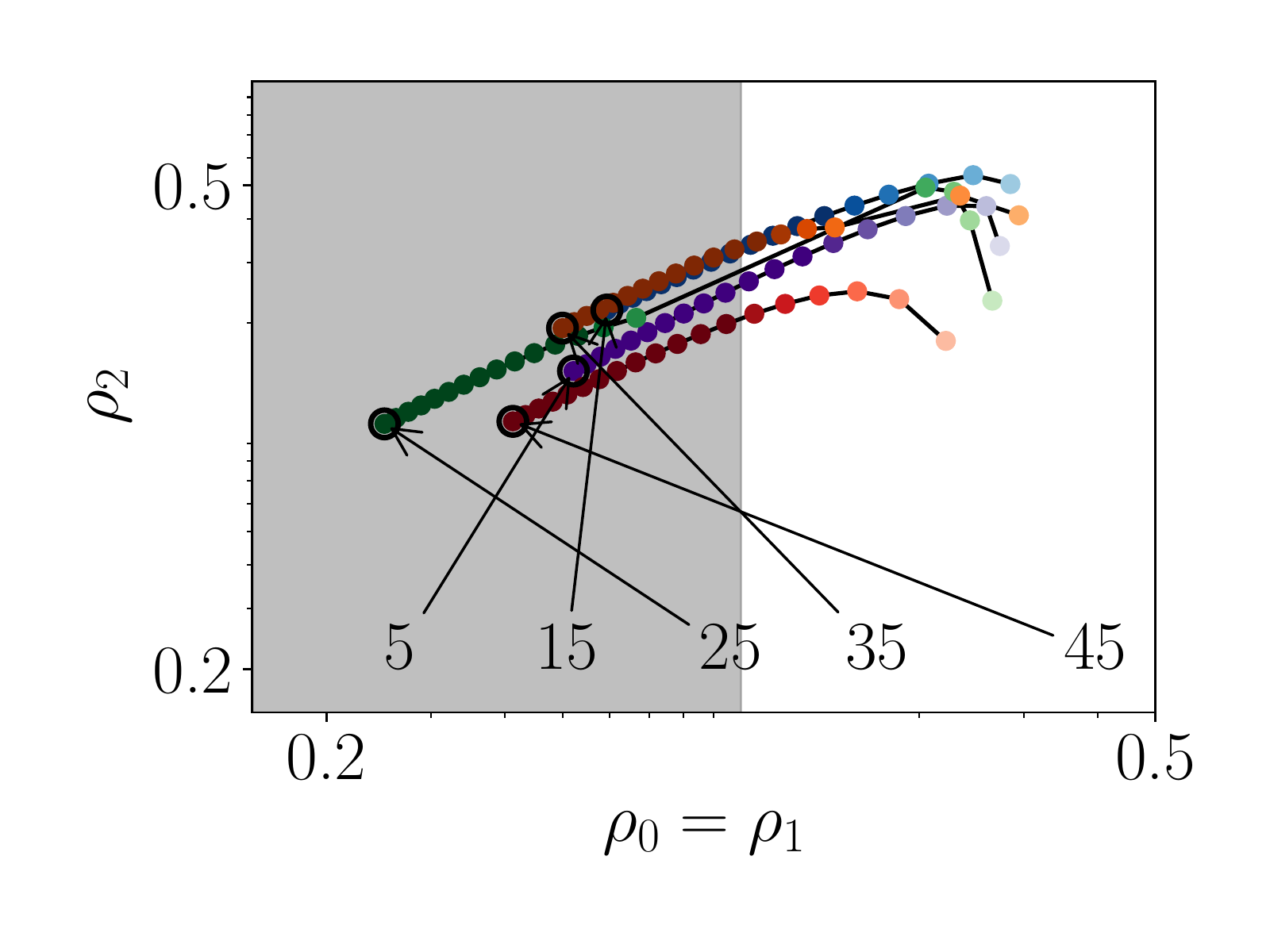}}
	\subfloat[Mat\'{e}rn ($\nu=2.5$)]{\includegraphics[height=2.5cm]{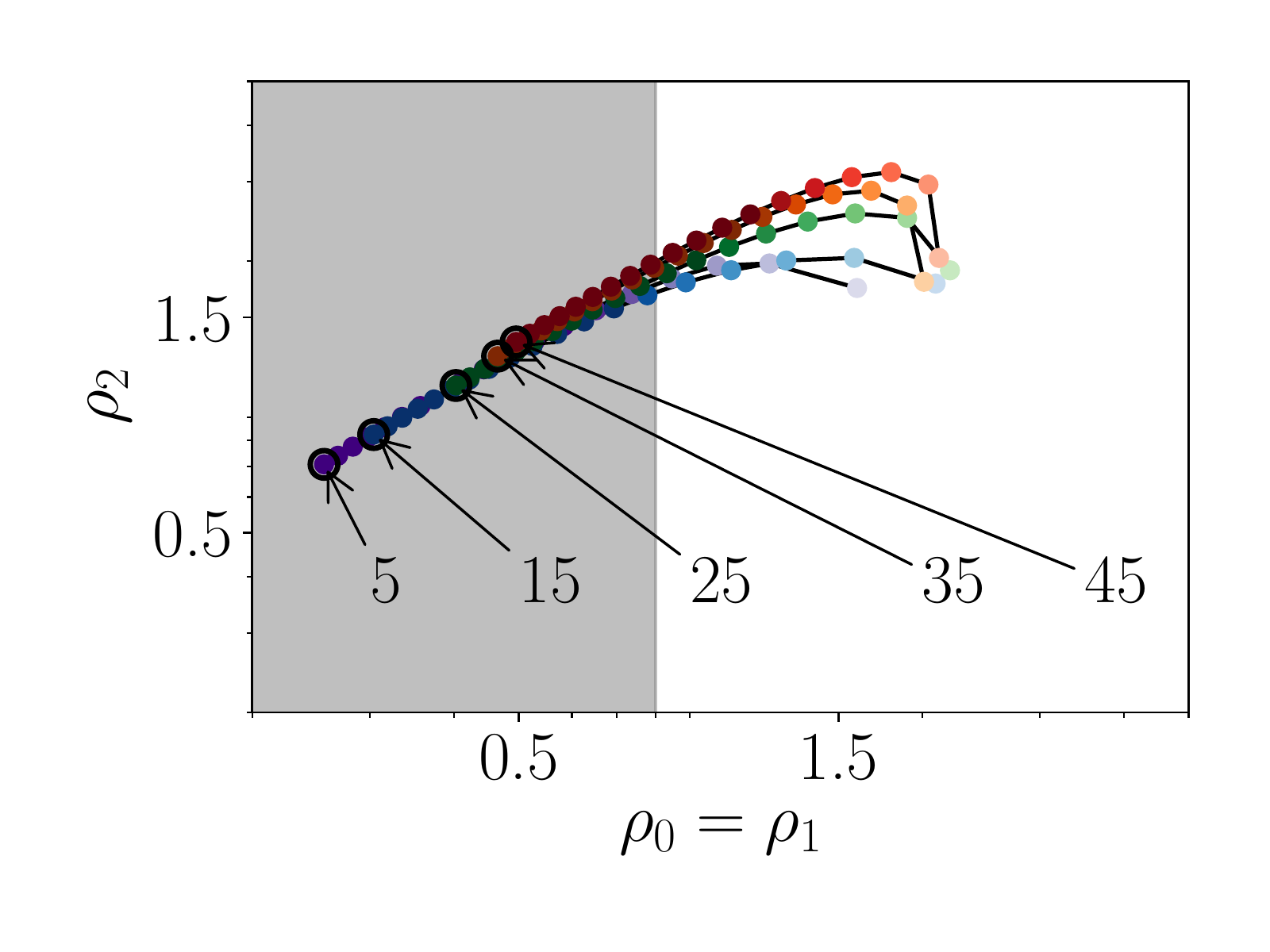}}
         \subfloat[]{\includegraphics[height=2.4cm]{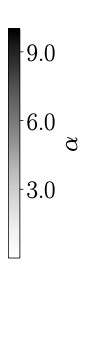}}
	\caption[Trajectories of optimal  models in hyperparameter space for different datasets.]{Trajectories of optimal models in hyperparameter space for different datasets (each dataset is represented by a different colour and the dots follow their respective best minima on the LML landscapes).  The end of the trajectory (where the label is given) is going towards larger Morse parameters and each trajectory has points,  since the progression is smooth,  ordered for increasing $\alpha$ along itself.  The colourbar is given as greys since it has shared by all training sets.}
	\label{fig:following_different_minima}
\end{figure}

In figure \ref{fig:following_different_minima},  each progression in hyperparameter space seem to have an initial curved trajectory followed by a linear trajectory.  This linear regime was already observed in figure \ref{fig:hyp-MAE-morse} as a consequence of the Morse transform limit (see equation \ref{eq:morse-trans-limit}).  If one considers the end of the trajectory of the Mat\'{e}rn GPs,  one sees that the optimised length scales of GPs with the same Morse projection increases with the training set size.  The trend is less clear for the RBF GPs where trajectories are not as well-behaved.
\par
\begin{figure}[H]
	\centering
	\tiny \hspace{0.03\textwidth} MAE: 3.06 mHa  \hspace{0.06\textwidth} MAE: 0.73 mHa  \hspace{0.06\textwidth} MAE: 0.39 mHa \hspace{0.03\textwidth} \par
	\subfloat[$N=29$]{\includegraphics[width=0.15\textwidth]{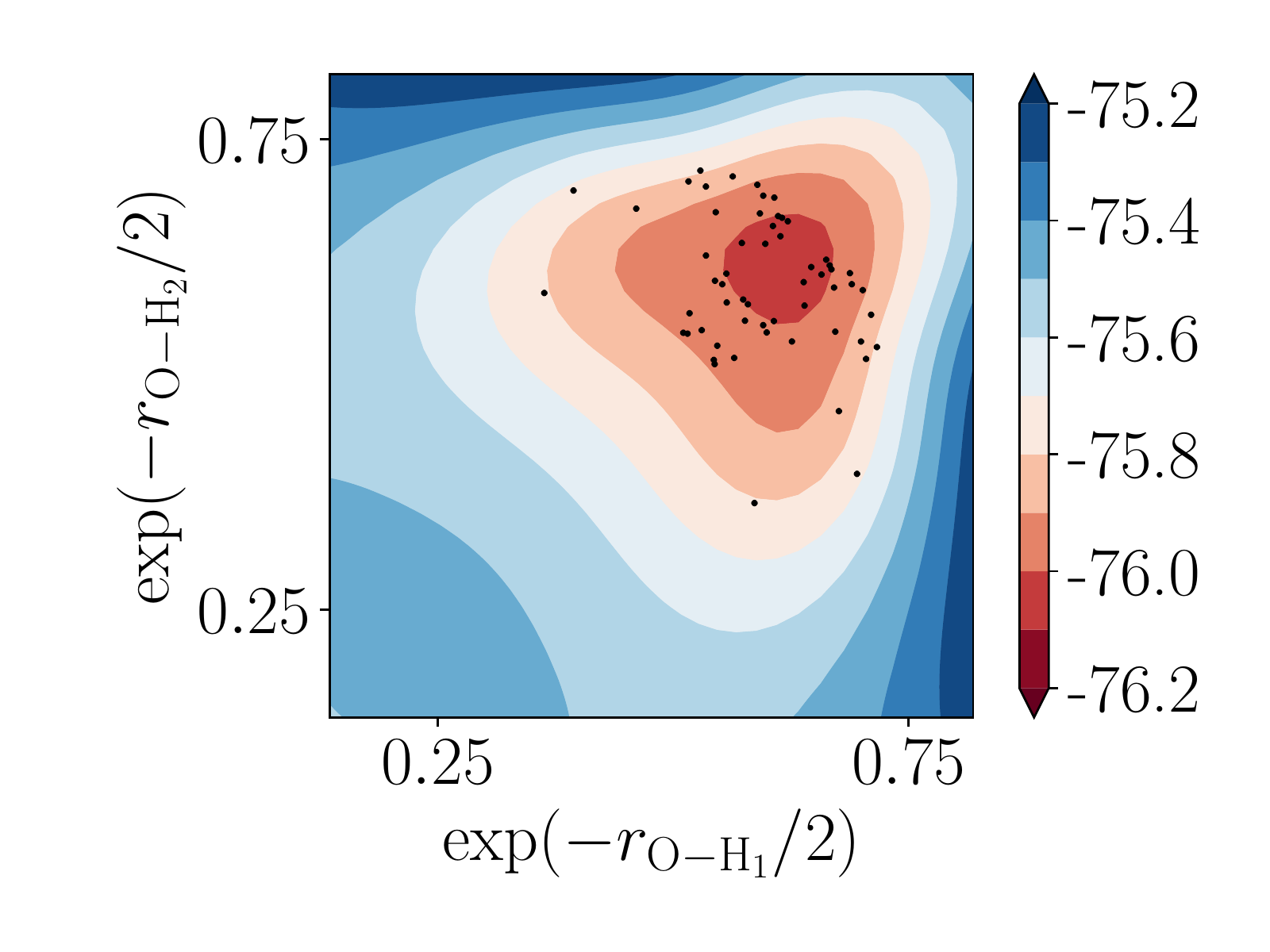}}
	\hfill
	\subfloat[$N=49$]{\includegraphics[width=0.15\textwidth]{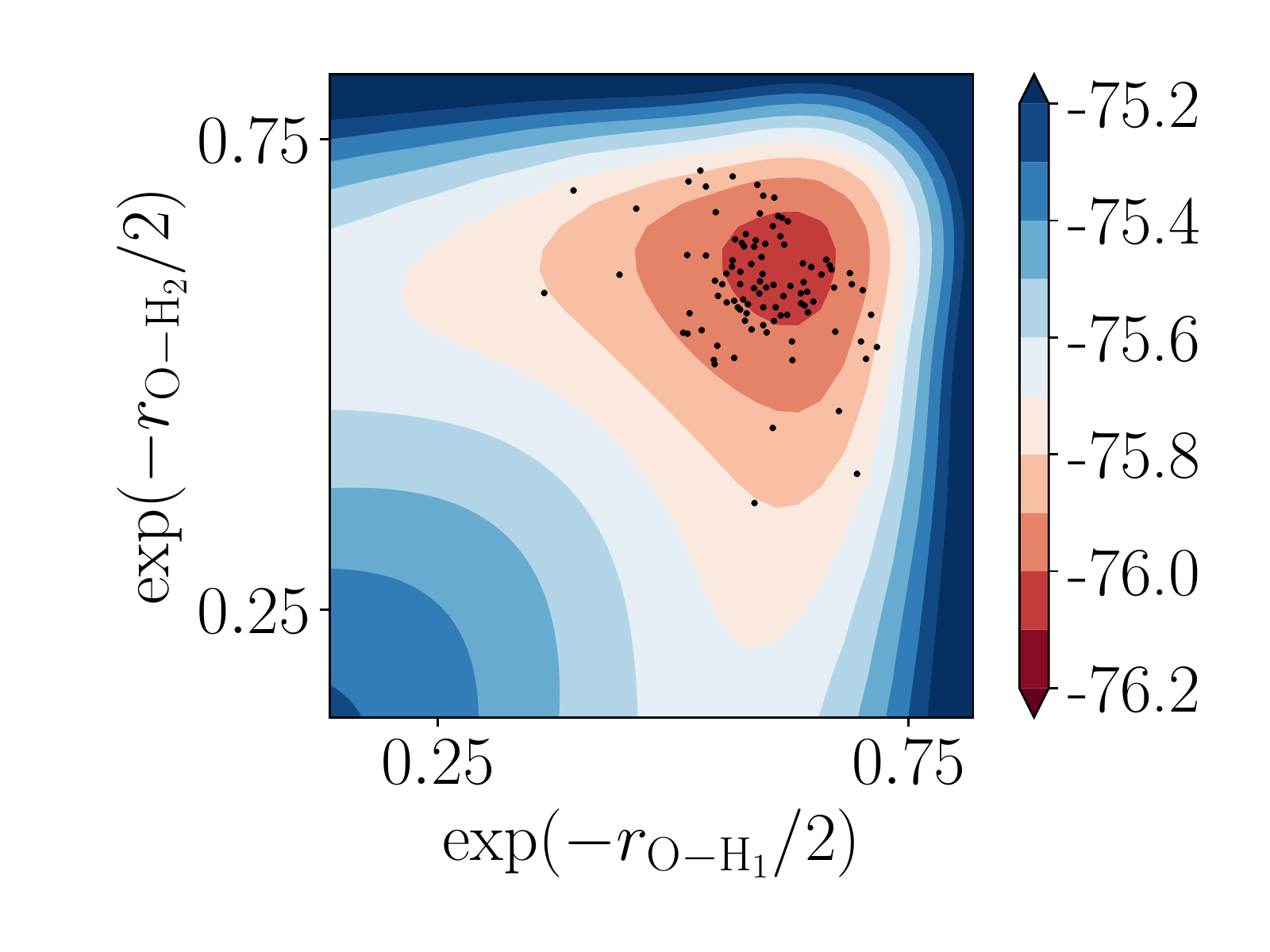}}
	\hfill
	\subfloat[$N=69$]{\includegraphics[width=0.15\textwidth]{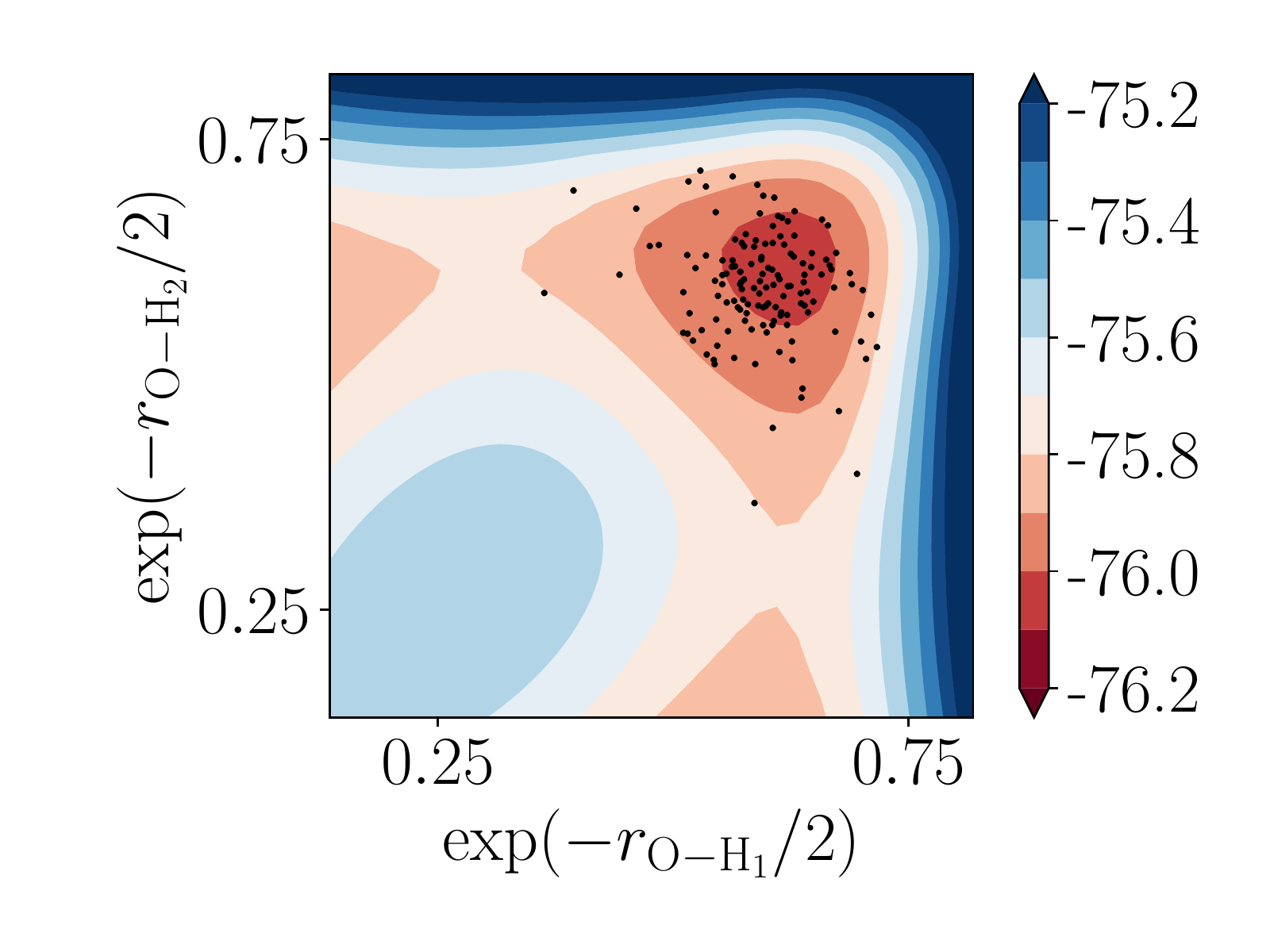}}
	\hfill
	\caption[Latent function of different Morse projections for different training data set sizes.]{Resulting PES,  projected on the Morse transformed O-H internuclear distances,  for Mat\'{e}rn kernels trained on Morse transformed spaces with parameters $\alpha=$2.0 for different training dataset sizes.}
	\label{fig:morse_n45_contours}
\end{figure}

Some latent functions of the Mat\'{e}rn GPs are plotted in figure \ref{fig:morse_n45_contours}.  Despite the length scale growing larger,  the PESs seem to strongly ``oscillate'',  as if they had a short length scale.  This is particularly seen away from the data as seen in panel (e) and (f).

If one thinks of the optimisation in a Bayesian sense,  one would expect the opposite where minima on the LML become more well-defined as new data,  if it still agrees with the hyperparameters of the model of that particular minimum,  is added. 

\subsection{Optimisable Morse kernels for changing data}
Gaussian processes trained with a MorseMat\'{e}rn ($\nu=2.5$) kernel and a MorseRBF kernel are trained on the same datasets and compared to the MAE($\alpha$) trends of figure \ref{fig:morse-different-sets}.  These results do not use the full GMIN implementation and use a basic sklearn\autocite{scikit-learn} L-BFGS approach.  All reported minima have projected gradients converged to 10$^{-2}$,  which is not as tight a convergence criterion as LML minima that were found using the GMIN implementation.  Table \ref{tab:dataset-morserbf} summaries the results for both the Morse-transformed kernels with the training data ranging from the initial set to the fully ``merged'' one in increments of 5 data points.

\end{multicols}
\vspace{0.4cm}
\addcontentsline{lof}{figure}{T 5.1  Summary of hyperparameters and MAEs for MorseRBF Gaussian processes.}
\begin{longtable}{@{\extracolsep{\fill}} c c c c c }
\toprule
$N$ & MorseRBF: $\alpha$ &  MorseRBF: $\rho_0=\rho_1$ &  MorseMat\'{e}rn: $\alpha$ &  MorseMat\'{e}rn: $\rho_0=\rho_1$ \\
\midrule \vspace{2pt}
 29 &  0.534 &  0.41  &   0.580 &   2.08  \\[0.4cm]
 34 &  0.774 &  0.62  &   0.960 &   0.17  \\[0.4cm]
 39 &  0.804 &  0.39  &   0.804 &   0.86  \\[0.4cm]
 44 &  0.937 &  0.22  &   \color{green}0.514 &   \color{green}3.05  \\[0.4cm]
 49 &  0.802 &  0.18  &   \color{green}0.539 &   \color{green}0.88  \\[0.4cm]
 54 &  0.806 &  0.12  &   \color{green}0.569 &   \color{green}0.55  \\[0.4cm]
 59 &  0.440 &  0.42  &   \color{green}0.993 &   \color{green}2.40  \\[0.4cm]
 64 &  \color{green}0.391 & \color{green} 0.32  &   0.622 &   2.27  \\[0.4cm]
 69 &  0.949 &  0.23  &   \color{green}0.448 &   \color{green}0.15  \\[0.4cm]
 74 &  0.750 &  0.50  &   0.525 &   0.72  \\[0.4cm]
 79 &  0.251 &  0.12  &   0.721 &   0.30  \\[0.4cm]
\midrule[\heavyrulewidth]
\caption{Summary of Morse parameters and length scales for optimised Morse kernels with an increasing size of training set (number of samples given by $N$).  When the Morse kernel performs better than its ``standard'' kernel counterpart,  the values are written in green.  One can see that the improvement is very rarely seen for the RBF kernel side and only seen about half the time for the kernel Mat\'{e}rn.}
\label{tab:dataset-morserbf}
\end{longtable}
\begin{multicols}{2}

There does not seem to be a smooth transition as data is added in terms of an optimal $\alpha$ and there does not seem to a consistently better method for choosing the Morse parameters in the Bayesian GP framework.
\par
Since $N$,  the size of the training set,  is not a continuous variable that changes the LML smoothly,  it is not necessary that the change in $\alpha$ is smooth.  One expects the $N$ parameter to produce smooth changes of the LML minima.  It is clear that MorseMat\'{e}rn kernels,  like the MorseRBF kernels,  tend to favour small $\alpha$ values when optimised through the LML but it is hard to understand the reason for the progression seen in the tables above.  These optimal values are not similar to the usual values used for Morse projections in the literature but it does not in terms of MAE discredit those choices.

\section{Conclusion}
Optimising,  with a Bayesian approach,  the feature space of the GP to produce more performant latent functions,  in terms of MAE,  is not straight forward.  The LML is made more complex by the additional DOF and there is a strong correlation between the hyperparameters.  Different transforms might not necessarily suffer from this but,  for the Morse transform,  the relation between the Morse parameter and the length scales is evident. 
\par
For the Morse transform,  since the limit of its parameters tending to a certain value ($\alpha \to \infty$ here) give a linear transform,  optimising the transform parameters can also inform us on the ``usefulness'' of the transform.  The curves of figure \ref{fig:MAE-vs-none},  for example,  show that the transform is actually producing less performant GP models for this system.
\par
The ``best-fit'' approach also produced some interesting results regarding the choice of testing set to produce the MAE curves one minimises.  A target function is assumed to have an optimal Morse parameter to project it to a ``simpler'' surface\footnote{This is not guaranteed to a GP easier to train on that surface.}.  However,  a different testing set can significantly affect the result of the minimisation (this cannot be interpreted in the Bayesian approach since there is no testing set in the LML minimisation).  This should not be the case if the testing set is ``complete'',  in the sense that new samples are drawn from the same distribution.  It will be the case if the distribution change\footnote{In the results presented in figure \ref{fig:morse-different-sets},  it was the temperature of the Boltzmann distribution that changed between the distributions. } which suggests that one should use testing data that is suitable for the intended use of the GP model. 

\section*{Acknowledgments}
I would like to thank the Royal Society for funding as well as the Wales group of the University of Cambridge for providing access to the GMIN suite\autocite{GMIN}.  Moreover,  I would like to thank Angelos Michaelides and Albert Partay Bart\`{o}k for fruitful discussion during my PhD viva that improved this work.

\printbibliography
\end{multicols}
\end{document}